\documentclass{article}

\usepackage{amssymb,amsfonts,amsmath}
\usepackage{cite,enumerate,float}
\usepackage{url}
\usepackage{color}
\usepackage{tikz}
\usetikzlibrary{arrows,backgrounds}

\usepackage[utf8]{inputenc}
\usepackage[english]{babel}

\def\be{\begin{eqnarray}}
\def\ee{\end{eqnarray}}
\allowdisplaybreaks[1]
\def\nn{\nonumber}

\def\tr{{\rm tr}}

\def\CR{{\cal R}}
\def\CK{{\cal K}}
\def\CH{{\cal H}}
\def\Crb{{\rm Cab}_r(\beta)}

\definecolor{red}{rgb}{1,0,0}
\definecolor{orange}{rgb}{1,0.5,0}
\definecolor{violet}{rgb}{0.7,0,1}

\hoffset= - 1.0in         

\usepackage{mathtools,xcolor,array}
\usetikzlibrary{arrows.meta,calc,decorations.markings,decorations.pathreplacing}
\newcommand{\cl}{\operatorname{cl}}
\newcommand{\HH}{\mathsf H}
\definecolor{strand}{RGB}{27,79,113}
\definecolor{accent}{RGB}{164,69,43}
\definecolor{quiet}{RGB}{103,112,119}
\tikzset{wire/.style={draw=strand,line width=.85pt,line cap=round},
  boundary/.style={draw=quiet!55,densely dashed,line width=.45pt},
  lab/.style={font=\small},
  arr/.style={-{Stealth[length=1.8mm]},line width=.6pt},
  flow/.style={postaction={decorate},decoration={markings,
    mark=at position .68 with {\arrow{Stealth[length=1.5mm]}}}}}
\newcommand{\braidpic}[3]{%
  \foreach \j in {1,...,#1}{\draw[wire] ({.48*(\j-1)},0)--({.48*(\j-1)},#3);}
  \foreach \g [count=\k] in {#2}{%
    \pgfmathtruncatemacro{\gi}{abs(\g)}
    \pgfmathsetmacro{\xx}{.48*(\gi-1)}
    \pgfmathsetmacro{\yy}{.32*\k-.14}
    \draw[white,line width=2.2pt] (\xx,\yy-.15)--(\xx,\yy+.15);
    \draw[white,line width=2.2pt] (\xx+.48,\yy-.15)--(\xx+.48,\yy+.15);
    \ifnum\g>0
      \draw[wire] (\xx+.48,\yy-.15)--(\xx,\yy+.15);
      \draw[white,line width=3.4pt] (\xx,\yy-.15)--(\xx+.48,\yy+.15);
      \draw[wire] (\xx,\yy-.15)--(\xx+.48,\yy+.15);
    \else
      \draw[wire] (\xx,\yy-.15)--(\xx+.48,\yy+.15);
      \draw[white,line width=3.4pt] (\xx+.48,\yy-.15)--(\xx,\yy+.15);
      \draw[wire] (\xx+.48,\yy-.15)--(\xx,\yy+.15);
    \fi
  }
}

\newcommand{\braidpics}[3]{%
  \foreach \j in {1,...,#1}{\draw[wire] ({.48*(\j-1)},0)--({.48*(\j-1)},#3);}
  \foreach \g [count=\k] in {#2}{%
    \pgfmathtruncatemacro{\gi}{abs(\g)}
    \pgfmathsetmacro{\xx}{.48*(\gi-1)}
    \pgfmathsetmacro{\yy}{.22*\k-.09}
    \draw[white,line width=2.2pt] (\xx,\yy-.11)--(\xx,\yy+.11);
    \draw[white,line width=2.2pt] (\xx+.48,\yy-.11)--(\xx+.48,\yy+.11);
    \ifnum\g>0
      \draw[wire] (\xx+.48,\yy-.11)--(\xx,\yy+.11);
      \draw[white,line width=3pt] (\xx,\yy-.11)--(\xx+.48,\yy+.11);
      \draw[wire] (\xx,\yy-.11)--(\xx+.48,\yy+.11);
    \else
      \draw[wire] (\xx,\yy-.11)--(\xx+.48,\yy+.11);
      \draw[white,line width=3pt] (\xx+.48,\yy-.11)--(\xx,\yy+.11);
      \draw[wire] (\xx+.48,\yy-.11)--(\xx,\yy+.11);
    \fi
  }
}
\newcommand{\xterm}[2]{%
  \pgfmathsetmacro{\H}{1.6}
  \foreach \k in {1,...,#1}{\ifnum\k<#1 \draw[wire] ({.5*(\k-1)},0)--({.5*\k},\H);\fi}
  \draw[white,line width=3.4pt] ({.5*(#1-1)},0)--(0,\H);
  \draw[wire,draw=accent] ({.5*(#1-1)},0)--(0,\H);
  \foreach \k in {1,...,#1}{\ifnum\k<#1 \ifnum\k>#2 \else
    \pgfmathsetmacro{\t}{(#1-\k)/#1}
    \pgfmathsetmacro{\xa}{.5*(\k-1)+.5*(\t-.09)} \pgfmathsetmacro{\ya}{\H*(\t-.09)}
    \pgfmathsetmacro{\xb}{.5*(\k-1)+.5*(\t+.09)} \pgfmathsetmacro{\yb}{\H*(\t+.09)}
    \draw[white,line width=3.4pt] (\xa,\ya)--(\xb,\yb); \draw[wire] (\xa,\ya)--(\xb,\yb);
  \fi\fi}
}
\newcommand{\cabpic}[1]{%
  \foreach \x in {0,.5,1.5,2}{\draw[wire] (\x,-.3)--(\x,3.6);}
  \draw[fill=strand!9,draw=strand] (-.25,1.2) rectangle (2.25,2.1); \node[font=\small] at (1,1.65) {${\rm Cab}_2(\beta)$};
  \foreach \e/\y/\l in {#1}{%
    \pgfmathsetmacro{\xl}{1.5*(\e-1)-.2}
    \draw[fill=white,draw=strand] (\xl,\y) rectangle (\xl+.9,\y+.45); \node[font=\small] at (\xl+.45,\y+.22) {$\l$};}
  \foreach \x/\d in {2/.25,1.5/.5,.5/.75,0/1}{
   \draw[wire] (\x,3.6) -- (\x,{3.6+\d}) arc (180:90:.3) -- ({2.5+\d},{3.9+\d}) arc (90:0:.3) -- (2.8+\d,{3.6+\d})
     -- (2.8+\d,{-.3-\d}) arc (0:-90:.3) -- ({\x+.3},{-.6-\d}) arc (-90:-180:.3) -- (\x,-.3);}
}

\newcommand{\closearcs}[2]{%
  \foreach \j in {1,...,#1}{
    \pgfmathsetmacro{\x}{.5*(\j-1)}
    \pgfmathsetmacro{\d}{.25*(#1-\j+1)}
    \draw[wire] (\x,#2) -- (\x,{#2+\d}) arc (180:90:.25) -- ({.5*(#1-1)+.3+\d},{#2+\d+.25}) arc (90:0:.25)
      -- ({.5*(#1-1)+.55+\d},{#2+\d}) -- ({.5*(#1-1)+.55+\d},{-\d}) arc (0:-90:.25) -- ({\x+.25},{-\d-.25}) arc (-90:-180:.25) -- (\x,0);}
}

\begin{document}

\centerline{\Large{What a knot sees at a root of unity}}

\bigskip

\centerline{\bf Ya.Kononov and A.Morozov$^*$}

\bigskip

\centerline{$^*$ITEP, Moscow, Russia}

\bigskip

\centerline{ABSTRACT}

\bigskip

{\footnotesize
HOMFLY-PT polynomials describe analytic continuation of Wilson averages in Chern-Simons theory
to arbitrary complex values of two independent parameters $q$ and $A$.
They also depend on the knot and representation of the gauge group.
However, when $q^2$ is a primitive $m$-th root of unity, e.g. $q=\pm e^{i\pi/m}$,
these polynomials exhibit a universal (knot-independent) factorization:
$H_R$ becomes the product of the polynomials for the $m$-core of the Young diagram $R$
and of one and the same factor $H_{[m]}$ for every $m$-ribbon of $R$.
In all the examples which we computed this factor is just the special polynomial at $A^m$.
At $q=\pm1$ factorization follows from the singularity of the cabled HOMFLY,
which is the same for all representations.
At $m>1$ the relevant singularity is in the projector, and what survives of it is a finite sum of braids $X_m$ --
the $m$-th power sum inside the cable, which can be freely moved through the other strands.
For the proof we use the Murnaghan-Nakayama rule and the Adams operation.
Discovery of this hidden structure explains the old factorization puzzle.
}

\section{Introduction}

$3d$ Chern-Simons theory \cite{CS74,Wit89,MS10} provides the simplest non-trivial example of Yang-Mills theory,
which is exactly solvable and where one can reliably address the most puzzling physical problem.
Wilson averages in this case are also known as knot polynomials,
and they depend polynomially (!)
on peculiar non-perturbative variables $q=e^{\frac{2\pi i}{k+N}}$ and $A=q^N$.
Most interesting is their dependence on representation $R$ --
e.g. in $4d$ we expect deconfinement for representations from adjoint sector
(presumably governed by the universal Vogel theory \cite{Vog99,Vog11,MMM16,KMS25}),
and confinement otherwise.
As to knot polynomials, in the weak-coupling limit $q=1$
they become nearly independent of the representation $R$, see (\ref{spefact}) below.
But one can go even further -- factorization property (\ref{spefact})
has a mysterious generalization to {\it integer}\,(!) inverse couplings $k$,
see (\ref{Rpm}) and (\ref{RMfact}).
Connection of this exact result in $3d$ and intuitive expectations in $4d$ is still unknown.

Meanwhile, we address it in $3d$, where the problem also did not attract enough attention.
We reduce factorization statements to a three-step procedure, illustrated by a simple example.
The general proof is in sec.\ref{proofs}.

\bigskip

\noindent{\bf The statement.}
As noted in \cite{DMMSS13}, for the reduced $R$-colored HOMFLY of any knot ${\cal K}$
\be
H^{\cal K}_R(A,q=1) =  \left(H^{\cal K}_{\Box}(A,q=1)\right)^{|R|}
\label{spefact}
\ee
i.e. the colored {\it special} HOMFLY are factorized and all reduce just to (powers of) the fundamental one.

At roots of unity, when $q^2$ is a primitive $m$-th root of unity, $m>1$
(we denote such $q$ by $\zeta$; the basic example is $\zeta=e^{i\pi/m}$), the following holds
(observed in \cite{KM15} and proved there for a few families of representations; for general $R$ it was a conjecture, which we prove
in sec.\ref{proofs}):
\be
\left.H_{R}^{\cal K} = H_{[m]}^{\cal K}\cdot H_{R-M}^{\cal K}\right|_{q^{2m}=1}
\ \ \ \ \ \forall \ \text{ connected skew diagram}\ M \ \text{of width one (no $2\times2$ block) with}\ |M|=m
\label{Rpm}
\ee

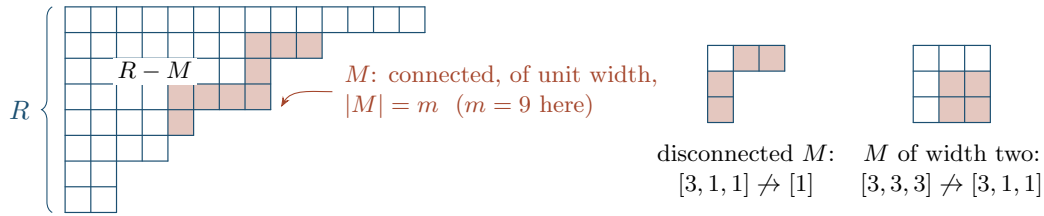
\begin{figure}[h]\centering
\begin{tikzpicture}[x=.34cm,y=.34cm]
\tikzset{grid/.style={draw=strand,line width=.45pt,line cap=round}}
\begin{scope}
\fill[accent!30] (7,-1) rectangle (8,-2);
\fill[accent!30] (8,-1) rectangle (9,-2);
\fill[accent!30] (9,-1) rectangle (10,-2);
\fill[accent!30] (7,-2) rectangle (8,-3);
\fill[accent!30] (4,-3) rectangle (5,-4);
\fill[accent!30] (5,-3) rectangle (6,-4);
\fill[accent!30] (6,-3) rectangle (7,-4);
\fill[accent!30] (7,-3) rectangle (8,-4);
\fill[accent!30] (4,-4) rectangle (5,-5);
\draw[grid] (0,0)--(14,0);
\draw[grid] (0,-1)--(14,-1);
\draw[grid] (0,-2)--(10,-2);
\draw[grid] (0,-3)--(8,-3);
\draw[grid] (0,-4)--(8,-4);
\draw[grid] (0,-5)--(5,-5);
\draw[grid] (0,-6)--(4,-6);
\draw[grid] (0,-7)--(2,-7);
\draw[grid] (0,-8)--(2,-8);
\draw[grid] (0,0)--(0,-8);
\draw[grid] (1,0)--(1,-8);
\draw[grid] (2,0)--(2,-8);
\draw[grid] (3,0)--(3,-6);
\draw[grid] (4,0)--(4,-6);
\draw[grid] (5,0)--(5,-5);
\draw[grid] (6,0)--(6,-4);
\draw[grid] (7,0)--(7,-4);
\draw[grid] (8,0)--(8,-4);
\draw[grid] (9,0)--(9,-2);
\draw[grid] (10,0)--(10,-2);
\draw[grid] (11,0)--(11,-1);
\draw[grid] (12,0)--(12,-1);
\draw[grid] (13,0)--(13,-1);
\draw[grid] (14,0)--(14,-1);
\draw[decorate,decoration={brace,amplitude=4pt},strand] (-.5,-8) -- (-.5,0) node[midway,left=5pt] {$R$};
\node[font=\small,fill=white,inner sep=1.5pt] at (3.5,-2.5) {$R-M$};
\node[font=\small,accent,anchor=west,align=left] at (10.5,-3.3) {$M$: connected, of unit width,\\ $|M|=m$ \ ($m=9$ here)};
\draw[accent,-{Stealth[length=4pt]}] (10.4,-3.3) to[out=180,in=60] (8.4,-3.9);
\end{scope}
\begin{scope}[shift={(25,-1.5)}]
\fill[accent!30] (1,0) rectangle (2,-1);
\fill[accent!30] (2,0) rectangle (3,-1);
\fill[accent!30] (0,-1) rectangle (1,-2);
\fill[accent!30] (0,-2) rectangle (1,-3);
\draw[grid] (0,0)--(3,0);
\draw[grid] (0,-1)--(3,-1);
\draw[grid] (0,-2)--(1,-2);
\draw[grid] (0,-3)--(1,-3);
\draw[grid] (0,0)--(0,-3);
\draw[grid] (1,0)--(1,-3);
\draw[grid] (2,0)--(2,-1);
\draw[grid] (3,0)--(3,-1);
\node[font=\small,anchor=north,align=center] at (1.5,-3.5) {disconnected $M$:\\ $[3,1,1]\not\to[1]$};
\end{scope}
\begin{scope}[shift={(33,-1.5)}]
\fill[accent!30] (1,-1) rectangle (2,-2);
\fill[accent!30] (2,-1) rectangle (3,-2);
\fill[accent!30] (1,-2) rectangle (2,-3);
\fill[accent!30] (2,-2) rectangle (3,-3);
\draw[grid] (0,0)--(3,0);
\draw[grid] (0,-1)--(3,-1);
\draw[grid] (0,-2)--(3,-2);
\draw[grid] (0,-3)--(3,-3);
\draw[grid] (0,0)--(0,-3);
\draw[grid] (1,0)--(1,-3);
\draw[grid] (2,0)--(2,-3);
\draw[grid] (3,0)--(3,-3);
\node[font=\small,anchor=north,align=center] at (1.5,-3.5) {$M$ of width two:\\ $[3,3,3]\not\to[3,1,1]$};
\end{scope}
\end{tikzpicture}
\caption{\footnotesize
The ribbon $M$ of the size $m$ at the ``boundary'' of Young diagram $R$,
also known as ``$m$-border strip'' \cite{Mur37,Nak40,Mac95,JK81}.
It should be of width one and connected, also $R-M$, obtained by removal of $M$, should
remain Young diagram. The two small pictures show what is {\it not} allowed (the counterexamples of the footnote).
Factorization property (\ref{RMfact}) includes the case when many such ribbons
are selected one after another (slice by slice) at the border of $R$.
}
\label{ribbon}
\end{figure}

Here $R$ is obtained from $R-M$ by adding a connected one-slice ribbon (border strip) $M$, as shown in Fig.\ref{ribbon}.
In (\ref{Rpm}) it is enough that $q^{2m}=1$; below we mostly assume that $q^2$ is a {\it primitive} $m$-th root of unity,
the non-primitive case is reduced to it in the Remark at the end of sec.\ref{proofs}:\footnote{
Both conditions in (\ref{Rpm}) are necessary.
The simple counterexamples are:

\bigskip

$\bullet$ at $q^8=1$: $H_{[311]}\neq H_{[4]} H_{[1]}$ and $H_{[311]}\neq H_{[2]}^2H_{[1]}$ --
connectedness is indeed needed
and

$\bullet$ at $q^8=1$: $H_{[333]}\neq H_{[4]}H_{[311]}$ and $H_{[333]}\neq H_{[22]}H_{[311]}$ --
width $2$ is too much. Note that $[3,3,3]$ does have removable $4$-ribbons of width one: $[3,3,3]\to[2,2,1]\to[1]$,
and accordingly $H_{[333]}=H_{[4]}^2H_{[1]}$ at $q^8=1$.
}

Repeating the procedure, we can remove $w$ slices (ribbons) of the size $m$ each,
leaving what we call a {\it core}.
Then
\be
\boxed{
\left.H^{\cal K}_{R}   =
\left(H_{[m]}^{\cal K}\right)^{w}  \cdot H_{{\rm core}_m(R)}^{\cal K}\ \right|_{q=\zeta}
}
\ \ \ \ \ \ \ {\rm in\ particular}\ \ \ \ \
\left.H^{\cal K}_{[wm+b]}  = \left(H_{[m]}^{\cal K}\right)^w H_{[b]}^{\cal K}\ \right|_{q=\zeta},\ \ \ 0\le b<m
\label{RMfact}
\ee
The result does not depend on the order of removals.
Moreover, the factor $H_{[m]}$ itself does not seem to be a new object: as already noted in \cite{KM15} for single-hook diagrams,
at $q=\zeta$ it coincides with the {\it special} polynomial (\ref{spefact}) with $A$ replaced by $A^m$,
\be
\left.H_{[m]}^{\cal K}(A,q)\right|_{q=\zeta}=\sigma^{\cal K}(A^m),\ \ \ \ \ \ \ \sigma^{\cal K}(A):=H^{\cal K}_\Box(A,q=1),
\nn\\
{\rm so\ that}\ \ \ \ \ \
\left.H^{\cal K}_{R}\right|_{q=\zeta}=\sigma^{\cal K}(A^m)^{w}\cdot \left.H_{{\rm core}_m(R)}^{\cal K}\right|_{q=\zeta}.
\label{sigmam}
\ee
In other words, every ribbon contributes just the special polynomial at $A^m$, and the whole $q$-dependence is in the core.
For example, for the trefoil $\sigma(A)=2A^{-2}-A^{-4}$, and indeed
$H_{[2]}(A,i)=2A^{-4}-A^{-8}$, $H_{[3]}(A,e^{i\pi/3})=2A^{-6}-A^{-12}$.
We checked (\ref{sigmam}) for the trefoil up to $m=6$, for the figure-eight up to $m=5$, for $5_1$ and $5_2$ up to $m=4$
and for $8_{19}$ and $9_{34}$ up to $m=3$,
but we do not have a proof. Note that for generic $R$ there are still infinitely many different polynomials at $q=\zeta$ --
one for every $m$-core, e.g. for $[1],[2,1],[3,2,1],\ldots$ at $m=2$.

A remark on the physical point of Chern-Simons theory, $A=q^N$. There $q^{2m}=1$ implies $A^{2m}=1$, i.e. $\{A^m\}=0$: the
quantum dimension of the band, $\{A^m\}/\{q\}$, vanishes, the normalization in (\ref{RMfact}) degenerates ($0/0$), and
$\sigma(A^m)=\sigma(\pm1)=1$. Thus (\ref{RMfact}) and (\ref{sigmam}) are statements about the analytic continuation in $A$, at generic $A$
and $q=\zeta$; at the physical point itself they say that the ribbons drop out -- the familiar decoupling, in Chern-Simons theory at
integer level, of the representations with vanishing quantum dimension. Everywhere below $A$ is generic.

\section{Notation, conventions and the three steps \label{not}}

We will use some standard abbreviations, usual for consideration of quantum algebras:\\
$\{x\}:=x-x^{-1}$,  \ \ \  $z:=\{q\}$, \ \ \ $D:=\frac{\{A\}}{\{q\}}, \ \ \
[n]:=\frac{\{q^n\}}{\{q\}}$, \ \ \ $\zeta$ is any $q$ with $q^2$ a primitive $m$-th root of unity, $m>1$ (e.g. $\zeta=e^{i\pi/m}$); the case $m=1$, $q=\pm1$, is treated separately in sec.\ref{z0}. Then $[m]_\zeta=0$, while $z_\zeta\neq 0$ and $[j]_\zeta\neq0$ unless $j$ is divisible by $m$.

We also need and use some basic facts about skein theory \cite{Mor02a,AM98,Ais97,Luk05,Zhu13,Zhu23}, which we briefly summarize now.

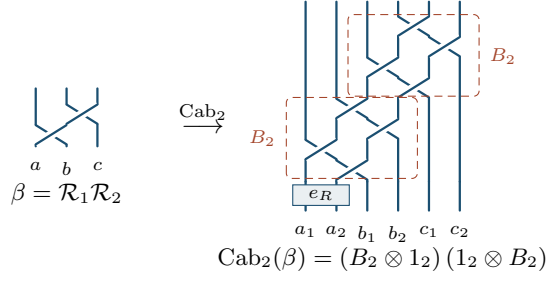
\begin{figure}[!t]\centering
\begin{tikzpicture}[scale=.85]
\begin{scope}
\braidpic{3}{1,2}{.9}
\foreach \x/\l in {0/a,.48/b,.96/c}{\node[font=\scriptsize,below] at (\x,-.1) {$\l$};}
\node[font=\small] at (.48,-.75) {$\beta=\CR_1\CR_2$};
\end{scope}
\node at (2.6,.45) {$\stackrel{{\rm Cab}_2}{\longrightarrow}$};
\begin{scope}[shift={(4.2,-1)}]
\foreach \j in {1,...,6}{\draw[wire] ({.48*(\j-1)},0)--({.48*(\j-1)},.45);}
\draw[fill=strand!9,draw=strand] (-.2,.08) rectangle (.68,.4); \node[font=\scriptsize] at (.24,.24) {$e_R$};
\begin{scope}[shift={(0,.45)}]\braidpic{6}{2,1,3,2,4,3,5,4}{2.8}\end{scope}
\foreach \x/\l in {0/a_1,.48/a_2,.96/b_1,1.44/b_2,1.92/c_1,2.4/c_2}{\node[font=\scriptsize,below] at (\x,-.1) {$\l$};}
\draw[accent,densely dashed,rounded corners=2pt] (-.3,.5) rectangle (1.74,1.76); \node[font=\scriptsize,accent,left] at (-.32,1.13) {$B_2$};
\draw[accent,densely dashed,rounded corners=2pt] (.66,1.78) rectangle (2.7,3.05); \node[font=\scriptsize,accent,right] at (2.72,2.4) {$B_2$};
\node[font=\small] at (1.2,-.75) {${\rm Cab}_2(\beta)=(B_2\otimes1_2)\,(1_2\otimes B_2)$};
\end{scope}
\end{tikzpicture}
\caption{\footnotesize Cabling: every strand of $\beta$ is replaced by $r$ parallel strands and every crossing by the $r\times r$ block $B_r$
(here $r=2$, $s=3$); the projector $e_R$ is inserted on one of the cabled strands.}
\label{fig:cabling}
\end{figure}

\begin{figure}[!t]\centering
\begin{tikzpicture}[scale=.76]
\begin{scope}
\braidpic{4}{2,1,3,2}{1.5}
\node[font=\scriptsize,below] at (0,-.1) {$a_1$}; \node[font=\scriptsize,below] at (.48,-.1) {$a_2$};
\node[font=\scriptsize,below] at (.96,-.1) {$b_1$}; \node[font=\scriptsize,below] at (1.44,-.1) {$b_2$};
\node[font=\scriptsize,above] at (0,1.6) {$b_1$}; \node[font=\scriptsize,above] at (.48,1.6) {$b_2$};
\node[font=\scriptsize,above] at (.96,1.6) {$a_1$}; \node[font=\scriptsize,above] at (1.44,1.6) {$a_2$};
\node[lab] at (.72,-.75) {$B_2=\CR_2\CR_1\CR_3\CR_2$};
\end{scope}
\node at (3.2,.75) {$=$};
\begin{scope}[shift={(4.3,-.3)}]
\foreach \x in {0,.4}{\draw[wire] (\x,0)--(\x,.3);}
\foreach \x in {1.2,1.6}{\draw[wire] (\x,0)--(\x,.3);}
\draw[wire] (1.2,.3)--(0,1.8); \draw[wire] (1.6,.3)--(.4,1.8);
\draw[white,line width=5pt] (0,.3)--(1.2,1.8); \draw[white,line width=5pt] (.4,.3)--(1.6,1.8);
\draw[wire] (0,.3)--(1.2,1.8); \draw[wire] (.4,.3)--(1.6,1.8);
\foreach \x in {0,.4,1.2,1.6}{\draw[wire] (\x,1.8)--(\x,2.1);}
\node[lab] at (.8,-.6) {cabled strand $a$ over $b$};
\end{scope}
\begin{scope}[shift={(8.4,-.3)}]
\braidpics{6}{3,2,1,4,3,2,5,4,3}{2.1}
\foreach \x/\l in {0/a_1,.48/a_2,.96/a_3,1.44/b_1,1.92/b_2,2.4/b_3}{\node[font=\scriptsize,below] at (\x,-.1) {$\l$};}
\foreach \x/\l in {0/b_1,.48/b_2,.96/b_3,1.44/a_1,1.92/a_2,2.4/a_3}{\node[font=\scriptsize,above] at (\x,2.2) {$\l$};}
\node[lab,align=center,anchor=north] at (1.2,-.5) {$B_3=(\CR_3\CR_2\CR_1)\cdot$\\ $(\CR_4\CR_3\CR_2)(\CR_5\CR_4\CR_3)$};
\end{scope}
\begin{scope}[shift={(13.2,-.3)}]
\foreach \x in {0,.4}{\draw[wire] (\x,-.3)--(\x,0);}
\draw[fill=strand!9,draw=strand] (-.2,0) rectangle (.6,.45); \node[font=\small] at (.2,.22) {$Y$};
\foreach \x in {1.2,1.6}{\draw[wire] (\x,-.3)--(\x,.6);}
\foreach \x in {0,.4}{\draw[wire] (\x,.45)--(\x,.6);}
\draw[wire] (1.2,.6)--(0,1.8); \draw[wire] (1.6,.6)--(.4,1.8);
\draw[white,line width=5pt] (0,.6)--(1.2,1.8); \draw[white,line width=5pt] (.4,.6)--(1.6,1.8);
\draw[wire] (0,.6)--(1.2,1.8); \draw[wire] (.4,.6)--(1.6,1.8);
\foreach \x in {0,.4,1.2,1.6}{\draw[wire] (\x,1.8)--(\x,2.4);}
\node at (2.4,1) {$=$};
\begin{scope}[shift={(3.2,0)}]
\foreach \x in {0,.4,1.2,1.6}{\draw[wire] (\x,-.3)--(\x,.4);}
\draw[wire] (1.2,.4)--(0,1.6); \draw[wire] (1.6,.4)--(.4,1.6);
\draw[white,line width=5pt] (0,.4)--(1.2,1.6); \draw[white,line width=5pt] (.4,.4)--(1.6,1.6);
\draw[wire] (0,.4)--(1.2,1.6); \draw[wire] (.4,.4)--(1.6,1.6);
\foreach \x in {0,.4}{\draw[wire] (\x,1.6)--(\x,2.4);}
\foreach \x in {1.2,1.6}{\draw[wire] (\x,1.6)--(\x,1.8);}
\draw[fill=strand!9,draw=strand] (1,1.8) rectangle (1.8,2.25); \node[font=\small] at (1.4,2.02) {$Y$};
\foreach \x in {1.2,1.6}{\draw[wire] (\x,2.25)--(\x,2.4);}
\end{scope}
\node[lab] at (2.4,-.6) {$(Y\otimes1_r)\,B_r=B_r\,(1_r\otimes Y)$};
\end{scope}
\end{tikzpicture}
\caption{\footnotesize Left: the cable blocks (\ref{block}): $B_2$ as a braid and as ``cabled strand $a$ passes over cabled strand $b$'', and $B_3$.
Right: the block commutes with any box $Y$ on a cabled strand, (\ref{slide}).}
\label{fig:block}
\end{figure}
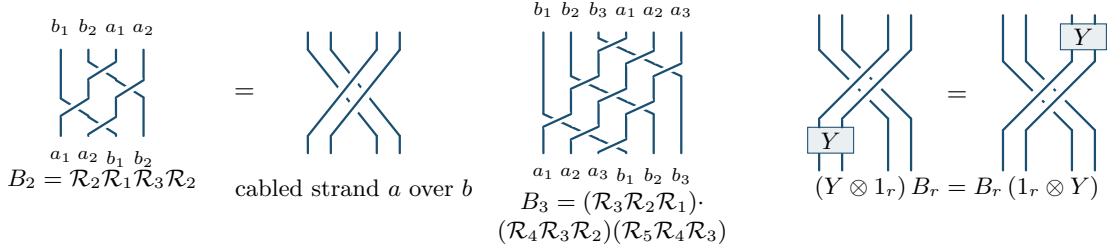

We consider a knot $\CK$, defined as a closure of a braid $\beta^\CK$ without extra crossings,
and knot polynomials, defined as $\CH^\CK_R \approx \tr_q \beta$,
with $\beta$ represented by a product of  $\CR$-matrices for a quantum group $U_q(G)$.
We will often omit index $\CK$ from $\beta^\CK$, and use notation $\CK$ and $\beta$ on equal footing.
The ``color'' $R$ refers not to $N$ in $G=sl_N$, but to representation of $G$,
and there are two possible formalisms.
In one the $\cal R$-matrices depend on $R$, in another -- which we will actually use --
the $\cal R$-matrices belong to the simplest representation (fundamental of the $sl_N$),
and instead $\beta^\CK$ is promoted to a cable $\Crb$, in which every strand of $\beta^\CK$ is replaced by $r=|R|$ parallel strands, where $|R|$ denotes the size
(number of boxes) in the Young diagram $R$, which labels irreducible finite-dimensional representations of $sl_N$.
In this formalism a projector $e_R$ is used to select the contribution of particular $R$.
The precise notation is
\be
H_R^{\cal K}=\theta_R^{-n_\beta}\,\frac{\tr_q\big((e_R\otimes 1)\,{\rm Cab}_r\beta\big)}{\tr_q\,e_R},
\ \ \ \ \ \ \ \theta_R=A^{|R|}q^{2\varkappa_R},\ \ \ \varkappa_R=\sum_{(i,j)\in R}(j-i),
\label{Hdef}
\ee
where $\beta$ has $s$ strands, its cable $\Crb$ has $rs$, and the projector acts on $r$ of the strands within the cable
and is inserted at any place in $\Crb$.
$n_\beta$ is the writhe of $\beta$, i.e. the algebraic number of its crossings,
and the cabled braid is, of course, the corresponding product of $\CR$-matrices.
Actually every crossing of original $\beta$ is substituted by a ``cable vertex'' or an $r\times r$ block,
see Figs.\ref{fig:cabling} and \ref{fig:block}:
\be
B_r:=\prod_{j=1}^{r}\Big(\CR_{r+j-1}\CR_{r+j-2}\cdots\CR_j\Big),
 \ \ {\rm e.g.} \ \ B_2=\CR_2\CR_1\CR_3\CR_2,  \ \ \ B_3=(\CR_3\CR_2\CR_1)(\CR_4\CR_3\CR_2)(\CR_5\CR_4\CR_3),
\ \ \ldots
\label{block}
\ee

\begin{figure}[H]\centering
\begin{tikzpicture}[scale=.8]
\begin{scope}
\foreach \x in {0,.5,1}{\draw[wire] (\x,0)--(\x,2.4);}
\draw[fill=strand!9,draw=strand] (-.2,.9) rectangle (.7,1.5); \node[font=\small] at (.25,1.2) {$X$};
\closearcs{3}{2.4}
\node at (3,1.2) {$=\ D\ \cdot$};
\end{scope}
\begin{scope}[shift={(4,0)}]
\foreach \x in {0,.5}{\draw[wire] (\x,0)--(\x,2.4);}
\draw[fill=strand!9,draw=strand] (-.2,.9) rectangle (.7,1.5); \node[font=\small] at (.25,1.2) {$X$};
\closearcs{2}{2.4}
\end{scope}
\begin{scope}[shift={(8.5,0)}]
\foreach \x in {0,.5,1}{\draw[wire] (\x,0)--(\x,.4);}
\draw[fill=strand!9,draw=strand] (-.2,.4) rectangle (.7,.9); \node[font=\small] at (.25,.65) {$X$};
\draw[wire] (0,.9)--(0,2.4); \draw[wire] (1,.4)--(1,1.1); \draw[wire] (.5,.9)--(.5,1.1);
\draw[wire] (1,1.1)--(.5,1.6); \draw[white,line width=3pt] (.5,1.1)--(1,1.6); \draw[wire] (.5,1.1)--(1,1.6);
\draw[wire] (1,1.6)--(1,2.4); \draw[wire] (.5,1.6)--(.5,1.8);
\draw[fill=strand!9,draw=strand] (-.2,1.8) rectangle (.7,2.3); \node[font=\small] at (.25,2.05) {$Y$};
\draw[wire] (.5,2.3)--(.5,2.4);
\closearcs{3}{2.4}
\node at (3,1.2) {$=\ A\ \cdot$};
\end{scope}
\begin{scope}[shift={(12.5,0)}]
\foreach \x in {0,.5}{\draw[wire] (\x,0)--(\x,2.4);}
\draw[fill=strand!9,draw=strand] (-.2,.5) rectangle (.7,1); \node[font=\small] at (.25,.75) {$X$};
\draw[fill=strand!9,draw=strand] (-.2,1.5) rectangle (.7,2); \node[font=\small] at (.25,1.75) {$Y$};
\closearcs{2}{2.4}
\end{scope}
\node[lab] at (2,-1.3) {$\tr_q^{(3)}(X)=D\,\tr_q^{(2)}(X)$};
\node[lab] at (10.5,-1.3) {$\tr_q^{(3)}(X\CR_2Y)=A\,\tr_q^{(2)}(XY)$};
\end{tikzpicture}
\caption{\footnotesize The two rules (\ref{Markov}) for $s=2$. Left: the third strand is not crossed by anything and closes
to a separate circle. Right: the third strand crosses strand 2 once and closes into a curl on it; the curl is removed
at the price of $A$.}
\label{fig:markov}
\end{figure}
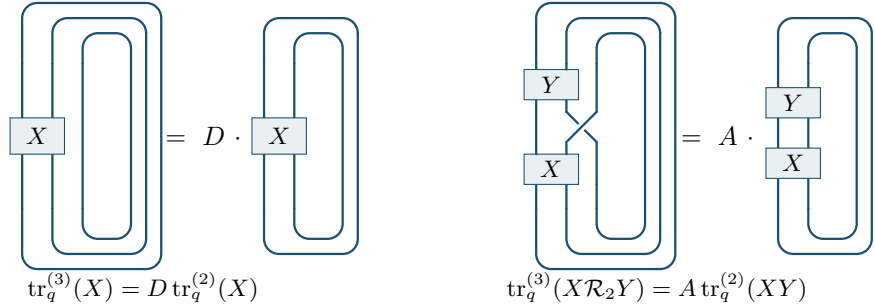

$H_R^\beta$ actually is a topological invariant and depends only on knot $\CK$,
and not on the choice of $\beta^\CK$, $s$ and the position of the projector $e_R$.
This is guaranteed by the properties of $\CR$-matrices:

\begin{itemize}
\item{}
Braid relations:
\be
\CR_i\CR_{i+1}\CR_i=\CR_{i+1}\CR_i\CR_{i+1}
\ \ {\rm and} \ \ \CR_i\CR_j=\CR_j\CR_i \ \ {\rm for} \ \ |i-j|>1,
\label{YB}
\ee
i.e. the statement that isotopic braids are equal.
In particular, any box $Y$ on a cabled strand can be pushed through a cable vertex (Fig.\ref{fig:block}, right):
\be
(Y\otimes 1_r)\,B_r\ =\ B_r\,(1_r\otimes Y).
\label{slide}
\ee

\item{}
Consistency with the quantum trace (the closure of an $s$-strand braid).
{\bf Quantum trace} $\tr_q^{(s)}$  is fixed by normalization conditions
\be
\tr_q^{(s+1)} X = D\cdot \tr_q^{(s)} X, \ \ \ \ \
\tr_q^{(s+1)} (X\CR_s Y) = A\cdot \tr_q^{(s)} (XY), \ \ \ \ \ X,Y\ {\rm on}\ s\ {\rm strands}
\label{Markov}
\ee
i.e. a free strand gives $D$ and a curl gives $A$, see Fig.\ref{fig:markov}.
Quantum trace is cyclic,  i.e.
\be
\tr_q(XY)=\tr_q(YX)
\label{xyclic}
\ee
for {\it any} two braids or combinations of braids $X,Y$ on the same $n$ strands.
It means that the closure has no beginning.
Also it is multiplicative for braids on disjoint strands, i.e.
if a braid $X$ on $n$ strands and a braid $Y$ on $m$ strands are placed side by side, then
\be
\tr_q^{(n+m)}(X\otimes Y)=\tr_q^{(n)}(X)\cdot\tr_q^{(m)}(Y),
\label{multr}
\ee
\begin{center}
\begin{tikzpicture}[scale=.5]
\begin{scope}
\foreach \x in {0,.5,1,1.5,2}{\draw[wire] (\x,0)--(\x,2.4);}
\draw[fill=strand!9,draw=strand] (-.2,.9) rectangle (1.2,1.5); \node[font=\small] at (.5,1.2) {$X$};
\draw[fill=strand!9,draw=strand] (1.3,.9) rectangle (2.2,1.5); \node[font=\small] at (1.75,1.2) {$Y$};
\closearcs{5}{2.4}
\node at (4.6,1.2) {$=$};
\end{scope}
\begin{scope}[shift={(5.6,0)}]
\foreach \x in {0,.5,1}{\draw[wire] (\x,0)--(\x,2.4);}
\draw[fill=strand!9,draw=strand] (-.2,.9) rectangle (1.2,1.5); \node[font=\small] at (.5,1.2) {$X$};
\closearcs{3}{2.4}
\node at (3.3,1.2) {$\times$};
\end{scope}
\begin{scope}[shift={(9.6,0)}]
\foreach \x in {0,.5}{\draw[wire] (\x,0)--(\x,2.4);}
\draw[fill=strand!9,draw=strand] (-.2,.9) rectangle (.7,1.5); \node[font=\small] at (.25,1.2) {$Y$};
\closearcs{2}{2.4}
\end{scope}
\end{tikzpicture}
\end{center}
i.e. the closure of two braids placed side by side is a disjoint union of their closures.
A word in the generators $\CR_i$, or a linear combination of words, like $X_2=2\CR_1-z$ below, is an element of the braid algebra on any number $n\ge i+1$ of strands,
acting on the first ones -- the superscript in $\tr_q^{(n)}$ then shows $n$, e.g. $\tr_q^{(3)}(X_2)$.
For boxes like $Y$ or $e_R$ we always write the identity factor explicitly, $Y\otimes1$.
There are no denominators in these rules, therefore the traces can be calculated right at $q=\zeta$.
Note that (\ref{xyclic}) only moves a box along the same strand through the closure; to pass it through a crossing one needs (\ref{YB}).

\item{} Skein relation:
\be
\CR - \CR^{-1} = z:=\{q\}=q-q^{-1}, \ \ \ {\rm i.e.}\ \ \CR^2=1+z\CR
\label{skein}
\ee
It is the main advantage of the cabling method, it holds only for the fundamental representation
of $G=sl_N$.

\end{itemize}

The {\it dis}advantage of cabling is that the cabled knot is not just a collection of $r$ independent knots,
but an $r$-component {\it link}.
Factorization is based on the possibility to efficiently separate these linked knots,
which is possible only at roots of unity.
Among other things, having $r$ components means that $\CH^{\Crb}\sim z^{-r}$ is actually singular at $q=1$
(this leading pole is the same for linked and unlinked components; linking shows up in the subleading terms).
In fact, exactly this singularity can be used to explain the simplest factorization (\ref{spefact})
of {\it special} HOMFLY, see sec.\ref{z0}.
But at roots of unity the story is a little bit lengthier and requires some more notions and arguments.

\bigskip

The central role in the skein approach to colored knot polynomials is played by {\bf projectors} $e_R$
onto particular representations $R$.
They are made of ${\cal R}$-matrices, and are also known as {\it idempotents}, $e_R^2=e_R$,
of the Hecke algebra (the algebra of braids with the skein relation $\CR_i-\CR_i^{-1}=z$ imposed).
According to \cite{AM14},
\be
e_{[2]}=\frac{q^{-1}+\CR_1}{[2]}, \nn \\
e_{[1,1]} = \frac{q- \CR_1}{[2]}, \nn \\ \nn \\
e_{[3]} = \frac{1+q\CR_1+q\CR_2+q^2\CR_1\CR_2+q^2\CR_2\CR_1+q^3\CR_1\CR_2\CR_1}{q^3[2][3]},\nn \\
e_{\underline{[2,1]}}=\frac{q^2[2]+q^3[2]\CR_1- \CR_2 - q(\CR_1\CR_2+\CR_2\CR_1)
- q^2\CR_1\CR_2\CR_1}{q^2[2][3]}, \nn \\
e_{\overline{[2,1]}}=\frac{q[2]-[2]\CR_1+q^3\CR_2 - q^2(\CR_1\CR_2+\CR_2\CR_1)
+ q\CR_1\CR_2\CR_1}{q[2][3]}, \nn \\
e_{[1,1,1]} = \frac{q^3-q^2(\CR_1+\CR_2)+q(\CR_1\CR_2+\CR_2\CR_1) -\CR_1\CR_2\CR_1 }{[2][3]},\nn\\
 \ \ \ \ {\rm so\ that}\ \ e_{\underline{[2,1]}}+e_{\overline{[2,1]}} =
 \frac{2+(q-q^{-1})(\CR_1+\CR_2)-(\CR_1\CR_2+\CR_2\CR_1)}{[3]},
\nn \\ \nn \\
e_{[4]} = \frac{1}{q^6[2][3][4]}\Big(1+q(\CR_1+\CR_2+\CR_3) + q^2(\CR_1\CR_2+\CR_2\CR_1+\CR_2\CR_3+\CR_3\CR_2+\CR_1\CR_3)
+ \nn \\
+ q^3(\CR_1\CR_2\CR_1+\CR_2\CR_3\CR_2+\CR_1\CR_2\CR_3+\CR_2\CR_1\CR_3+\CR_3\CR_1\CR_2 + \CR_3\CR_2\CR_1)
+ \nn \\
+q^4(\CR_1\CR_2\CR_1\CR_3+\CR_1\CR_2\CR_3\CR_2+\CR_1\CR_3\CR_2\CR_1+\CR_2\CR_1\CR_3\CR_2+\CR_2\CR_3\CR_2\CR_1)
+ \nn \\
+ q^5(\CR_1\CR_2\CR_1\CR_3\CR_2+\CR_1\CR_2\CR_3\CR_2\CR_1 + \CR_2\CR_1\CR_3\CR_2\CR_1)
+ q^6\CR_1\CR_2\CR_1\CR_3\CR_2\CR_1\Big),
\nn \\
\ldots
\label{projectors}
\ee
For symmetric representations there is a closed formula,
$e_{[r]}=\frac{1}{q^{r(r-1)/2}[r]!}\sum_{\pi\in S_r}q^{\ell(\pi)}\CR_\pi$, characterized by
$\CR_ie_{[r]}=qe_{[r]}$ for all $i=1,\ldots,r-1$ (the $q$-symmetrizer).
Here $\CR_\pi$ denote products of $\CR$-matrices, one for each permutation $\pi$.
For other $R$ the coefficients are
more complicated, see $e_{[2,1]}$ in (\ref{projectors}), but we will need only their general properties.
At $q=1$ only the identity term of $e_R$ is important, while at $q=\zeta$ one needs the whole $e_R$,
because $[m]$ can appear in its denominator.

\begin{figure}[H]\centering
\begin{tikzpicture}[scale=.55]
\begin{scope}\cabpic{1/0/Y,1/.62/Y'}\node[font=\small] at (1.4,-2.5) {$YY'$: $Y$ below $Y'$};\end{scope}
\node at (6.6,1.6) {$=$};
\begin{scope}[shift={(8.2,0)}]\cabpic{1/.4/Y',2/.4/Y}\node[font=\small] at (1.4,-2.5) {one round};\end{scope}
\node at (14.8,1.6) {$=$};
\begin{scope}[shift={(16.4,0)}]\cabpic{1/0/Y',1/.62/Y}\node[font=\small] at (1.4,-2.5) {two rounds: $Y'Y$};\end{scope}
\end{tikzpicture}
\caption{\footnotesize The cyclic rule (\ref{cyc}) for the 2-cable of a 2-strand braid: after two rounds through the closure
$Y$ returns to the same cabled strand, now above $Y'$.}
\label{fig:cyc}
\end{figure}
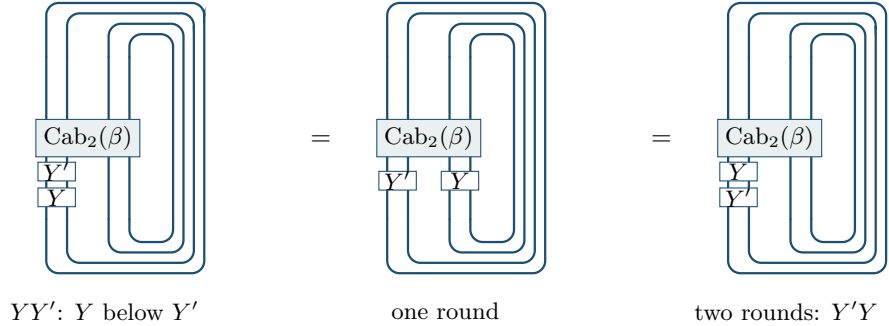

Like any other box on a cabled strand, projectors pass through the cable vertices, see (\ref{slide}):
\be
(e_R\otimes 1_r)\,B_r=B_r\,(1_r\otimes e_R),
\label{projslide}
\ee
so that $e_R$ can be put on any cabled strand and anywhere in the braid:
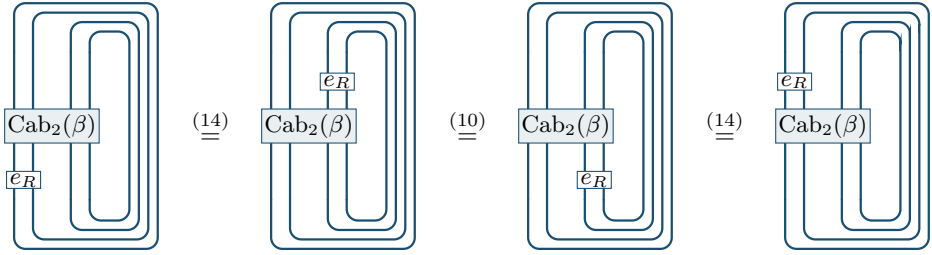
\begin{center}
\begin{tikzpicture}[scale=.5]
\begin{scope}\cabpic{1/0/e_R}\node at (5.2,1.65) {$\stackrel{(\ref{projslide})}{=}$};\end{scope}
\begin{scope}[shift={(6.8,0)}]\cabpic{2/2.6/e_R}\node at (5.2,1.65) {$\stackrel{(\ref{xyclic})}{=}$};\end{scope}
\begin{scope}[shift={(13.6,0)}]\cabpic{2/0/e_R}\node at (5.2,1.65) {$\stackrel{(\ref{projslide})}{=}$};\end{scope}
\begin{scope}[shift={(20.4,0)}]\cabpic{1/2.6/e_R}\end{scope}
\end{tikzpicture}
\end{center}
(here $\beta=\CR$, so that ${\rm Cab}_2(\beta)=B_2$ exchanges the two cabled strands). Together with cyclicity (\ref{xyclic}) this gives
what we call the {\bf cyclic rule inside a cable}: for two boxes $Y,Y'$ on the same cabled strand
\be
\tr_q\big((YY'\otimes 1)\,{\rm Cab}_r\beta\big)=\tr_q\big((Y'Y\otimes 1)\,{\rm Cab}_r\beta\big)
\ \ \ \ \ \ {\rm for\ any}\ \beta\ {\rm whose\ closure\ is\ a\ knot}.
\label{cyc}
\ee
To see this (we read braids from bottom to top, so that in $YY'$ the box $Y$ is below $Y'$), push $Y$ down through the closure by (\ref{xyclic}).
It appears at the top of the same column, above the cable, and can be slid down along its cabled strand through all the crossings by (\ref{slide}).
It comes out at the bottom of the {\it next} cabled strand, where it does not meet $Y'$.
Repeating this $s$ times, we return to the first cabled strand (the closure is a knot), but now $Y$ is above $Y'$.
Cyclicity (\ref{xyclic}) alone is not enough: it moves $Y$ through the closure, but leaves it on the same position, see Fig.\ref{fig:cyc}.

Rule (\ref{cyc}) allows cyclic permutations of $\CR$-matrices inside $e_R$, after which one can use (\ref{YB}) and (\ref{skein}).
We write $Y\simeq Y'$ when $Y$ and $Y'$ give the same trace, being inserted into the cable of any knot.
For example, $\CR_2=g\,\CR_1\,g^{-1}$ with $g=\CR_1\CR_2$ (this is the braid relation), hence $\CR_2\simeq\CR_1g^{-1}g=\CR_1$;
similarly $\CR_2\CR_1\simeq\CR_1\CR_2$ and $\CR_1\CR_2\CR_1\simeq\CR_2\CR_1\CR_1\stackrel{(\ref{skein})}{=}\CR_2+z\CR_2\CR_1\simeq\CR_1+z\CR_1\CR_2$.
Thus on three strands there are only three independent insertions, $1$, $\CR_1$ and $\CR_1\CR_2$ -- as many as cycle types of permutations
of three elements, $1^3$, $21$, $3$; this is the first sign of the symmetric functions which appear below.
In this way one can simplify $e_R$ to a much shorter $\tilde e_R\simeq e_R$ -- which, however, is no longer an idempotent.
Besides the explicit form (\ref{projectors}), we need only the following properties of projectors:
\be
e_R^2=e_R,\ \ \ \ \ \ \CR_ie_{[r]}=e_{[r]}\CR_i=q\,e_{[r]} .
\label{Pprop}
\ee
As to the framing factor $\theta_R$ in (\ref{Hdef}), we need only that
\be
\theta_{R}=\theta_{[m]}\,\theta_{R-M}\ \ \ \ {\rm for}\ q^{2m}=1,
\label{Fprop}
\ee
which is true because $\varkappa_R$ is the sum of contents $j-i$ of all boxes, and an $m$-ribbon $M$ contains $m$ consecutive contents.

\begin{figure}[!t]\centering
\begin{tikzpicture}[x=.34cm,y=.34cm]
\tikzset{grid/.style={draw=strand,line width=.45pt,line cap=round}}
\newcommand{\yd}[2]{
  \foreach \i/\j/\c in {#2}{\fill[\c] (\j-1,-\i+1) rectangle (\j,-\i);}
  \foreach \i/\l in {#1}{\foreach \j in {1,...,\l}{\draw[grid] (\j-1,-\i+1) rectangle (\j,-\i);}}}
\begin{scope}
\yd{1/2,2/2}{1/2/accent!40,2/2/accent!40,1/1/strand!22,2/1/strand!22}
\node[font=\small] at (1,-3.1) {$[2,2]$};
\node at (3.6,-1) {$\to$};
\begin{scope}[shift={(5,0)}]\yd{1/1,2/1}{1/1/strand!22,2/1/strand!22}\node[font=\small] at (.5,-3.1) {$[1,1]$};\end{scope}
\node at (7.6,-1) {$\to\ \varnothing$};
\node[font=\small,anchor=west] at (10,-1) {signs $(-1)(-1)=+1$};
\end{scope}
\begin{scope}[shift={(0,-6)}]
\yd{1/2,2/2}{2/1/accent!40,2/2/accent!40,1/1/strand!22,1/2/strand!22}
\node[font=\small] at (1,-3.1) {$[2,2]$};
\node at (3.6,-1) {$\to$};
\begin{scope}[shift={(5,0)}]\yd{1/2}{1/1/strand!22,1/2/strand!22}\node[font=\small] at (1,-3.1) {$[2]$};\end{scope}
\node at (7.6,-1) {$\to\ \varnothing$};
\node[font=\small,anchor=west] at (10,-1) {signs $(+1)(+1)=+1$};
\end{scope}
\node[font=\small,anchor=west] at (21,-4) {$\nu_{[2,2]}=2,\ \ \eta=+1,\ \ \varphi_{[2,2]}(2,2)=2$};
\draw[quiet,line width=.4pt] (20.4,-1)--(20.4,-7);
\begin{scope}[shift={(0,-12)}]
\yd{1/3,2/1}{1/2/accent!40,1/3/accent!40,1/1/strand!22,2/1/strand!22}
\node[font=\small] at (1.5,-3.1) {$[3,1]$};
\node at (4.1,-1) {$\to$};
\begin{scope}[shift={(5.5,0)}]\yd{1/1,2/1}{1/1/strand!22,2/1/strand!22}\node[font=\small] at (.5,-3.1) {$[1,1]$};\end{scope}
\node at (8.1,-1) {$\to\ \varnothing$};
\node[font=\small,anchor=west] at (10,-1) {signs $(+1)(-1)=-1$};
\node[font=\small,anchor=west] at (21,-1) {$\nu_{[3,1]}=1,\ \ \eta=-1,\ \ \varphi_{[3,1]}(2,2)=-1$};
\end{scope}
\end{tikzpicture}
\caption{\footnotesize Removal of $2$-ribbons (dominoes) from $[2,2]$ and $[3,1]$. The first domino removed is dark, the sign of a
ribbon is $(-1)^{{\rm height}-1}$. The number of removal sequences is $\nu_R$, their common sign is $\eta$, and
$\eta\,\nu_R=\varphi_R(m^w)$ by the Murnaghan-Nakayama rule.}
\label{fig:dominoes}
\end{figure}
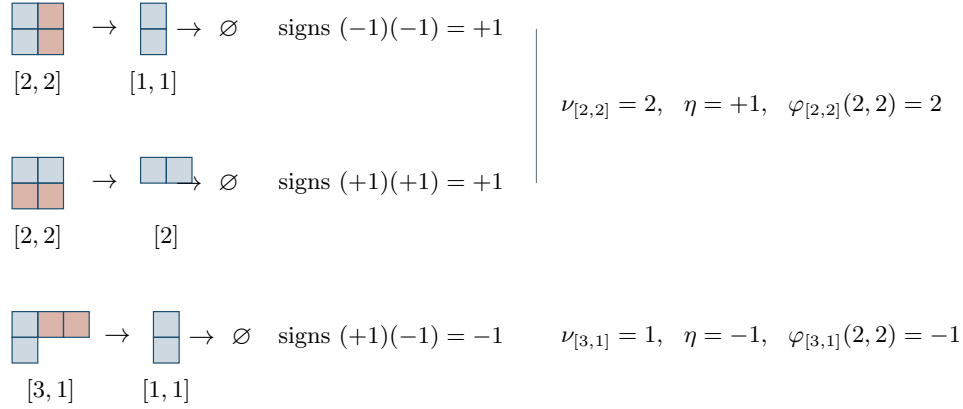

At $q=\zeta$, inside a cable, the projector $e_R$ reduces to $e_{{\rm core}_m(R)}$ times a power of
\be
X_m=\sum_{i=0}^{m-1}\ \underbrace{\CR^{-1}_{m-1}\cdots\CR^{-1}_{i+1}}_{m-1-i\ {\rm negative}}\ \underbrace{\CR_i\cdots\CR_1}_{i\ {\rm positive}},
\label{Xmpre}
\ee
e.g.  $X_1=1_1,\ \ \ \ X_2=\CR_1^{-1}+\CR_1,\ \ \ \ X_3=\CR_2^{-1}\CR_1^{-1}+\CR_2^{-1}\CR_1+\CR_2\CR_1, \ldots$
In every term one strand passes across all the other $m-1$: in the $i$-th term it goes over $m-1-i$ of them and under the remaining $i$ (Fig.\ref{fig:Xm}).
\begin{figure}[!t]\centering
\begin{tikzpicture}[scale=.62]
\node at (-1.1,.8) {$X_2\ =$};
\begin{scope}\xterm{2}{0}\end{scope}
\node at (1.2,.8) {$+$};
\begin{scope}[shift={(1.9,0)}]\xterm{2}{1}\end{scope}
\node at (4.4,.8) {$X_3\ =$};
\begin{scope}[shift={(5.5,0)}]\xterm{3}{0}\end{scope}
\node at (7.2,.8) {$+$};
\begin{scope}[shift={(7.9,0)}]\xterm{3}{1}\end{scope}
\node at (9.6,.8) {$+$};
\begin{scope}[shift={(10.3,0)}]\xterm{3}{2}\end{scope}
\node[lab] at (0.3,-.5) {$\CR_1^{-1}$}; \node[lab] at (2.2,-.5) {$\CR_1$};
\node[lab] at (6,-.5) {$\CR_2^{-1}\CR_1^{-1}$}; \node[lab] at (8.4,-.5) {$\CR_2^{-1}\CR_1$}; \node[lab] at (10.8,-.5) {$\CR_2\CR_1$};
\end{tikzpicture}
\caption{\footnotesize The winding sums (\ref{Xmpre}) for $m=2,3$. One strand (shown in color) passes across the other $m-1$;
the terms differ by how many of the crossings are positive. The permutation of every term is the cyclic shift, so that each term,
closed in the tube around the knot, is a single curve winding $m$ times around the tube -- hence the name. Braids are read from bottom to top, so that in $\CR_2^{-1}\CR_1$ the crossing $\CR_2^{-1}$ is the lower one.}
\label{fig:Xm}
\end{figure}
More precisely,
\be
\boxed{\ [m]^w\,e_R\ \stackrel{q=\zeta}{\simeq}\ \gamma_R\ X_m^{\otimes w}\otimes e_{{\rm core}_m(R)}\ }
\ \ \ \ \ \ \ \ \gamma_R=\frac{\eta\,\nu_R}{m^w\,w!},
\label{residue}
\ee
where $\nu_R$ is the number of ways to remove the $w$ ribbons one after another, and $\eta=\pm1$ is a sign, the same for all of them.
The equality $\simeq$ is inside the cable of any knot, see (\ref{cyc}), and is understood as the limit $q\to\zeta$,
since at $q=\zeta$ the Hecke algebra is not semisimple.
The constant $\gamma_R$ is the same for the unknot and cancels in (\ref{Hdef}).
The factor $[m]^w$ in the l.h.s. is needed, because $e_R$ may be singular at $q=\zeta$. The precise statement about the order of the
pole concerns the closure $Q_R=\cl(e_R)$, i.e. what the knot sees: by Lemma 3 of sec.\ref{proofs} it has a pole of order exactly $w$ at $[m]=0$.
A representative $F_R\simeq e_R$ of the same closure, which is regular whenever $Q_R$ is -- in particular for the core -- is given after
Lemma 3, and in (\ref{residue}) $e_{\rm core}$ may be read as $F_{\rm core}$.
Note that $[m]^we_R$ is not an idempotent, and $X_m$ is not an idempotent either: e.g. at $q=i$ we have $\frac12X_2=\CR_1-i$ and $(\CR_1-i)^2=0$.
Two examples with $m=2$ and $w=2$, where $\nu_R$ and $\eta$ are non-trivial, are shown in Fig.\ref{fig:dominoes}.
For $R=[2,2]$ the hook lengths are $3,2,2,1$, so $w=2$ and the core is empty. The two dominoes can be removed in two orders,
$[2,2]\to[1,1]\to\varnothing$ (two vertical dominoes, sign $(-1)^2$) and $[2,2]\to[2]\to\varnothing$ (two horizontal ones, sign $+1$):
$\nu_{[2,2]}=2$, $\eta=+1$, in agreement with $\varphi_{[2,2]}(2,2)=2$, the character of the two-dimensional representation of $S_4$ on the
class $(12)(34)$. Hence $[2]^2e_{[2,2]}\stackrel{q=i}{\simeq}\frac14X_2\otimes X_2$ and $H_{[2,2]}=H_{[2]}^2$ at $q=i$.
For $R=[3,1]$ (hook lengths $4,2,1,1$) there is one way, $[3,1]\to[1,1]\to\varnothing$, the second domino is vertical, so $\nu=1$, $\eta=-1$,
in agreement with $\varphi_{[3,1]}(2,2)=-1$; and $H_{[3,1]}=H_{[2]}^2$ at $q=i$ as well.
For symmetric representations $[r+m]$ with $0\le r<m$ the core is $[r]$, and (\ref{residue}) is just
\be
[m]\,e_{[r+m]}\ \stackrel{q=\zeta}{\simeq}\ \tfrac1m\,X_m\otimes e_{[r]},\ \ \ \ \ \ \ \ \ \ {\rm e.g.}\ \ \ \
[2]e_{[3]}\ \stackrel{q=i}{\simeq}\ \tfrac12 X_2\otimes 1_1=[2]e_{[2]}\otimes 1_1 .
\ee

Let us check the last example. From (\ref{projectors}), $q^3[2][3]\,e_{[3]}=N_3=
1+q(\CR_1+\CR_2)+q^2(\CR_1\CR_2+\CR_2\CR_1)+q^3\CR_1\CR_2\CR_1$, and the coefficient $q^2$ of the term $\CR_1\CR_2$,
which connects all three strands, does not vanish at any $q$, in particular at $q=i$: as an element of the Hecke algebra, $[2]e_{[3]}$
does not reduce to $[2]e_{[2]}\otimes1_1$.
But inside a cable, by (\ref{cyc}), $\CR_2\simeq\CR_1$, $\CR_2\CR_1\simeq\CR_1\CR_2$ and $\CR_1\CR_2\CR_1\simeq\CR_2\CR_1\CR_1
\stackrel{(\ref{skein})}{=}\CR_2+z\CR_2\CR_1\simeq\CR_1+z\CR_1\CR_2$, i.e. the three terms, which connect all three strands into a single
component, become the same, with the {\it total} coefficient $q^2+q^2+q^3z=q^3[2]$:
\be
N_3\ \simeq\ 1+q(2+q^2)\,\CR_1+q^3[2]\,\CR_1\CR_2\ \ \ \ {\rm for\ all}\ q,
\nn\\
{\rm hence}\ \ \ \ \
[2]e_{[3]}=\frac{N_3}{q^3[3]}\ \simeq\ \frac{1+q(2+q^2)\CR_1}{q^3[3]}+\frac{[2]}{[3]}\,\CR_1\CR_2
\ \stackrel{q=i}{=}\ \CR_1-i\ =\ \tfrac12X_2\otimes1_1 .
\label{N3colA}
\ee
Thus the connecting term does not disappear from $e_{[3]}$, but inside the cable it gets the factor $[2]$
(the other coefficients at $q=i$ are $q(2+q^2)=i$ and $q^3[3]=i$). The same happens in general: the terms which connect a ribbon with the rest are proportional to $[m]$,
and what remains is $X_m$.

\bigskip

Factorization (\ref{RMfact}) of colored HOMFLY at roots of unity is now based
on the 3-step transformation of $H^\beta_R$.

{\bf (A)} First, substitute $[m]^we_R \longrightarrow \gamma_R\,X_m^{\otimes w} \otimes e_{{\rm core}_m(R)}$ according to (\ref{residue}).
For $q=1$ this means just substitution of any $e_R$ by (a constant times) unity, $1^{(r)}$.

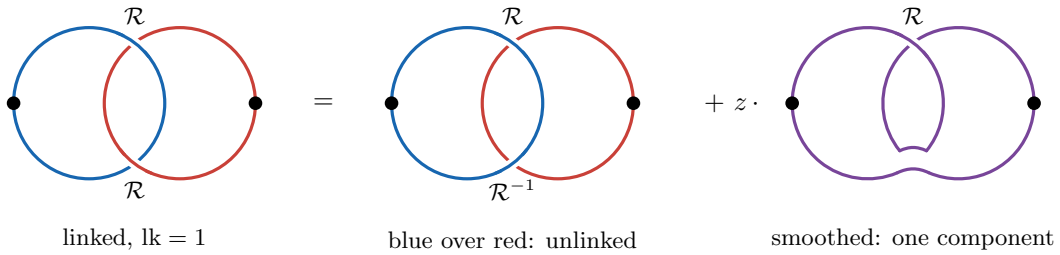
\begin{figure}[!t]\centering\begingroup
\definecolor{cbblue}{rgb}{.09,.40,.69}
\definecolor{cbred}{rgb}{.79,.25,.22}
\definecolor{cbpurple}{rgb}{.47,.27,.60}
\begin{tikzpicture}[x=1cm,y=1cm,line cap=round,line join=round]
\newcommand{\rings}[2]{
  \draw[cbred,line width=1.2pt] (1.2,0) circle (1);
  \draw[cbblue,line width=1.2pt] (0,0) circle (1);
  \ifnum#1=1 \draw[white,line width=4pt] (0,0)++(47:1)arc(47:59:1); \draw[cbblue,line width=1.2pt] (0,0)++(38:1)arc(38:68:1);
  \else \draw[white,line width=4pt] (1.2,0)++(121:1)arc(121:133:1); \draw[cbred,line width=1.2pt] (1.2,0)++(112:1)arc(112:142:1);\fi
  \ifnum#2=1 \draw[white,line width=4pt] (0,0)++(-59:1)arc(-59:-47:1); \draw[cbblue,line width=1.2pt] (0,0)++(-68:1)arc(-68:-38:1);
  \else \draw[white,line width=4pt] (1.2,0)++(227:1)arc(227:239:1); \draw[cbred,line width=1.2pt] (1.2,0)++(218:1)arc(218:248:1);\fi
  \fill (-1,0) circle (.09); \fill (2.2,0) circle (.09);
}
\begin{scope}
\rings{1}{0}
\node[font=\small] at (.6,1.15) {$\CR$}; \node[font=\small] at (.6,-1.15) {$\CR$};
\node[font=\small] at (.6,-1.8) {linked, ${\rm lk}=1$};
\end{scope}
\node at (3.1,0) {$=$};
\begin{scope}[shift={(5.0,0)}]
\rings{1}{1}
\node[font=\small] at (.6,1.15) {$\CR$}; \node[font=\small] at (.6,-1.15) {$\CR^{-1}$};
\node[font=\small] at (.6,-1.8) {blue over red: unlinked};
\end{scope}
\node at (8.5,0) {$+\ z\,\cdot$};
\begin{scope}[shift={(10.3,0)}]
\draw[cbpurple,line width=1.2pt] (0,0)++(-39:1) arc (-39:293:1);
\draw[cbpurple,line width=1.2pt] (1.2,0)++(247:1) arc (247:579:1);
\draw[cbpurple,line width=1.2pt] (0,0)++(293:1) to[out=25,in=155] ($(1.2,0)+(247:1)$);
\draw[cbpurple,line width=1.2pt] (0,0)++(-39:1) to[out=155,in=25] ($(1.2,0)+(219:1)$);
\draw[white,line width=4pt] (0,0)++(47:1)arc(47:59:1); \draw[cbpurple,line width=1.2pt] (0,0)++(38:1)arc(38:68:1);
\fill (-1,0) circle (.09); \fill (2.2,0) circle (.09);
\node[font=\small] at (.6,1.15) {$\CR$};
\node[font=\small] at (.6,-1.8) {smoothed: one component};
\end{scope}
\end{tikzpicture}\endgroup
\caption{\footnotesize Lifting blue over red near one crossing: $\CR\to\CR^{-1}$ in the lower mixed crossing unlinks the copies,
at the price of $z$ times the smoothed diagram.}
\label{fig:lifting}
\end{figure}

{\bf (B)} Second, lift one connected component of $\Crb$ over the others
(blue over red in Fig.\ref{fig:lifting}).
This is accompanied by an extra contribution with $z$ from skein relation,
which is absent for $q=1$, when $z=0$.
Spectacularly, it is also canceled for $q=\zeta$, if the contributions from projectors
are taken into account.
Procedure ``blue over red'' is {\it exact} both for $q=1$ and for $q=\zeta$ -- for the normalized quantities (\ref{Hdef}): at $q=1$
it is the equality of the leading singular terms of numerator and denominator, see sec.\ref{z0}, and at $q=\zeta$ an identity for the traces
multiplied by $[m]^w$.

{\bf (C)} Third, after linked knots are lifted one over the other, i.e. are {\it un}linked,
it remains to separate them.
This looks like an obvious procedure, still at the formal level it implies the multiple use of
YB relations (\ref{YB}).

\bigskip

In the remaining part of the text we illustrate how the (ABC) procedure actually works.
We begin from the case $q=1$ in sec.\ref{z0}, where the steps (A) and (B) are trivial, as we already explained,
and consider it in more detail.
This completes the story about (\ref{spefact});
the proof in this case was actually outlined already in \cite{AM14}.

As to (\ref{RMfact}), sec.\ref{qzeta} provides illustrations, which are exact in every example, while for the general case we need
some facts about symmetric functions, see sec.\ref{proofs}. Before that we introduce the language in which (A) and (B) are
naturally formulated, and explain why they work: (A) is point (ii) below, (B) is point (iii).

\medskip
\noindent{\bf Definitions.} The algebra generated by $\CR_1,\ldots,\CR_{r-1}$ with the relations (\ref{YB}) and (\ref{skein}) is the Hecke
algebra $H_r$. For generic $q$ it is a deformation of the group algebra of $S_r$ with the same representation theory: $H_r$ is a direct sum of
matrix algebras, one for each Young diagram $R$ with $|R|=r$, the character of the irreducible representation $R$ is the matrix trace
$\chi_R$ in the $R$-th block, and $e_R$ of (\ref{projectors}) is a rank-one projector in this block, so that $\chi_R(e_R)=1$ and $\chi_{R'}(e_R)=0$
for $R'\neq R$. This is what ``projector onto the representation $R$'' in (\ref{Hdef}) means. (At $q=\zeta$ semisimplicity is lost, and $e_R$ does
not exist, see the discussion after (\ref{residue}).)

By the cyclic rule (\ref{cyc}), the trace $\tr_q\big((Y\otimes1){\rm Cab}_r\beta\big)$ is a linear function of the box $Y\in H_r$,
which does not change under $YY'\to Y'Y$. On a matrix algebra every such function is proportional to the matrix trace, so that
\be
\boxed{\ \tr_q\big((Y\otimes1)\,{\rm Cab}_r\beta\big)=\sum_{R:\ |R|=r}\chi_R(Y)\ {\cal H}_R^{\cal K}\ }\ \ \ \ \ \ \ \ \ \ \
{\cal H}_R^{\cal K}:=\tr_q\big((e_R\otimes1)\,{\rm Cab}_r\beta\big),
\label{charexp}
\ee
where the coefficients ${\cal H}_R$ are found by putting $Y=e_R$ (Fig.\ref{fig:closure}).
Thus the knot sees only the $p(r)$ numbers $\chi_R(Y)$ ($p(r)$ is the number of Young diagrams with $r$ boxes): two boxes with $\chi_R(Y)=\chi_R(Y')$ for all $R$ give the same trace for every knot,
and this is the meaning of $Y\simeq Y'$ (the cyclic moves of (\ref{cyc}) preserve the characters, so they are a sufficient condition).
This criterion is valid for generic $q$. At $q=\zeta$ it can not be used literally: e.g. at $q=i$ all characters of $X_2$ vanish,
$\chi_{[2]}(X_2)=-\chi_{[1,1]}(X_2)=[2]=0$, while $\tr_q^{(2)}X_2=(A+A^{-1})D\neq0$ -- the r.h.s. of (\ref{charexp}) becomes $0\cdot\infty$,
because ${\cal H}_{[2]}$ and ${\cal H}_{[1,1]}$ are singular there. At $q=\zeta$ we therefore understand $Y\simeq Y'$ either as
the result of the cyclic moves (\ref{cyc}), which are valid for any $q$, or as the limit $q\to\zeta$ of the equivalence of regular families $Y(q)\simeq Y'(q)$. This set of characters we call
the {\bf closure} of $Y$ and write it as a formal sum over $R$ with some symbols $Q_R$:
\be
\cl(Y):=\sum_R\chi_R(Y)\,Q_R,\ \ \ \ \ \ \ \ \ \ \cl(e_R)=Q_R,\ \ \ \ \ \ h_k:=\cl(e_{[k]})=Q_{[k]} .
\label{cldef}
\ee
For $r=2$ everything can be done by hand: $H_2$ is spanned by $1$ and $\CR_1$, the two representations are one-dimensional,
$\chi_{[2]}(\CR_1)=q$ and $\chi_{[1,1]}(\CR_1)=-q^{-1}$ (the two roots of (\ref{skein})), and
\be
\begin{array}{c|c|c|l}
Y & \chi_{[2]}(Y) & \chi_{[1,1]}(Y) & \cl(Y)\\ \hline
1_2 & 1 & 1 & Q_{[2]}+Q_{[1,1]}=h_1^2\\
\CR_1 & q & -q^{-1} & q\,Q_{[2]}-q^{-1}Q_{[1,1]}\\
\CR_1^{-1} & q^{-1} & -q & q^{-1}Q_{[2]}-q\,Q_{[1,1]}\\
X_2=\CR_1+\CR_1^{-1} & [2] & -[2] & [2]\,(Q_{[2]}-Q_{[1,1]})=[2]\,(2h_2-h_1^2)=[2]P_2
\end{array}
\ee
(the last column uses $h_1^2=Q_{[2]}+Q_{[1,1]}$, i.e. $Q_{[1,1]}=h_1^2-h_2$). The table shows at once what $\cl$ is, that $X_2$ is
``the power sum $P_2$ up to $[2]$'', and that $P_2$ has a pole at $[2]=0$ while $X_2$ does not.
\begin{figure}[!t]\centering
\begin{tikzpicture}[scale=.7]
\begin{scope}\cabpic{1/.4/Y}\end{scope}
\node at (6.4,1.65) {$=\ \ \displaystyle\sum_{R:\,|R|=r}\chi_R(Y)\ \cdot$};
\begin{scope}[shift={(10.2,0)}]\cabpic{1/.4/e_R}\end{scope}
\node[anchor=west] at (14.6,1.65) {$=\ \ \displaystyle\sum_{R:\,|R|=r}\chi_R(Y)\,{\cal H}_R^{\cal K}$};
\end{tikzpicture}
\caption{\footnotesize The character expansion (\ref{charexp}): the knot with any box $Y$ inside the cable is a combination of the
same knot with the projectors $e_R$, with the coefficients $\chi_R(Y)$. The closure $\cl(Y)$ is the same expansion before the knot is chosen,
with $Q_R$ standing for the picture with $e_R$.}
\label{fig:closure}
\end{figure}
Geometrically, $\cl(Y)$ is the closed braid $\hat Y$ inside the tube around the knot, {\it before} the tube is knotted,
and (\ref{charexp}) says that the knot feels only $\cl(Y)$: whatever is cancelled inside the unknotted tube is cancelled inside any knot.
The closures, unlike the traces, can be multiplied: two boxes $Y$ and $Y'$, on $r$ and $r'$ strands of the same cabled strand,
placed side by side form a box $Y\otimes Y'$ on $r+r'$ strands, and $\cl(Y)\cl(Y'):=\cl(Y\otimes Y')$. Since $\chi_R(Y\otimes Y')$ is given by
the Littlewood-Richardson rule \cite{Mur37,Nak40,Mac95,JK81}, this is the product of the representation ring, $Q_SQ_{S'}=\sum_RN^R_{SS'}Q_R$ (the theorem of
Aiston-Morton and Lukac \cite{AM98,Luk05}, Lemma 1 in sec.\ref{proofs}). It is {\it not} the product of traces: for the trefoil
${\cal H}(1_2)={\cal H}_{[2]}+{\cal H}_{[1,1]}\neq{\cal H}_{[1]}^2$, see point (i) in sec.\ref{qzeta}; the two coincide only for the unknot.

Thus the closures form the ring of symmetric functions, with $Q_R$ being the Schur functions. Besides $Q_R$ we need two more bases of this ring.
The complete symmetric functions $h_k=Q_{[k]}=\cl(e_{[k]})$ are the closures of the symmetrizers -- the only projectors with a simple
explicit form (\ref{projectors}); they connect the braids with the {\it power sums} $P_m$, which are defined by Newton's formula
\be
P_1:=h_1,\ \ \ P_2:=2h_2-h_1^2,\ \ \ P_3:=3h_3-3h_2h_1+h_1^3,\ \ldots,
\ \ \ \ \ {\rm in\ general}\ \ \
\sum_{k\ge0}h_ku^k=\exp\Big(\sum_{m\ge1}\frac{P_m}{m}u^m\Big).
\label{hP}
\ee
The power sums are the basis in which everything we need is simple.

\noindent (i) $X_m$ is a single element of this basis (Morton \cite{Mor02b}):
\be
\cl(X_m)=[m]\,P_m,\ \ \ \ \ \ {\rm i.e.}\ \ \ \ \ \chi_R(X_m)=(-1)^k[m]\ \ {\rm for\ hooks}\ R=[m-k,1^k],\ \ \ \ \chi_R(X_m)=0\ \ {\rm otherwise}.
\label{XP}
\ee
For $m=2$ this is the identity $[2](2e_{[2]}-1)=\CR_1+\CR_1^{-1}$, by (\ref{projectors}) and (\ref{skein}); for $m\ge3$ it holds after the cyclic moves
(\ref{cyc}), see (\ref{X23}) below. Since $X_m$ is a sum of $m$ braids, its closure is regular at $[m]=0$; hence $P_m$ and all $P_{km}$ have a simple
pole there, while $P_j$ with $j$ not divisible by $m$ are regular.

\noindent (ii) The Schur functions are expressed through the power sums by the Frobenius formula
\be
Q_R=\sum_{\lambda\vdash|R|}\frac{\varphi_R(\lambda)}{z_\lambda}\,P_\lambda,\ \ \ \ \ \ P_\lambda=\prod_iP_{\lambda_i},\ \ \ \ z_\lambda=\prod_j j^{a_j}a_j!
\label{frob}
\ee
($\varphi_R(\lambda)$ is the character of the symmetric group $S_{|R|}$ on the conjugacy class of cycle type $\lambda$, not to be confused with the Hecke character $\chi_R(Y)$; $a_j$ is the number of parts of $\lambda$ equal to $j$). Together with (i) this
tells where the poles of $Q_R=\cl(e_R)$ are -- this is step (A), see below.

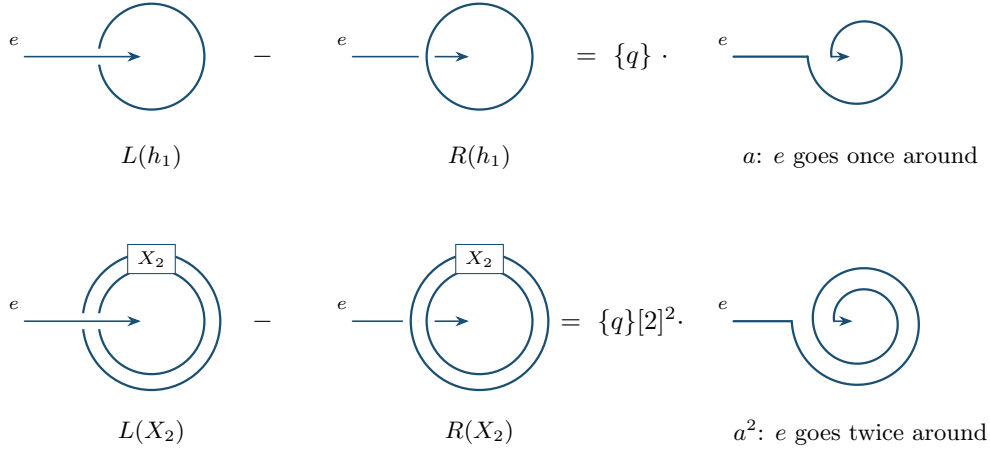
\begin{figure}[!t]\centering
\begin{tikzpicture}[scale=.7]
\newcommand{\loopone}{\draw[wire] (0,0) circle (1);}
\newcommand{\looptwo}{\draw[wire] (0,0) circle (1); \draw[wire] (0,0) circle (1.3);
  \draw[fill=white,draw=strand] (-.45,.9) rectangle (.45,1.45); \node[font=\scriptsize] at (0,1.17) {$X_2$};}
\newcommand{\eover}{\draw[white,line width=6pt] (-2.4,0)--(-.2,0); \draw[wire,arr] (-2.4,0)--(-.2,0); \node[font=\scriptsize] at (-2.6,.3) {$e$};}
\newcommand{\eunderone}{\draw[wire,arr] (-2.4,0)--(-.2,0); \node[font=\scriptsize] at (-2.6,.3) {$e$};
  \draw[white,line width=6pt] (165:1) arc (165:195:1); \draw[wire] (0,0) circle (1);}
\newcommand{\eundertwo}{\draw[wire,arr] (-2.4,0)--(-.2,0); \node[font=\scriptsize] at (-2.6,.3) {$e$};
  \draw[white,line width=6pt] (168:1) arc (168:192:1); \draw[white,line width=6pt] (170:1.3) arc (170:190:1.3); \looptwo}
\begin{scope}
\loopone \eover \node[lab] at (0,-1.9) {$L(h_1)$};
\end{scope}
\node at (2.1,0) {$-$};
\begin{scope}[shift={(6.2,0)}]
\eunderone \node[lab] at (0,-1.9) {$R(h_1)$};
\end{scope}
\node at (8.9,0) {$=\ \{q\}\ \cdot$};
\begin{scope}[shift={(13.4,0)}]
\draw[wire] (-2.4,0)--(-1,0);
\draw[wire,domain=180:540,samples=120,variable=\t] plot ({(1-0.45*(\t-180)/360)*cos(\t)},{(1-0.45*(\t-180)/360)*sin(\t)});
\draw[wire,arr] (-.55,0)--(-.2,0);
\node[font=\scriptsize] at (-2.6,.3) {$e$};
\node[lab] at (0,-1.9) {$a$: $e$ goes once around};
\end{scope}
\begin{scope}[shift={(0,-5)}]
\looptwo \eover
\node[lab] at (0,-2.1) {$L(X_2)$};
\end{scope}
\node at (2.1,-5) {$-$};
\begin{scope}[shift={(6.2,-5)}]
\eundertwo
\node[lab] at (0,-2.1) {$R(X_2)$};
\end{scope}
\node at (8.9,-5) {$=\ \{q\}[2]^2\cdot$};
\begin{scope}[shift={(13.4,-5)}]
\draw[wire] (-2.4,0)--(-1.3,0);
\draw[wire,domain=180:900,samples=240,variable=\t] plot ({(1.3-0.4*(\t-180)/360)*cos(\t)},{(1.3-0.4*(\t-180)/360)*sin(\t)});
\draw[wire,arr] (-.5,0)--(-.15,0);
\node[font=\scriptsize] at (-2.6,.3) {$e$};
\node[lab] at (0,-2.1) {$a^2$: $e$ goes twice around};
\end{scope}
\end{tikzpicture}
\caption{\footnotesize Top: eq.(\ref{skein1}); the circle is the closure of a cabled strand (the knot, schematically), and the smoothing
$a$ is $e$ going once around it. Bottom: eq.(\ref{adams}) for $m=2$; in $a^2$ the strand $e$ goes twice around. In sec.\ref{proofs} the
same $L$, $R$, $a$ are drawn in the cut tube, Fig.\ref{fig:tube}: the circle is the tube before cutting.}
\label{fig:loop}
\end{figure}

\noindent (iii) {\it Passing through the band.} Let a strand $e$ of the cable cross a closed insertion $Y$ sitting on another cabled strand;
$L(Y)$ denotes the picture with $e$ passing over $Y$, and $R(Y)$ -- with $e$ passing under it (Fig.\ref{fig:loop}).
For a single closed strand, $Y=h_1$, there is one crossing, and the skein relation (\ref{skein}) gives
\be
L(h_1)-R(h_1)=\{q\}\,a ,
\label{skein1}
\ee
where $a$ is the smoothing of the crossing: the strand $e$, instead of crossing, turns and goes once around the knot along the former
closed strand, which is absorbed into $e$ (Fig.\ref{fig:loop}, top). For the band $Y=X_m$ there are $m$ crossings, and the answer is
\be
L(X_m)-R(X_m)=\{q\}[m]^2\,a^m ,
\label{adams}
\ee
where $a^m$ is the strand $e$ going $m$ times around the knot (Fig.\ref{fig:loop}, bottom). This is Lemma 4 of sec.\ref{proofs}, where the
case $m=2$ is first worked out by hand, eq.(\ref{LR2}); see also the example with a ring below.
At $q=\zeta$ the right hand side vanishes: {\it the band is transparent}. This is step (B).

\noindent{\it Why $\{q\}[m]^2$: the Adams operation.} One factor $[m]$ is trivial, $X_m=[m]P_m$ by (\ref{XP}). The rest is
$\{q\}[m]=\{q^m\}$, i.e.
\be
L(P_m)-R(P_m)=\{q^m\}\,a^m
\label{adamsP}
\ee
-- exactly (\ref{skein1}) with $q\to q^m$ and $a\to a^m$. The meaning: every term of $P_m=\cl(X_m)/[m]$ is a {\it single} strand wound $m$ times
around the knot (every term of $X_m$ is a cyclic shift, Fig.\ref{fig:Xm}, so its closure is one curve), and (\ref{adamsP}) says that
the combination $P_m$ of the $m$ times wound strands is crossed by the skein relation with $q^m$: its effective skein parameter is $\{q^m\}$,
which vanishes at $q^{2m}=1$ for the same reason as $\{q\}$ vanishes at $q^2=1$. This is a property of the whole combination $P_m$, not of its separate terms (see the example below). In the ring of symmetric functions the map $P_1\mapsto P_m$, or in general $P_j\mapsto P_{jm}$,
is the Adams operation $\psi^m$ (on the variables, $x_i\mapsto x_i^m$) -- hence the name. For the unknot the trace of $P_m$ is indeed
obtained by the substitution $(A,q)\to(A^m,q^m)$, $\tr_q\,\cl(P_m)=\{A^m\}/\{q^m\}$; for other knots this is {\it not} a general rule --
whether something like it holds for $H_{[m]}$ at $q=\zeta$ is exactly the conjecture (\ref{sigmam}). Note also that $P_m$ is a linear
combination of $m$ closed braids, (\ref{XP}), not a single diagram; ``wound $m$ times'' refers to each term.

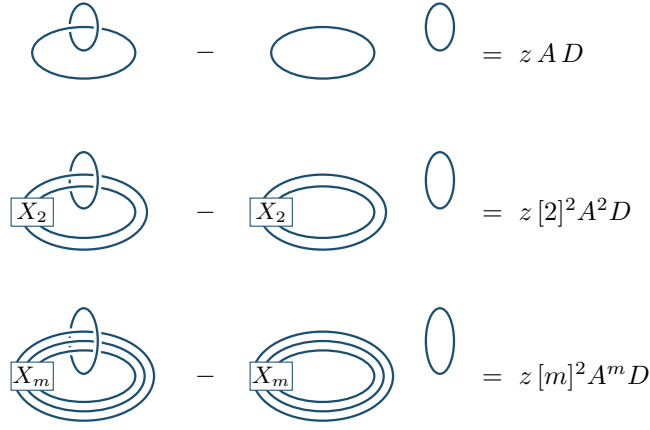
\begin{figure}[!htb]\centering
\begin{tikzpicture}[scale=.62]
\newcommand{\bandring}[5]{%
  \ifnum#3=1
    \draw[wire] (0,#4)++(90:.3 and #5) arc (90:270:.3 and #5);                 
    \foreach \ra/\rb in {#1}{\draw[white,line width=3pt] (0,0) ellipse [x radius=\ra, y radius=\rb]; \draw[wire] (0,0) ellipse [x radius=\ra, y radius=\rb];}
    \draw[white,line width=3pt] (0,#4)++(-90:.3 and #5) arc (-90:90:.3 and #5); 
    \draw[wire] (0,#4)++(-90:.3 and #5) arc (-90:90:.3 and #5);
  \else
    \foreach \ra/\rb in {#1}{\draw[wire] (0,0) ellipse [x radius=\ra, y radius=\rb];}
    \draw[wire] (2.5,#4) ellipse [x radius=.3, y radius=#5];
  \fi
  \ifx\relax#2\relax\else
  \draw[fill=white,draw=strand] (-1.55,-.3) rectangle (-.65,.3); \node[font=\small] at (-1.1,0) {#2};\fi
}
\newcommand{\row}[6]{
\begin{scope}[shift={(0,#3)}]
\bandring{#1}{#2}{1}{#5}{#6}
\node at (2.6,0) {$-$};
\begin{scope}[shift={(5.1,0)}]\bandring{#1}{#2}{0}{#5}{#6}\end{scope}
\node[anchor=west] at (8.3,0) {$=\ #4$};
\end{scope}}
\row{1.1/.55}{}{0}{z\,A\,D}{.55}{.5}
\row{1.1/.55,1.35/.8}{$X_2$}{-3.4}{z\,[2]^2A^2D}{.68}{.6}
\row{1.1/.55,1.3/.75,1.5/.95}{$X_m$}{-6.9}{z\,[m]^2A^mD}{.75}{.7}
\end{tikzpicture}
\caption{\footnotesize A ring on a closed strand, on the band $X_2$ and on the band $X_m$, minus the ring taken off.}
\label{fig:ring}
\end{figure}

\noindent{\it Example: a ring on the band.} The simplest closed version of (\ref{adams}) is a trace. Consider a closed band of $m$ parallel strands with $X_m$ inserted
somewhere, and let $e$ be a ring, linked once with the band; compare it with the same picture without the ring (Fig.\ref{fig:ring}).
For a single strand, $m=1$ and $X_1=1$, this is just the Hopf link versus two unlinked circles, and
\be
\langle{\rm ring\ on\ a\ strand}\rangle-\langle{\rm ring}\rangle\langle{\rm strand}\rangle=\tr_q(\CR^2)-\tr_q(1_2)
\stackrel{(\ref{skein})}{=}\tr_q(1+z\CR)-D^2\stackrel{(\ref{Markov})}{=}z\,A\,D,
\ee
which is $O(1)$ at $z\to0$, i.e. two orders of $z$ smaller than the traces themselves, $\sim D^2\sim z^{-2}$: the normalized difference
is $zAD/D^2=z^2A/\{A\}\to0$, so at $q^2=1$ (generic $A$) the ring can be taken off -- this is the comparison of orders of poles of sec.\ref{z0} -- while
for $q^2\neq1$ it can not. Note that the unnormalized difference $zAD=A\{A\}$ itself does not vanish at $z=0$.
For the band $X_2$ the ring is the third strand, going around the first two, $\CR_2\CR_1\CR_1\CR_2$, and the calculation
is not much longer. With $X_2=2\CR_1-z$, see (\ref{skein}),
\be
\tr_q\big(X_2\,\CR_2\CR_1^2\CR_2\big)&\stackrel{(\ref{skein})}{=}&\tr_q\big(X_2\CR_2^2\big)+z\,\tr_q\big(X_2\CR_2\CR_1\CR_2\big)
\stackrel{(\ref{skein}),(\ref{YB})}{=}\tr_q(X_2)+z\,\tr_q(X_2\CR_2)+z\,\tr_q\big(X_2\CR_1\CR_2\CR_1\big),
\nn\\
\tr_q^{(3)}(X_2)&\stackrel{(\ref{Markov})}{=}&(A+A^{-1})D^2,\ \ \ \ \ \ \ \tr_q(X_2\CR_2)\stackrel{(\ref{Markov})}{=}A(A+A^{-1})D,
\nn\\
\tr_q\big(X_2\CR_1\CR_2\CR_1\big)&\stackrel{(\ref{skein})}{=}&\tr_q\big(2\CR_2\CR_1+z\CR_1\CR_2\CR_1\big)
\stackrel{(\ref{YB}),(\ref{Markov})}{=}2A^2D+zA(D^2+zAD),
\nn
\ee
and, putting all together, the difference is $zAD\big(A+A^{-1}+2A+zD+z^2A\big)=zAD\cdot A\,(4+z^2)$, since $zD=A-A^{-1}$;
with $4+z^2=[2]^2$ we get
\be
\langle{\rm ring\ on\ }X_2\rangle-\langle{\rm ring}\rangle\langle X_2\rangle=
\tr_q\big(X_2\,\CR_2\CR_1^2\CR_2\big)-\tr_q^{(3)}(X_2)=z\,[2]^2\,A^2D.
\label{ring2}
\ee

The same calculation on $m+1$ strands gives the general formula, which is (\ref{adams}) under the trace, with $\langle a^m\rangle=A^mD$:
\be
\boxed{\ \langle{\rm ring\ on\ }X_m\rangle-\langle{\rm ring}\rangle\langle X_m\rangle=z\,[m]^2\,A^mD\ }
\label{ringm}
\ee

Thus at $q=\zeta$ the ring can be taken off the band $X_m$. Note that this is true only for the whole sum $X_m$: if instead
of $X_2$ we put on the band just $\CR$, the r.h.s. of (\ref{ring2}) would be $zAD(3A-A^{-1}+z^2A)$, which does not vanish at $q=i$, and for $\CR^{-1}$ it is
$zAD(A+A^{-1})$. Only their sum contains the factor $[2]^2=4+z^2$.
Also important is that the band is closed: as elements of the braid algebra, the two sides are different, see (\ref{hecke2}),
and the difference vanishes only under the trace.
Finally, the ring can be substituted by any strand which crosses the band, e.g. by a strand of another copy of the knot, if only
it crosses the band as a whole (above or below both its strands) and the band is closed around the knot. Just in this
form (\ref{ringm}) will be used in sec.\ref{qzeta}; the proof of (\ref{adams}) in sec.\ref{proofs} is a skein calculation
of the same type as (\ref{ring2}), but it uses also the property (\ref{Pprop}) of projectors.

\medskip
\noindent{\it A heuristic remark.} Step (B) also gives a hint why at $q=\zeta$ $H_{[m]}$ is the special polynomial at $A^m$, eq.(\ref{sigmam}):
what remains of $e_{[m]}$ is a $P_m$-colored strand, and it behaves like the fundamental one with $(A,q)\to(A^m,\pm1)$.

\section{Colored knot polynomials in the $z=0$ limit \label{z0}}

The idea is simple: a link with $l$ components has a pole $z^{-l}$, and in the $r$-cable the maximal $l$ is $r$,
so that everything with less components can be neglected. For a 2-strand knot the $r$-cable has $2r$ strands.

\subsection{Unknot}

\be
\tr_q^{(2)} 1 = D^2, \ \ \ \ \ \tr_q^{(2)} \CR = AD
\ee
\be
\tr_q^{(2)} e_{[2]} = \tr_q^{(2)} \frac{q^{-1}+\CR}{[2]}
= D\cdot \frac{q^{-1}D+A  }{[2]} = \frac{\{A\}\{Aq\}}{[2]\{q\}^2} = D_{[2]}, \nn\\
\tr_q^{(2)} e_{[1,1]} = \tr_q^{(2)} \frac{q-\CR}{[2]}
=  D\cdot \frac{qD- A  }{[2]}= \frac{\{A\}\{Aq^{-1}\}}{[2]\{q\}^2} = D_{[1,1]}
\ee

\subsection{Singularities at $z=0$}

\be
{\cal H}(L)=O(z^{-l}) \ \ \ {\rm for\ an}\ l{\rm -component\ link}\ L,
\label{polebound}
\ee
and for a knot the pole $z^{-1}$ comes only from the factor $D$ in front of the reduced polynomial.
This is easily proved by induction in $l$. Change the mixed crossings until the components can be pulled apart.
Every change costs $z$ times an $(l-1)$-component link, i.e. $O(z^{2-l})$, and after separation we get a product of $l$ knots, which is
$O(z^{-l})$. Therefore at $z=0$ in the $r$-cable only the contributions with all the $r$ components survive.

\subsection{2-strand knot in the fundamental representation}

Non-reduced HOMFLY
\be
{\cal H}_{\Box}^{{\rm torus}_{2,2n+1}} = \tr_q^{(2)} \CR^{2n+1}
= q^{2n+1}D_{[2]}-q^{-(2n+1)}D_{[1,1]}
= \frac{[2n+2]A -[2n]A^{-1}}{[2]}\,D, \nn\\
H_{\Box}^{{\rm torus}_{2,2n+1}} = A^{-(2n+1)}\,\frac{[2n+2]A -[2n]A^{-1}}{[2]}
\ \stackrel{q=1}{=}\ (n+1)A^{-2n}-nA^{-2n-2}
\ee
The skein relation gives
\be
\tr_q^{(2)} \CR^{2n+1}
= \tr_q^{(2)} \CR^{2n-1} + z\cdot \tr_q^{(2)} \CR^{2n}
\ee
The second term does not vanish at $z=0$, because $\tr_q\CR^{2n}$ is a 2-component link, i.e. it is $O(z^{-2})$ by (\ref{polebound}).
Thus a smoothing can be neglected only when it decreases the number of components.

\subsection{Cyclic symmetry of $\tr_q$ for braids}

For any braids $X,Y$ (products of $\CR_i^{\pm1}$)
\be
\tr_q  XY = \tr_q YX
\ee
since the weight $\Omega^{\otimes s}$ in $\tr_q={\rm Tr}\,\Omega^{\otimes s}(\cdot)$ commutes with all $\CR_i$, cf. (\ref{xyclic}).
This means that conjugation by a braid does not change the trace, and exact braid identities (Yang-Baxter and far commutativity)
can be freely used inside $\tr_q$, also at $q=\zeta$.

\subsection{Knots in symmetric representation $[r]$}

Since the $r$-cable is $O(z^{-r})$, from (\ref{polebound}) we get two simple rules:

(a) In $e_{[r]}=\sum_{\pi\in S_r}c_\pi(q)\CR_\pi$ all $r$ components are preserved only by the identity term.
All other $\CR_\pi$ join some of the copies and can be neglected. The coefficient $c^{(R)}_1(1)\neq0$ is the same
for the knot and for the unknot, so it cancels in (\ref{Hdef}).

(b) A mixed crossing can be substituted by its inverse, $\CR=\CR^{-1}+z$, since the correction joins two copies and is
$z\cdot O(z^{-(r-1)})=O(z^{2-r})$. For a self-crossing this is not allowed: its smoothing splits a copy into two.

At $q=\zeta$ rule (a) will be substituted by (\ref{residue}), with $1_1\to X_m$, and rule (b) -- by the passage relation (\ref{pass}).

\subsection{Knots in symmetric representation $[2]$}

4-strand formula:
\be
H_{[2]}^{{\rm torus}_{2,2n+1}} = \theta_{[2]}^{-(2n+1)}\,
\frac{\tr_q^{(4)} \left(e_{[2]}\cdot
\Big(\CR_2(\CR_1\CR_3)\CR_2\Big)^{2n+1}\right)}{\tr_q^{(2)}e_{[2]}}
\ee
In the block $\CR_2(\CR_1\CR_3)\CR_2$ both $\CR_2$ are mixed crossings, while $\CR_1$ and $\CR_3$ are self-crossings.
According to (b), the first $\CR_2$ can be changed for $\CR_2^{-1}$, and then $\CR_2^{-1}(\CR_1\CR_3)\CR_2$ is just conjugated $\CR_1\CR_3$.
According to (a), $e_{[2]}\to \frac{q^{-1}}{[2]}\cdot 1$. Then
\be
\tr_q^{(4)} \left(e_{[2]}\cdot
\Big(\CR_2(\CR_1\CR_3)\CR_2\Big)^{2n+1}\right)
= \frac{q^{-1}}{[2]}\tr_q^{(4)}  \Big(\CR_2^{-1}(\CR_1\CR_3)\CR_2  \Big)^{2n+1}+O(z^{-1})
=\nn\\
= \frac{q^{-1}}{[2]}\tr_q^{(4)}  ( \CR_1\CR_3)^{2n+1} + O(z^{-1})
= \frac{q^{-1}}{[2]}\tr_q^{(2)}  \CR_1 ^{2n+1}\cdot \tr_q^{(2)}  \CR_3 ^{2n+1} + O(z^{-1})
= \frac{q^{-1}}{[2]}\Big( {\cal H}_{[1]}^{{\rm torus}_{2,2n+1}} \Big)^2 + O(z^{-1})
\ee
Division by $\tr_q^{(2)}e_{[2]}=\frac{q^{-1}}{[2]}D^2+O(z^{-1})$ and by $\theta_{[2]}^{2n+1}=A^{2(2n+1)}$
gives $H_{[2]}=H_{[1]}^2$ at $q=1$.

Why $\CR_2^2\to 1$, but neither $\CR_1^2\to1$ nor $\CR_3^2\to1$?
Because $\CR_2$ is mixed and $\CR_1,\CR_3$ are self-crossings: the $z\CR_2$ correction decreases the number of
components, the $z\CR_1$ or $z\CR_3$ corrections would increase it and must be kept.

\subsection{2-strand knots in symmetric representation $[3]$}
6-strand formula:
\be
H_{[3]}^{{\rm torus}_{2,2n+1}} =\theta_{[3]}^{-(2n+1)}\,
\frac{\tr_q^{(6)}  \left(e_{[3]}\cdot \Big(\CR_3(\CR_2\CR_4)(\CR_1\CR_3\CR_5)(\CR_2\CR_4) \CR_3\Big)^{2n+1}\right)}
{\tr_q^{(3)}e_{[3]}}
\ee
(this ``symmetric'' block is obtained from $B_3$ in (\ref{block}) by Yang-Baxter and far commutativity).
Here $\CR_3$ and $\CR_2,\CR_4$ are the six mixed crossings, and $\CR_1,\CR_3,\CR_5$ in the middle are the three self-crossings.
We invert the first $\CR_3$ and the first $(\CR_2\CR_4)$, which costs $O(z^{2})$ relatively, and after that the block is conjugated
$\CR_1\CR_3\CR_5$:
\be
\tr_q^{(6)}  \left(e_{[3]}\cdot \Big(\CR_3(\CR_2\CR_4)(\CR_1\CR_3\CR_5)(\CR_2\CR_4) \CR_3\Big)^{2n+1}\right)
=\nn\\
= \frac{1}{q^3[2][3]}\tr_q^{(6)}  ( \CR_1\CR_3\CR_5)^{2n+1} + O(z^{-2})
= \frac{1}{q^3[2][3]}\Big( {\cal H}_{[1]}^{{\rm torus}_{2,2n+1}} \Big)^3 + O(z^{-2})
\ee
and $H_{[3]}=H_{[1]}^3$ at $q=1$. It is instructive to write the same as a chain of formulas, since exactly the same chain will appear at $q=i$ in sec.\ref{full}:
\be
H_{[3]}^{{\rm torus}_{2,2n+1}} =\theta_{[3]}^{-(2n+1)}\,
\frac{\tr_q^{(6)}  \left(e_{[3]}\cdot \Big(\CR_3(\CR_2\CR_4)(\CR_1\CR_3\CR_5)(\CR_2\CR_4) \CR_3\Big)^{2n+1}\right)}
{\tr_q^{(3)}e_{[3]}}
\ \stackrel{(a)}{\longrightarrow}
\nn\\
\stackrel{(a)}{\longrightarrow}\
\theta_{[3]}^{-(2n+1)}\,
\frac{\tr_q^{(6)}  \left(1^{(3)} \cdot \Big(\CR_3(\CR_2\CR_4)(\CR_1\CR_3\CR_5)(\CR_2\CR_4) \CR_3\Big)^{2n+1}\right)}
{\tr_q^{(3)}1}
\ \stackrel{(b)}{\longrightarrow}
\nn \\
\stackrel{(b)}{\longrightarrow}\
\theta_{[3]}^{-(2n+1)}\,
\frac{\tr_q^{(6)}  \left(\Big(\CR_3^{-1}(\CR_2^{-1}\CR_4^{-1})(\CR_1\CR_3\CR_5)(\CR_2\CR_4) \CR_3\Big)^{2n+1}\right)}
{D^3}
\ \stackrel{(c)}{=}
\nn\\
\stackrel{(c)}{=}\
\theta_{[3]}^{-(2n+1)}\,
\frac{\tr_q^{(6)}  \left((\CR_1\CR_3\CR_5)^{2n+1}\right)}{D^3}
=\theta_{[3]}^{-(2n+1)}\left(\frac{\tr_q^{(2)}\CR^{2n+1}}{D}\right)^3
\stackrel{q=1}{=}\Big(H_{[1]}^{{\rm torus}_{2,2n+1}}\Big)^3
\ee
At step (a) $e_{[3]}\to\frac16\cdot1^{(3)}$, because all other terms of $e_{[3]}$ join the copies and are relatively $O(z)$; the factor $\frac16$ cancels
between numerator and denominator. At step (b) the three mixed crossings in the first half of the block are inverted, each at the relative price
$O(z^2)$, and the block becomes $G(\CR_1\CR_3\CR_5)G^{-1}$ with $G=\CR_3^{-1}\CR_2^{-1}\CR_4^{-1}$. Step (c) is exact: the conjugation disappears
under the trace, and $\CR_1,\CR_3,\CR_5$ act on different pairs of strands. At last, $\theta_{[3]}=A^3q^6\to A^3=\theta_{[1]}^3$.

\subsection{Other symmetric representations}

Everything is the same: all the $r(r-1)$ mixed crossings of $B_r$ are sorted, only the $r$ self-crossings $\CR_1\CR_3\cdots\CR_{2r-1}$ remain,
$e_{[r]}$ is substituted by its identity term, and $H_{[r]}=H_{[1]}^r$.

\subsection{Non-symmetric representations \label{nonsymz0}}

Again the same. In (a) we used only that $c^{(R)}_1(1)\neq0$ (this is true for any nonzero idempotent of ${\mathbb C}[S_r]$,
because $r!\,c_1$ is its trace in the regular representation) and that every $\CR_\pi$ with $\pi\neq1$ joins some copies.
This proves (\ref{spefact}) for arbitrary $R$.

\subsection{Other knots in the $z=0$ limit \label{bor}}

Now the idea is ``blue over red'' and then Yang-Baxter.
Let ${\cal K}=\hat\beta$, $\beta=\prod_\alpha\sigma_{i_\alpha}^{\epsilon_\alpha}$, and color the $r$ copies.
In every cable block apply skein relation $\CR=\CR^{-1}+z$ to change
-- at the negligible cost $O(z^{2-r})$ --
the mixed crossings so that copy 1 ("blue") passes over copy 2 ("red"),
copy 2 over copy 3, etc. Self-crossings are not touched, therefore every copy remains ${\cal K}$.
The sorted cable $S_r(\beta)$ is then an {\it exact} conjugate of $r$ separate copies of $\beta$,
\be
S_r(\beta)=Q\,\Big(\beta^{(1)}\beta^{(2)}\cdots\beta^{(r)}\Big)\,Q^{-1},
\label{YBconj}
\ee
where $Q$ is some shuffle braid, which regroups the strands copy by copy; (\ref{YBconj}) needs only Yang-Baxter and
far commutativity, but no skein relations. E.g. for $\beta=(\sigma_1\sigma_2)^2$, $r=2$: $Q=\CR_2^{-1}\CR_4^{-1}\CR_3^{-1}$
and $Q^{-1}S_2Q=(\CR_1\CR_2)^2(\CR_4\CR_5)^2$. By cyclicity the conjugation disappears from the trace,
\be
{\cal H}\big({\rm Cab}_r(\beta)\big)=\Big({\cal H}_{\Box}^{\cal K}\Big)^r + O(z^{2-r}),
\ \ \ \ \ \ \ \ \
H_R^{\cal K}(A,1)=\Big(H_{\Box}^{\cal K}(A,1)\Big)^{|R|}
\ee
where the second equality follows from sec.\ref{nonsymz0}.

\section{Roots of unity \label{qzeta}}

Now $z_\zeta\neq0$, and (\ref{polebound}) says nothing: corrections from crossings are not small,
and the cable itself is regular. Singular is only the projector, because of $[m]=0$ in its denominator.
This is illustrated by two facts.

(i) The naive identity for the cable is wrong: for the 2-cable of the trefoil, with
$\widetilde{\cal H}({\rm Cab}_2):=A^{-2n_\beta}\,\tr_q^{(4)}\big(B_2^{3}\big)$ (unlike (\ref{charexp}), including the framing factor, $n_\beta=3$),
at $q=1$ the ratio $\frac{\widetilde{\cal H}({\rm Cab}_2)}{D^2}=(H_\Box)^2$,
but at $q=i$ this is no longer true: $\frac{\widetilde{\cal H}({\rm Cab}_2)}{D^2}-(H_\Box)^2=-\frac{4(3A^4+A^2-1)}{A^6(A^2-1)}\neq0$.

(ii) Still $H_{[3]}^{3_1}(A,i)=H_{[2]}^{3_1}(A,i)H_{[1]}^{3_1}(A,i)$, as one can check by direct computation of the traces.

So the projector plays a role, it must be kept.
But in fact we need only a piece of it, which survives at $q=\zeta$.
And it turns out that this piece, $X_m$, is not only peculiar but perfectly adjusted to the next step, leading to factorization formulas.

Below we first write down what remains of the projector at $q=\zeta$, then go through all three steps for the trefoil in $[3]$ at $q=i$,
and then consider other representations and roots in sec.\ref{examples}.

\subsection{What remains of the projector}

At $q=\zeta\ne\pm1$ the cable has no pole, since $z\neq0$, and the only singularity is $[m]$ in the denominator of the projector. So first we need to understand what remains of the projector at $[m]=0$. Let us multiply the projectors (\ref{projectors}) by
$[m]$, to make them finite, and put $q=\zeta$:
\be
[2]e_{[2]}\ \stackrel{q=i}{=}\ \CR_1-i=\tfrac12\big(\CR_1+\CR_1^{-1}\big),\ \ \ \ \ \ \ \ \ \ \ \ \ \
[3]e_{[3]}\ \stackrel{q=e^{i\pi/3}}{\simeq}\ \tfrac13\big(\CR_2\CR_1+\CR_2^{-1}\CR_1+\CR_2^{-1}\CR_1^{-1}\big).
\label{X23}
\ee

The first formula is just an identity: $\CR^{-1}=\CR-z$ by (\ref{skein}), and $z=2i$ at $q=i$. The second one is true inside a cable, in the sense
of (\ref{cyc}), and is derived in sec.\ref{examples}. In the r.h.s. we have the {\it winding sums} $X_m$ of (\ref{Xmpre})
on $m$ strands (Fig.\ref{fig:Xm}): one strand crosses the other $m-1$, $i$ of them positively and the remaining $m-1-i$ negatively, and we sum over $i$. For $m=2$
it is just $\CR+\CR^{-1}$, the sum of two possible crossings of two strands.

\subsection{The trefoil in $[3]$ at $q=i$ \label{full}}

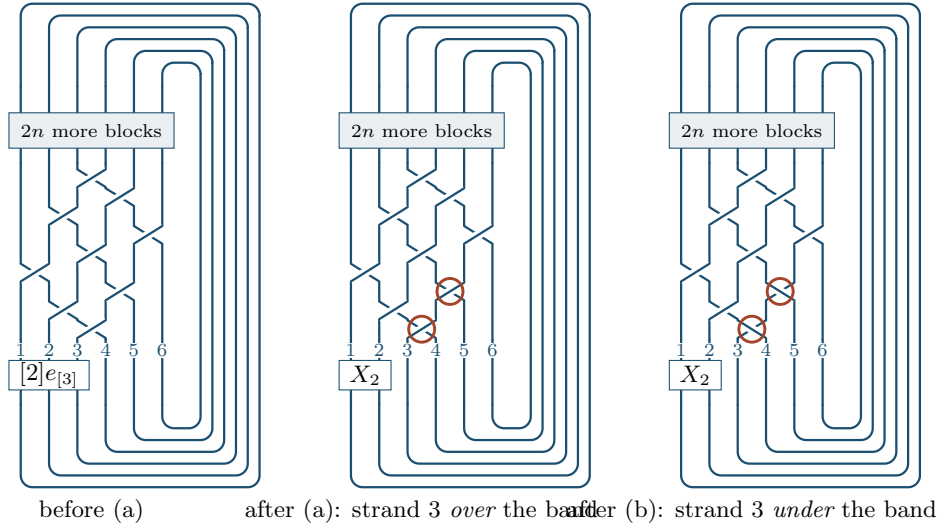
\begin{figure}[!t]\centering
\begin{tikzpicture}[scale=.78]
\newcommand{\cablepanel}[4]{%
  \foreach \j in {1,...,6}{\draw[wire] ({.48*(\j-1)},-1.1)--({.48*(\j-1)},0); \draw[wire] ({.48*(\j-1)},3.0)--({.48*(\j-1)},4.3);}
  \braidpic{6}{#1}{3.0}
  \draw[fill=strand!9,draw=strand] (-.2,3.25) rectangle (2.6,3.85); \node[font=\scriptsize] at (1.2,3.55) {$2n$ more blocks};
  \draw[fill=white,draw=strand] (-.2,-.85) rectangle ({.48*(#3-1)+.2},-.35); \node[font=\small] at ({.24*(#3-1)},-.6) {#2};
  \foreach \j in {1,...,6}{\pgfmathsetmacro{\x}{.48*(\j-1)} \pgfmathsetmacro{\d}{.2*(7-\j)}
    \draw[wire] (\x,4.3) -- (\x,{4.3+\d}) arc (180:90:.2) -- ({2.65+\d},{4.5+\d}) arc (90:0:.2) -- ({2.85+\d},{4.3+\d})
      -- ({2.85+\d},{-1.1-\d}) arc (0:-90:.2) -- ({\x+.2},{-1.3-\d}) arc (-90:-180:.2) -- (\x,-1.1);}
  \foreach \j in {1,...,6}{\node[font=\scriptsize,strand,fill=white,inner sep=.4pt] at ({.48*(\j-1)},-.18) {\j};}
  #4
}
\begin{scope}
\cablepanel{3,2,4,1,3,5,2,4,3}{$[2]e_{[3]}$}{3}{}
\node[font=\small] at (1.2,-2.9) {before (a)};
\end{scope}
\begin{scope}[shift={(5.6,0)}]
\cablepanel{3,2,4,1,3,5,2,4,3}{$X_2$}{2}{%
  \draw[accent,line width=1.1pt] (1.2,.18) circle (.22); \draw[accent,line width=1.1pt] (1.68,.82) circle (.22);}
\node[font=\small] at (1.2,-2.9) {after (a): strand 3 {\it over} the band};
\end{scope}
\begin{scope}[shift={(11.2,0)}]
\cablepanel{-3,2,-4,1,3,5,2,4,3}{$X_2$}{2}{%
  \draw[accent,line width=1.1pt] (1.2,.18) circle (.22); \draw[accent,line width=1.1pt] (1.68,.82) circle (.22);}
\node[font=\small] at (1.2,-2.9) {after (b): strand 3 {\it under} the band};
\end{scope}
\end{tikzpicture}
\caption{\footnotesize The 6-strand cable of $\CR^{2n+1}$ (one block shown) before (a), after (a) and after (b);
the circled crossings of strand 3 with the band are reversed in step (b).}
\label{fig:cable}
\end{figure}

Now we consider the first non-trivial example: the 2-strand torus knot, i.e. the closure of
$\beta=\CR^{2n+1}$ ($n=1$ for the trefoil), in representation $R=[3]$ at $m=2$, i.e. at $q=i$. The cable has six strands,
there are three copies of the knot, and $e_{[3]}$ is on the first three strands. Every crossing of $\beta$ is substituted by the $3\times3$
block
\be
B\ =\ B_3\ =\ (\CR_3\CR_2\CR_1)(\CR_4\CR_3\CR_2)(\CR_5\CR_4\CR_3)\ \stackrel{(\ref{YB})}{=}\ \CR_3(\CR_2\CR_4)(\CR_1\CR_3\CR_5)(\CR_2\CR_4)\CR_3,
\ee
where strands $1,2,3$ (the first cabled strand) pass over strands $4,5,6$ (the second one). The second form of the block
is adapted to the {\it copies}: the three crossings in the middle layer, $\CR_1$, $\CR_3$, $\CR_5$, are the crossings of
strand $k$ with strand $k+3$, $k=1,2,3$, i.e. the own crossing of the $k$-th copy of the knot -- the original crossing of $\beta$
seen in this copy -- while the six crossings in the outer layers, $\CR_3$ and $\CR_2\CR_4$ on each side, are the mixed crossings
between different copies. It is only the mixed crossings which have to be reversed in step (b), and once they are, the block
is conjugate, by the braid relations, to $\CR_1\CR_3\CR_5=\CR\otimes\CR\otimes\CR$ -- three separate copies of the crossing of
$\beta$ (step (c)). Both numerator and
denominator in (\ref{Hdef}) have a pole at $[2]=0$, because there is $[2][3]$ in the denominator of $e_{[3]}$, see (\ref{projectors}), and of its trace.
Therefore we multiply both by $[2]$. Then $[2]e_{[3]}=N_3/(q^3[3])$ is finite at $q=i$, and so is $\tr_q^{(3)}[2]e_{[3]}$, and the
whole calculation is the following chain, where every equality is exact at $q=i$:
\be
H_{[3]}^{{\rm torus}_{2,2n+1}} &=&\theta_{[3]}^{-(2n+1)}\,
\frac{\tr_q^{(6)}  \left([2]e_{[3]}\cdot B^{2n+1}\right)}{\tr_q^{(3)}\,[2]e_{[3]}}
\ \stackrel{(a)}{=}\
\theta_{[3]}^{-(2n+1)}\,
\frac{\tr_q^{(6)}  \left(\big([2]e_{[2]}\otimes 1\big)^{(3)} \cdot B^{2n+1}\right)}
{\tr_q^{(2)}\,[2]e_{[2]}\cdot D}
\ \stackrel{(b)}{=}
\nn \\
&\stackrel{(b)}{=}&
\theta_{[3]}^{-(2n+1)}\,
\frac{\tr_q^{(6)}  \left(\big([2]e_{[2]}\otimes 1\big)^{(3)} \cdot \Big(\CR_3^{-1}(\CR_2\CR_4^{-1})(\CR_1\CR_3\CR_5)(\CR_2\CR_4) \CR_3\Big)^{2n+1}\right)}
{\tr_q^{(2)}\,[2]e_{[2]}\cdot D}
\ \stackrel{(c)}{=}
\nn \\
&\stackrel{(c)}{=}&
\theta_{[3]}^{-(2n+1)}\,
\frac{\tr_q^{(4)}  \left([2]e_{[2]}\cdot B_2^{2n+1}\right)}{\tr_q^{(2)}\,[2]e_{[2]}}
\cdot\frac{\tr_q^{(2)}\CR^{2n+1}}{D}
\ \stackrel{(\ref{Fprop})}{=}\
H_{[2]}\,H_{[1]}
\label{chain3}
\ee
where $B_2=\CR_2(\CR_1\CR_3)\CR_2$ is the $2\times2$ block. The cable at different steps is shown in Fig.\ref{fig:cable}. Now let us
describe the three steps in more detail.

\medskip
\noindent{\it (a) The projector.} This was already done in (\ref{N3colA}): by the cyclic rule (\ref{cyc}) the six terms of $N_3$ reduce to three,
and at $q=i$ the last of them, $\CR_1\CR_2$, drops out, since its coefficient contains $[2]$. The other coefficients are $q(2+q^2)=i$ and $q^3[3]=i$, so that
\be
[2]e_{[3]}=\frac{N_3}{q^3[3]}\ \simeq\ \frac{1+i\CR_1}{i}=\CR_1-i\stackrel{(\ref{projectors})}{=}[2]e_{[2]}\otimes1_1\stackrel{(\ref{skein})}{=}\tfrac12X_2\otimes 1_1 .
\ee

Thus the third strand becomes free, while the first two form the band $X_2$:
\begin{center}
\begin{tikzpicture}[scale=.8]
\foreach \x in {0,.5,1}{\draw[wire] (\x,0)--(\x,2);}
\draw[fill=strand!9,draw=strand] (-.25,.7) rectangle (1.25,1.3); \node at (.5,1) {$N_3$};
\node at (2.3,1) {$\longrightarrow\ \ \tfrac{i}{2}$};
\begin{scope}[shift={(3.4,0)}]
\foreach \x in {0,.5,1}{\draw[wire] (\x,0)--(\x,2);}
\draw[fill=strand!9,draw=strand] (-.25,.7) rectangle (.75,1.3); \node at (.25,1) {$X_2$};
\end{scope}
\node at (5.6,1) {$=\ \tfrac{i}{2}\Big($};
\begin{scope}[shift={(6.6,.4)}]\draw[wire] (0,0)--(0,.35); \draw[wire] (.48,0)--(.48,.35); \draw[wire] (0,.85)--(0,1.2); \draw[wire] (.48,.85)--(.48,1.2);
\draw[wire] (.48,.35)--(0,.85); \draw[white,line width=3.4pt] (0,.35)--(.48,.85); \draw[wire] (0,.35)--(.48,.85);\end{scope}
\node at (7.9,1) {$+$};
\begin{scope}[shift={(8.5,.4)}]\draw[wire] (0,0)--(0,.35); \draw[wire] (.48,0)--(.48,.35); \draw[wire] (0,.85)--(0,1.2); \draw[wire] (.48,.85)--(.48,1.2);
\draw[wire] (0,.35)--(.48,.85); \draw[white,line width=3.4pt] (.48,.35)--(0,.85); \draw[wire] (.48,.35)--(0,.85);\end{scope}
\node at (9.7,1) {$\Big)$};
\draw[wire] (10.4,0)--(10.4,2);
\node[font=\small] at (10.4,-.4) {strand 3};
\node[font=\small] at (8.4,-.4) {strands 1,2};
\end{tikzpicture}
\end{center}

\noindent The denominator (the unknot) is treated in the same way,
but there one does not need even (\ref{cyc}): (\ref{Markov}) gives directly $\tr_q^{(3)}\CR_2=\tr_q^{(3)}\CR_1=AD^2$ etc., and
$\tr_q^{(3)}[2]e_{[3]}=D\,\tr_q^{(2)}[2]e_{[2]}$.

\medskip
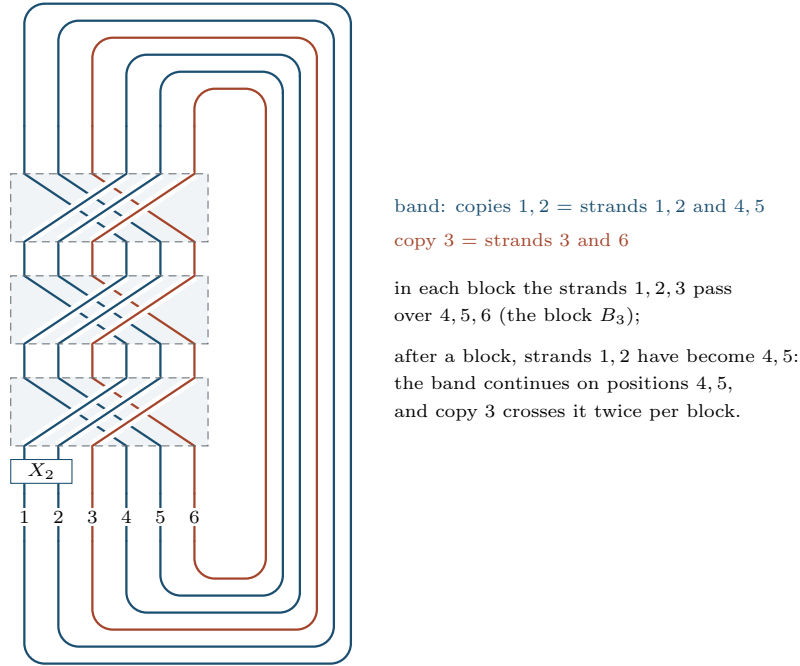
\begin{figure}[!t]\centering
\begin{tikzpicture}[scale=.9]
\tikzset{band/.style={draw=strand,line width=.85pt,line cap=round}, free/.style={draw=accent,line width=.85pt,line cap=round}}
\newcommand{\blockB}[1]{
  \fill[strand!6] (-.2,#1) rectangle (2.7,#1+1);
  \draw[band] (1.5,#1)--(0,#1+1); \draw[band] (2.0,#1)--(.5,#1+1); \draw[free] (2.5,#1)--(1.0,#1+1);
  \foreach \x/\st in {0/band,.5/band}{\draw[white,line width=4pt] (\x,#1)--(\x+1.5,#1+1); \draw[\st] (\x,#1)--(\x+1.5,#1+1);}
  \draw[white,line width=4pt] (1.0,#1)--(2.5,#1+1); \draw[free] (1.0,#1)--(2.5,#1+1);
  \draw[quiet,densely dashed,line width=.4pt] (-.2,#1) rectangle (2.7,#1+1);}
\foreach \y in {-.4,1.3,2.8,4.3}{
  \foreach \x in {0,.5,1.5,2.0}{\draw[band] (\x,\y)--(\x,\y+.7);}
  \foreach \x in {1.0,2.5}{\draw[free] (\x,\y)--(\x,\y+.7);}}
\foreach \x in {0,.5,1.5,2.0}{\draw[band] (\x,-1.1)--(\x,-.4);} \foreach \x in {1.0,2.5}{\draw[free] (\x,-1.1)--(\x,-.4);}
\blockB{.3}\blockB{1.8}\blockB{3.3}
\draw[fill=white,draw=strand] (-.2,-.25) rectangle (.7,.1); \node[font=\scriptsize] at (.25,-.07) {$X_2$};
\foreach \x/\d/\st in {2.5/.25/free,2.0/.5/band,1.5/.75/band,1.0/1.0/free,.5/1.25/band,0/1.5/band}{
  \draw[\st] (\x,5.0)--(\x,{5.0+\d}) arc (180:90:.3) -- ({3.0+\d},{5.3+\d}) arc (90:0:.3) -- ({3.3+\d},{5.0+\d})
   -- ({3.3+\d},{-1.1-\d}) arc (0:-90:.3) -- ({\x+.3},{-1.4-\d}) arc (-90:-180:.3) -- (\x,-1.1);}
\foreach \x/\l in {0/1,.5/2,1.0/3,1.5/4,2.0/5,2.5/6}{\node[font=\scriptsize,fill=white,inner sep=1pt] at (\x,-.75) {$\l$};}
\node[font=\scriptsize,strand,anchor=west] at (5.3,3.8) {band: copies $1,2$ = strands $1,2$ and $4,5$};
\node[font=\scriptsize,accent,anchor=west] at (5.3,3.3) {copy $3$ = strands $3$ and $6$};
\node[font=\scriptsize,anchor=west] at (5.3,2.6) {in each block the strands $1,2,3$ pass};
\node[font=\scriptsize,anchor=west] at (5.3,2.2) {over $4,5,6$ (the block $B_3$);};
\node[font=\scriptsize,anchor=west] at (5.3,1.6) {after a block, strands $1,2$ have become $4,5$:};
\node[font=\scriptsize,anchor=west] at (5.3,1.2) {the band continues on positions $4,5$,};
\node[font=\scriptsize,anchor=west] at (5.3,.8) {and copy $3$ crosses it twice per block.};
\end{tikzpicture}
\caption{\footnotesize The whole 6-strand cable of the trefoil ($\beta=\CR^3$, three blocks $B_3$), coloured by copies.
The band (copies $1,2$, blue) occupies positions $1,2$ below a block and $4,5$ above it; copy $3$ (red) crosses the band inside every block,
once over (from position $3$ to $6$) and once under (from $6$ to $3$).}
\label{fig:fullcable}
\end{figure}
\noindent{\it (b) The band is transparent.} The whole cable is shown in Fig.\ref{fig:fullcable}, one block in Fig.\ref{fig:cable}; consider one block $B$. The free strand $3$ goes over strands $4,5$
(this is the band after one turn around the knot) in the first two factors of $B=\CR_3\CR_4\cdot\tilde B$, where
$\tilde B=\CR_2(\CR_1\CR_3\CR_5)(\CR_2\CR_4)\CR_3$ (we used $\CR_2\CR_4=\CR_4\CR_2$). These two crossings are circled in the figure,
and just they are reversed in (\ref{chain3}). The first $\CR_2$ in the block is the crossing of strands $2$ and $4$,
i.e. it is inside the band, and it is {\it not} reversed -- this would change the answer by a finite amount. Denote by $X=X_2=2[2]e_{[2]}$
the box on strands $1,2$ and by $X^{(45)}$ the same box on strands $4,5$. The reversal needs four moves.

First, using (\ref{YB}) and (\ref{xyclic}), we bring the box $X$ to the needed place. The band slides through a block as a whole,
$(X\otimes1_4)B=BX^{(45)}$ and $X^{(45)}B=B(X\otimes1_4)$, so that
\be
\tr_q^{(6)}\Big((X\otimes1_4)\,B^{2n+1}\Big)=\tr_q^{(6)}\Big(B^{2n+1}X^{(45)}\Big)\stackrel{{\rm (\ref{xyclic})}}{=}\tr_q^{(6)}\Big(X^{(45)}B^{2n+1}\Big)
=\tr_q^{(6)}\Big(B^{2n}\cdot X^{(45)}\,\CR_3\CR_4\,\tilde B\Big):
\ee
and now $X$ is on strands $4,5$, right in front of the two crossings.

Second, from (\ref{skein}) we have $\CR_3\CR_4-\CR_3^{-1}\CR_4^{-1}=\CR_3\CR_4-(\CR_3-z)(\CR_4-z)=z(\CR_3+\CR_4)-z^2$, and $X^{(45)}\CR_4=2+z\CR_4$,
and $4+z^2=[2]^2$. Multiplying all this, we obtain for all $q$
\be
X^{(45)}\,\CR_3\CR_4\ =\ X^{(45)}\,\CR_3^{-1}\CR_4^{-1}\ +\ \frac z2\,X^{(45)}X^{(34)}\ +\ \frac{z[2]^2}{2},
\label{hecke2}
\ee
where $X^{(34)}=2\CR_3-z$ is $X_2$ on strands $3,4$, i.e. on the free strand and one strand of the band. The last term
vanishes at $q=i$, but the middle one does not, cf. the remark about closed bands after (\ref{ringm}): it is an element of the
braid algebra which is not proportional to $[2]$, and only its trace along the closed band is -- this is shown explicitly
in (\ref{LR2}) and the lines after it.

Third, by (\ref{lem4}) the trace of this middle term along the closed band is proportional to $[2]^2$, so that
\be
\tr_q^{(6)}\Big(B^{2n}X^{(45)}\CR_3\CR_4\tilde B\Big)-\tr_q^{(6)}\Big(B^{2n}X^{(45)}\CR_3^{-1}\CR_4^{-1}\tilde B\Big)
&=&\frac z2\,\tr_q^{(6)}\Big(B^{2n}\,X^{(45)}X^{(34)}\,\tilde B\Big)
\nn\\
&+&\frac{z[2]^2}{2}\,\tr_q^{(6)}\Big(B^{2n}\tilde B\Big)\ =\ z[2]^2\cdot(\ldots)\ \stackrel{q=i}{=}\ 0 .
\ee

This is (\ref{ringm}), where the ring is substituted by strand $3$, which crosses the band as a whole (above it before
the reversal and below after), and the rest of the cable is a fixed exterior. In Fig.\ref{fig:cable} the band is formed by strands $1,2\equiv4,5$,
closed around the knot, and strand $3$ goes through it just like the ring in Fig.\ref{fig:ring}. This is the only place where we need the band to be closed.

In terms of the cut tube (with top and bottom edges identified, and the band going around the knot) this can be written as
\be
L(X_2)-R(X_2)=z\,[2]^2\,a^2,
\label{pass}
\ee
where $a^2$ is a certain diagram (the crossing strand merged into the band and winding twice):
\begin{center}
\begin{tikzpicture}[scale=.8]
\begin{scope}
\draw[boundary] (-.3,-.3) rectangle (2.3,2.3);
\draw[wire] (.7,-.3)--(.7,2.3); \draw[wire] (1.3,-.3)--(1.3,2.3);
\draw[fill=white,draw=strand] (.4,1.4) rectangle (1.6,1.9); \node[font=\small] at (1,1.65) {$X_2$};
\draw[white,line width=4pt] (-.3,.7)--(2.3,.7); \draw[wire,arr] (-.3,.7)--(2.3,.7);
\node[lab] at (1,-.75) {$L(X_2)$: band below};
\end{scope}
\node at (3.6,1) {$-$};
\begin{scope}[shift={(5,0)}]
\draw[boundary] (-.3,-.3) rectangle (2.3,2.3);
\draw[wire,arr] (-.3,.7)--(2.3,.7);
\draw[white,line width=4pt] (.7,-.3)--(.7,2.3); \draw[white,line width=4pt] (1.3,-.3)--(1.3,2.3);
\draw[wire] (.7,-.3)--(.7,2.3); \draw[wire] (1.3,-.3)--(1.3,2.3);
\draw[fill=white,draw=strand] (.4,1.4) rectangle (1.6,1.9); \node[font=\small] at (1,1.65) {$X_2$};
\node[lab] at (1,-.75) {$R(X_2)$: band above};
\end{scope}
\node[anchor=west] at (7.9,1) {$=\ \ z[2]^2\cdot(\text{fixed diagram})\ \stackrel{q=i}{=}\ 0$};
\node[font=\footnotesize,align=center] at (5,2.9) {cut piece of the tube around the knot; top and bottom edges are the band\\ (strands 1,2 with $X_2$) continuing around the knot, left and right edges are strand 3 continuing};
\end{tikzpicture}
\end{center}

\noindent Fourth, we repeat all this. After the reversal the first block is $B'=\CR_3^{-1}\CR_4^{-1}\tilde B$, and still $X^{(45)}B'=B'(X\otimes1_4)$,
because in $B'$ the band crosses the other cabled strand as a whole. Therefore $X$ can be moved around once again, to
the next block $B$, and the three moves can be repeated. After $2n+1$ rounds all the blocks are $B'$, and we get the r.h.s.
of step (b) in (\ref{chain3}).

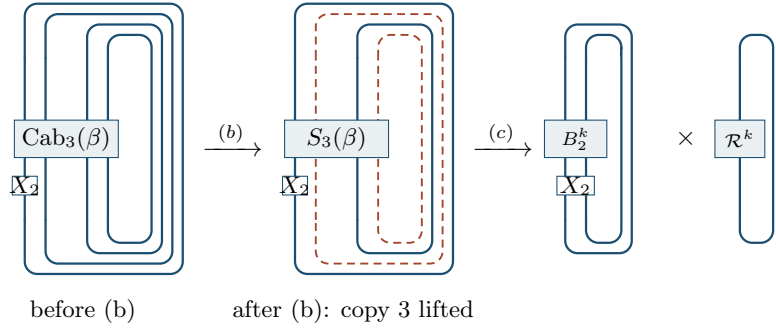
\begin{figure}[!t]\centering
\begin{tikzpicture}[scale=.55]
\tikzset{copythree/.style={draw=accent,line width=.7pt,densely dashed}}
\newcommand{\cabpicc}[3]{
  \foreach \x in {0,1.5}{\draw[wire] (\x,-.3)--(\x,3.6);}
  \foreach \x in {.5,2}{\draw[#2] (\x,-.3)--(\x,3.6);}
  \draw[fill=strand!9,draw=strand] (-.25,1.2) rectangle (2.25,2.1); \node[font=\small] at (1,1.65) {#1};
  \draw[fill=white,draw=strand] (-.3,.3) rectangle (.3,.75); \node[font=\small] at (0,.52) {#3};
  \foreach \x/\d/\st in {2/.25/#2,1.5/.5/wire,.5/.75/#2,0/1/wire}{
   \draw[\st] (\x,3.6) -- (\x,{3.6+\d}) arc (180:90:.3) -- ({2.5+\d},{3.9+\d}) arc (90:0:.3) -- (2.8+\d,{3.6+\d})
     -- (2.8+\d,{-.3-\d}) arc (0:-90:.3) -- ({\x+.3},{-.6-\d}) arc (-90:-180:.3) -- (\x,-.3);}
}
\begin{scope}\cabpicc{${\rm Cab}_3(\beta)$}{wire}{$X_2$}\node[font=\small] at (1.4,-2.5) {before (b)};\end{scope}
\node at (5,1.65) {$\xrightarrow{\ (b)\ }$};
\begin{scope}[shift={(6.5,0)}]\cabpicc{$S_3(\beta)$}{copythree}{$X_2$}\node[font=\small] at (1.4,-2.5) {after (b): copy 3 lifted};\end{scope}
\node at (11.5,1.65) {$\xrightarrow{\ (c)\ }$};
\begin{scope}[shift={(13,0)}]
\foreach \x in {0,.5}{\draw[wire] (\x,-.3)--(\x,3.6);}
\draw[fill=strand!9,draw=strand] (-.5,1.2) rectangle (1,2.1); \node[font=\scriptsize] at (.25,1.65) {$B_2^{k}$};
\draw[fill=white,draw=strand] (-.2,.3) rectangle (.7,.75); \node[font=\small] at (.25,.52) {$X_2$};
\foreach \x/\d in {.5/.25,0/.5}{\draw[wire] (\x,3.6) -- (\x,{3.6+\d}) arc (180:90:.3) -- ({.8+\d},{3.9+\d}) arc (90:0:.3) -- (1.1+\d,{3.6+\d}) -- (1.1+\d,{-.3-\d}) arc (0:-90:.3) -- ({\x+.3},{-.6-\d}) arc (-90:-180:.3) -- (\x,-.3);}
\node at (2.9,1.65) {$\times$};
\begin{scope}[shift={(4.2,0)}]
\draw[wire] (0,-.3)--(0,3.6);
\draw[fill=strand!9,draw=strand] (-.6,1.2) rectangle (.6,2.1); \node[font=\scriptsize] at (0,1.65) {$\CR^{k}$};
\draw[wire] (0,3.6) -- (0,3.85) arc (180:90:.3) -- (.5,4.15) arc (90:0:.3) -- (.8,3.85) -- (.8,-.55) arc (0:-90:.3) -- (.3,-.85) arc (-90:-180:.3) -- (0,-.3);
\end{scope}
\end{scope}
\end{tikzpicture}
\caption{\footnotesize Step (c), schematic, $k=2n+1$. Each cabled strand is drawn as a pair of lines: the band $X_2$ (solid, strands $1,2$ and $4,5$)
and copy 3 (strands $3$ and $6$; dashed after (b), when it passes below the band everywhere
and the block ${\rm Cab}_3$ becomes the sorted $S_3$); after (c) copy 3 is a separate 2-strand closure.}
\label{fig:sep}
\end{figure}

\medskip
\noindent{\it (c) Separation.} Schematically it is shown in Fig.\ref{fig:sep}.

\noindent Strand $3$, together with strand $6$ which is its continuation, now goes below strands $1,2,4,5$
everywhere, and by (\ref{YB}) it can be moved aside: $B'^{\,2n+1}=g\,\big(B_2^{2n+1}\otimes\CR^{2n+1}\big)\,g^{-1}$, where the braid $g$ involves only
strands $3,\ldots,6$ and thus commutes with $X\otimes 1_4$. The conjugation by $g$ disappears due to (\ref{xyclic}), the trace of
two separated pieces is the product of traces, and $[2]e_{[2]}\otimes 1$ on strands $1,2$ gives just the knot in representation $[2]$.
At last, $[2]$ cancels in the four-strand ratio, and the framing factors are the same by (\ref{Fprop}):
$\theta_{[3]}=A^3q^6=-A^3=(-A^2)\cdot A=\theta_{[2]}\theta_{[1]}$ at $q=i$, because the ribbon $\{(1,2),(1,3)\}$ of $[3]$ has contents
$1,2$ and $q^{2(1+2)}=q^6=q^2\cdot q^4$.

The difference from $q=1$ is that after step (a) we have $X_2$ instead of unity, and at step (b) only the crossings of strand $3$ with the band are reversed,
and this costs nothing.

\subsection{More examples \label{examples}}

Other cases of (\ref{RMfact}), i.e. other representations and the cubic root, are treated by the same three steps. The only difference is in step (a)
and in counting the factors $[m]$.
As before, $\simeq$ means the cyclic rule (\ref{cyc}), with the same three-strand moves as in step (a) above, and on four strands
also with $\CR_3\simeq\CR_1$, $\CR_2\CR_3\simeq\CR_1\CR_2$, $\CR_1\CR_3\simeq\CR_3\CR_1$, $\CR_2\CR_1\CR_3\simeq\CR_1\CR_2\CR_3\simeq\CR_1\CR_3\CR_2$
and $\CR_i\CR_{i+1}\CR_i\simeq\CR_i+z\CR_i\CR_{i+1}$ (each is a cyclic move followed by (\ref{skein}) or (\ref{YB})). With these moves
every projector reduces, exactly and for all $q$, to a combination of just a few words: $1$, $\CR_1$, $\CR_1\CR_2$, and on four strands
also $\CR_1\CR_3$ and $\CR_1\CR_2\CR_3$. Here $N_3:=q^3[2][3]\,e_{[3]}$ as in (\ref{N3colA}),
$N_{[1,1,1]}:=q^{-3}[2][3]\,e_{[1,1,1]}$ and $N_4:=q^6[2][3][4]\,e_{[4]}$ are the numerators in (\ref{projectors}):
\be
N_3&\simeq&1+q(2+q^2)\,\CR_1+q^3[2]\,\CR_1\CR_2,
\nn\\
N_{[1,1,1]}&\simeq& 1-q^{-1}(2+q^{-2})\,\CR_1+q^{-3}[2]\,\CR_1\CR_2,
\nn\\
{}[3]\,e_{\underline{[2,1]}}&\simeq&\frac{q^2+q^3\CR_1-q\CR_2-q^2\CR_1\CR_2}{q^2}\ \simeq\ 1+z\CR_1-\CR_1\CR_2\ \simeq\
\frac{q-\CR_1+q^2\CR_2-q\CR_1\CR_2}{q}\ \simeq\ [3]\,e_{\overline{[2,1]}},
\nn\\
N_4&\simeq&1+q(q^4+2q^2+3)\,\CR_1+q^4[3]\,\CR_1\CR_3+q^3[2](q^4+q^2+2)\,\CR_1\CR_2+q^6[2][3]\,\CR_1\CR_2\CR_3 .
\label{collapse}
\ee

Note that the two projectors for $[2,1]$, which look quite different in (\ref{projectors}), become
exactly the same inside a cable, for all $q$. Moreover, the factor $[2]$ disappears from the denominator, and the remaining $[3]$
is a simple pole, $[3]e_{[2,1]}\simeq$ regular, which cancels between the numerator and the denominator of (\ref{Hdef}) -- this is why
$H_{[2,1]}$ is finite at $[3]=0$, though $e_{[2,1]}$ and its trace are not. On the contrary, in $N_3$ and $N_4$ the factors $[2]$ and $[3]$ are still
there, in front of the terms which connect the strands into one component, and these terms will drop out.
Now consider three cases: $[2,1]$, $[1,1,1]$ and $[4]$ at $q=e^{i\pi/3}$, and $[4]$ at $q=i$.

\medskip
\noindent{\bf $[2,1]$ and $[1,1,1]$ at $q=e^{i\pi/3}$, $m=3$.} Here $[2]=1$, $[3]=0$, $q^3=-1$, $q^2=q-1$, and the coefficients in
(\ref{collapse}) become $q(2+q^2)=z=i\sqrt3$ and $q^{-1}(2+q^{-2})=-z$. Hence
\be
[3]e_{[3]}=\frac{N_3}{q^3[2]}\simeq-(1+z\CR_1-\CR_1\CR_2),\ \ \ \ \ \ \ \ \
[3]e_{[1,1,1]}=\frac{N_{[1,1,1]}}{q^{-3}[2]}\simeq-(1+z\CR_1-\CR_1\CR_2),
\nn\\
{}[3]e_{[2,1]}\simeq+(1+z\CR_1-\CR_1\CR_2).
\label{three}
\ee

On the other hand $X_3\simeq3\CR_1\CR_2-3z\CR_1+z^2$, and with $z^2=-3$ this is $-3(1+z\CR_1-\CR_1\CR_2)$. So $[3]e_{[3]}\simeq[3]e_{[1,1,1]}\simeq\frac13X_3$
and $[3]e_{[2,1]}\simeq-\frac13X_3$, and this is the second formula in (\ref{X23}). Each of the three representations is one $3$-ribbon
(of three different shapes) with empty core. There are no free strands, so that steps (b) and (c) are not needed, and
everything is done at step (a):
\be
H_{[2,1]}=\theta_{[2,1]}^{-(2n+1)}\frac{\tr_q^{(6)}\big([3]e_{[2,1]}\cdot B^{2n+1}\big)}{\tr_q^{(3)}[3]e_{[2,1]}}
\stackrel{(a)}{=}\theta_{[2,1]}^{-(2n+1)}\frac{-\tr_q^{(6)}\big([3]e_{[3]}\cdot B^{2n+1}\big)}{-\tr_q^{(3)}[3]e_{[3]}}
\stackrel{(\ref{Fprop})}{=}H_{[3]},
\nn\\
\theta_{[2,1]}=A^3q^{0},\ \ \ \ \theta_{[3]}=A^3q^{6},\ \ \ \ \theta_{[1,1,1]}=A^3q^{-6},
\ee
which coincide at $q^6=1$, because the contents of the three ribbons are $\{-1,0,1\}$, $\{0,1,2\}$ and $\{0,-1,-2\}$. The sign in (\ref{three}) cancels between numerator and denominator (step (a) is applied to the unknot as well,
with (\ref{Markov}) instead of (\ref{cyc})), and the factor $[3]$ cancels at the first step and does not appear anymore.

\medskip
\noindent{\bf $[4]$ at $q=e^{i\pi/3}$, $m=3$.} Now we have eight strands and four copies of the knot, and the block is $B_4$. Besides $[2]=1$, $[3]=0$ we have $[4]=-1$ and $q^6=1$, and in (\ref{collapse}) $q^4+2q^2+3=q^2+2$, $q^4+q^2+2=1$, so that
\be
[3]e_{[4]}=\frac{N_4}{q^6[2][4]}\ \simeq\ -\big(1+z\CR_1-\CR_1\CR_2\big)\ \simeq\ \tfrac13X_3\otimes1_1\ \simeq\ [3]e_{[3]}\otimes 1_1 .
\ee

Here we have one $3$-ribbon $\{(1,2),(1,3),(1,4)\}$ and the core $[1]$, i.e. strand $4$ becomes free and strands $1,2,3$ form the band $X_3$.
At step (b) in every block strand $4$ goes over the band (now it is strands $5,6,7$) in the first three crossings, $B_4=\CR_4\CR_5\CR_6\cdot\tilde B_4$,
and these crossings are reversed by (\ref{lem4}) with $m=3$. The analogue of (\ref{hecke2}) is again an identity in the braid algebra, and the traces of the extra
terms are proportional to $[3]^2$. Step (c) is the same as before, and finally
\be
H_{[4]}\stackrel{(a)}{=}\theta_{[4]}^{-(2n+1)}\frac{\tr_q^{(8)}\big(([3]e_{[3]}\otimes1)\cdot B_4^{2n+1}\big)}{\tr_q^{(3)}[3]e_{[3]}\cdot D}
\stackrel{(b),(c)}{=}
\nn\\
\stackrel{(b),(c)}{=}\theta_{[4]}^{-(2n+1)}\frac{\tr_q^{(6)}\big([3]e_{[3]}\cdot B_3^{2n+1}\big)}{\tr_q^{(3)}[3]e_{[3]}}\cdot\frac{\tr_q^{(2)}\CR^{2n+1}}{D}
\stackrel{(\ref{Fprop})}{=}H_{[3]}H_{[1]},\ \ \ \ \ \theta_{[4]}=A^4q^{12}=A^4=\theta_{[3]}\theta_{[1]} .
\ee

\medskip
\noindent{\bf $[4]$ at $q=i$, $m=2$.} Here $[2]=0$, $[3]=-1$ and $[4]=[2](q^2+q^{-2})=-2[2]$, so that $[2][3][4]=2[2]^2$ and $q^6=-1$. Now
$e_{[4]}$ has a double pole, and we multiply it by $[2]^2$. In (\ref{collapse}) $q(q^4+2q^2+3)=2i$, $q^4[3]=-1$, and the terms with
$[2]$ drop out:
\be
[2]^2e_{[4]}=\frac{N_4}{q^6[3][4]/[2]}=-\frac{N_4}{2}\ \simeq\ -\tfrac12\big(1+i(\CR_1+\CR_3)-\CR_1\CR_3\big)
\ \stackrel{(\ref{skein})}{=}\ \tfrac12(\CR_1-i)(\CR_3-i)=\tfrac12\,[2]e_{[2]}\otimes[2]e_{[2]}
\label{a4i}
\ee

This form is not unique, since $\CR_1\simeq\CR_3$, and we have chosen the one, symmetric in the two bands, with
$\frac{q(q^4+2q^2+3)}2(\CR_1+\CR_3)$ instead of the $\CR_1$-term in (\ref{collapse}). The factor $\frac12=\frac1{w!}$ is $\gamma_{[4]}$, and the sign is just the one needed for the product of two $[2]e_{[2]}$. Thus there are two
$2$-ribbons, $\{(1,3),(1,4)\}$ and $\{(1,1),(1,2)\}$, the core is empty, and we have two bands $X_2$, on strands $1,2$ and on strands $3,4$.
At step (b) in every block both strands $3$ and $4$ go over the first band (strands $5,6$) as a whole, this gives four crossings,
$B_4=(\CR_4\CR_3)(\CR_5\CR_4)\cdots$, and we pass the two strands below the band by (\ref{lem4}) one by one,
the other one being considered as a part of the ``fixed exterior''. The crossings of strands $3,4$ with $7,8$ (inside the second
band) and of $1,2$ with $5,6$ (inside the first one) are not touched. At step (c) the second band, which is now below
the first one everywhere, is pulled out by (\ref{YB}), and the trace factorizes into two four-strand traces:
\be
H_{[4]}\stackrel{(a)}{=}\theta_{[4]}^{-(2n+1)}\frac{\frac12\tr_q^{(8)}\big(([2]e_{[2]}\otimes[2]e_{[2]})\cdot B_4^{2n+1}\big)}{\frac12\big(\tr_q^{(2)}[2]e_{[2]}\big)^2}
\stackrel{(b),(c)}{=}
\nn\\
\stackrel{(b),(c)}{=}\theta_{[4]}^{-(2n+1)}\Bigg(\frac{\tr_q^{(4)}\big([2]e_{[2]}\cdot B_2^{2n+1}\big)}{\tr_q^{(2)}[2]e_{[2]}}\Bigg)^2
\stackrel{(\ref{Fprop})}{=}H_{[2]}^2,\ \ \ \ \ \ \theta_{[4]}=A^4q^{12}=A^4=(A^2q^2)^2=\theta_{[2]}^2 .
\ee

The two factors $[2]$ cancel separately, one in each four-strand ratio. We see from (\ref{a4i}) that
at $[m]=0$ inside a cable the projector $e_{[wm]}$ is, up to a constant, the product of $w$ projectors $e_{[m]}$.

\section{The general case \label{proofs}}

The examples of sec.\ref{qzeta} are exact, but they are just examples.
The general statement follows from the fact that closures of boxes in the cable are symmetric functions:
closures of projectors are Schur functions, and the winding sums $X_m$ are power sums $P_m$.
Then step (A) is the Murnaghan-Nakayama rule, step (B) is the Adams operation $P_j\to P_{jm}$,
and step (C) is the same as at $q=1$.
The $m$-cores, which appear in this way, are well known in the modular representation theory of $S_n$
and of Hecke algebras at roots of unity.

Below we formulate this as a sequence of lemmas, so that the reader interested only in the result can skip the rest of this section.

\medskip

In what follows we prove (\ref{residue}) and (\ref{adams}) for arbitrary $R$ and $m$. We use the results of \cite{Mor02a,AM98,Ais97,Luk05,Zhu13,Zhu23,Mor02b,Mur37,Nak40,Mac95,JK81}. As in sec.\ref{not}, $\cl$ denotes the closure,
and equalities of closures are understood up to $\simeq$.

\medskip
\noindent{\bf Lemma 1 (closures are symmetric functions, \cite{AM98,Luk05}).} {\it The ring of closures of boxes on cabled strands
is isomorphic to the ring of symmetric functions, $h_k=\cl(e_{[k]})$ being the complete symmetric functions and
$Q_R=\cl(e_R)$ the Schur functions. In particular (\ref{hP}) and (\ref{frob}) hold, with the characters $\varphi_R$ of the
symmetric group $S_{|R|}$, for all $q$ at which $e_R$ is defined.} At $q=\zeta$ we use them only in the limit $q\to\zeta$,
as identities between rational functions of $q$, and only for elements which are regular at $q=\zeta$, see the remark after (\ref{residue}).

\noindent This is the theorem of Aiston-Morton and Lukac: the closure of the Young idempotent $e_R$ in the skein of the annulus
is the Schur function of the $h_k$, and the product of the skein -- juxtaposition of closed braids in the tube -- is the product of symmetric functions.
In the language of (\ref{charexp}),(\ref{cldef}): the characters of $Y\otimes Y'$ are given by the Littlewood-Richardson rule \cite{Mur37,Nak40,Mac95,JK81}, so that the ring of closures
is the representation ring of the Hecke algebras, which is the ring of symmetric functions. What we need from it is that the closed projectors are multiplied as usual, e.g.
$h_1^2=Q_{[2]}+Q_{[1,1]}$, $h_2h_1=Q_{[3]}+Q_{[2,1]}$, so that the Frobenius formula (\ref{frob}) can be used
as an identity between closures.

\medskip
\noindent{\bf Lemma 2 (the winding sum is the power sum, \cite{Mor02b}).} $\ \cl(X_m)=[m]\,P_m$ {\it for all $m$.}

\noindent For $m=2$ this is the identity $[2](2e_{[2]}-1)=\CR_1+\CR_1^{-1}$ in the Hecke algebra itself, and for $m=3$ it is
(\ref{X23}), obtained by the cyclic moves (\ref{cyc}). In general it is Morton's formula for the power sums in the skein of the
annulus: $P_m$ is $\frac1{[m]}$ times the sum of the closures of the $m$ braids in which one strand winds around the other
$m-1$, crossing $i$ of them positively and the rest negatively, $i=0,\ldots,m-1$ -- which is (\ref{Xmpre}). Hence the poles described in sec.\ref{not}: simple for $P_{km}$, none for $P_j$ with $m\nmid j$ (for $Q_R$ the pole is in general of order $w$, see Lemma 3, and for $h_k$ it is of order $\lfloor k/m\rfloor$). Also, in the power-sum basis $P_\lambda$, which is a basis of
the ring of symmetric functions over the field of coefficients, the order of the pole of an element at $[m]=0$ is at most the
maximal number of factors $P_{km}$ in the terms of its expansion, with equality when the terms with this maximal number do not cancel
(i.e. when their coefficients do not sum up to $O([m])$).

\medskip
\noindent{\bf Lemma 3 (what survives of the projector).} {\it For every $R$, at $q=\zeta$, i.e. when $q^2$ is a primitive $m$-th root of unity,}
\be
[m]^w\,\cl(e_R)\ \simeq\ \frac{\eta\,\nu_R}{m^ww!}\ \cl(X_m)^w\,\cl(e_{{\rm core}_m(R)}) ,
\label{lem3}
\ee
{\it where $w$ is the $m$-weight of $R$, $\nu_R$ the number of sequences of $w$ removable $m$-ribbons leading from $R$ to its core,
and $\eta=\pm1$ the common sign of these sequences. This is (\ref{residue}), with $\gamma_R=\eta\,\nu_R/(m^ww!)$.}

\noindent{\it Proof.} Expand $Q_R$ by (\ref{frob}). By Lemma 2 a term $\varphi_R(\lambda)P_\lambda/z_\lambda$ has a pole at
$[m]=0$ of order equal to the number of parts of $\lambda$ divisible by $m$. Write $\lambda=(m^k,\lambda')$ with $\lambda'$ having no
part equal to $m$. The Murnaghan-Nakayama rule \cite{Mur37,Nak40,Mac95,JK81} computes $\varphi_R(\lambda)$ by removing from $R$ ribbons of lengths
$\lambda_1,\lambda_2,\ldots$ in turn, each with the sign $(-1)^{{\rm height}-1}$, where the height of a ribbon is the number of rows it occupies (so that a horizontal ribbon has sign $+1$). Removing the $k$ ribbons of length $m$ first,
$\varphi_R(m^k,\lambda')=\sum_{S}\pm\varphi_S(\lambda')$, the sum over all diagrams $S$ reachable from $R$ by $k$ removals. For $k>w$ there
are none, since after $w$ removals one arrives at the core, which has no $m$-ribbon; and a core has no ribbon of length $km$
either, because a diagram is an $m$-core if and only if none of its hook lengths is divisible by $m$,\footnote{The hook length of a box
is the number of boxes to its right in the same row, plus the number of boxes below it in the same column, plus one; a removable
ribbon of length $\ell$ exists if and only if some hook length equals $\ell$. This is immediate on the
$m$-abacus \cite{Mur37,Nak40,Mac95,JK81}: a removable $km$-ribbon is a bead at some position $b$ with the position $b-km$ empty, and then some bead
among $b,b-m,\ldots,b-(k-1)m$ has an empty position $m$ below it, i.e. there is a removable $m$-ribbon.} so that
$\varphi_{{\rm core}}(\lambda')=0$ whenever $\lambda'$ has a part divisible by $m$. More generally, removing a ribbon of length $km$
lowers the $m$-weight by $k$, so $\varphi_R(\lambda)=0$ whenever the parts of $\lambda$ divisible by $m$ sum up to more than $wm$;
hence a non-vanishing term has at most $w$ such parts, and exactly $w$ only if all of them are equal to $m$. Hence the terms of maximal pole order $w$ are
exactly $\lambda=(m^w,\mu)$ with $\mu$ having no part divisible by $m$, and for them $S={\rm core}_m(R)$ for every sequence, with one
and the same sign \cite{Mur37,Nak40,Mac95,JK81}: $\varphi_R(m^w,\mu)=\eta\,\nu_R\varphi_{{\rm core}_m(R)}(\mu)$. Since $z_{(m^w,\mu)}=m^ww!\,z_\mu$,
\be
Q_R=\frac{\eta\,\nu_R}{m^ww!}\,P_m^w\sum_{\mu}\frac{\varphi_{{\rm core}_m(R)}(\mu)}{z_\mu}P_\mu\ +\ \ldots
=\frac{\eta\,\nu_R}{m^ww!}\,P_m^w\,Q_{{\rm core}_m(R)}+\ldots,\nn\\ \ldots=({\rm poles\ of\ order}\le w-1),
\ee
where the sum over $\mu$ could be extended to all $\mu\vdash|{\rm core}|$ because the missing terms vanish. Multiplying by $[m]^w$,
using $[m]P_m=\cl(X_m)$ and taking the limit $[m]\to0$, we get (\ref{lem3}), since the corrections are $O([m])$.
The same argument for the unknot gives the same constant, which therefore cancels in (\ref{Hdef}). $\Box$

\medskip
\noindent{\it A regular representative.} Lemma 3 is a statement about closures. For $e_R$ itself, as an element of the Hecke algebra,
the order of the pole at $[m]=0$ depends on the choice of the rank-one idempotent in the block: e.g. $e_{\underline{[2,1]}}$ of
(\ref{projectors}) has $[2]$ in the denominator, although $[2,1]$ is a $2$-core. A choice which is free of such accidents is
\be
F_R:=\frac{E_R}{d_R},\ \ \ \ \ \ \ \cl(F_R)=\cl(e_R)=Q_R,
\ee
where $E_R$ is the central idempotent of the $R$-th block (the sum of $d_R$ orthogonal rank-one ones, $d_R$ the number of standard
tableaux of shape $R$, so that $\chi_R(E_R)=d_R$). $F_R$ is not an idempotent, but $F_R\simeq e_R$, and the ratio (\ref{Hdef}) is the same
for $e_R$ and $F_R$. The denominator of $F_R$ is the product of quantum hook lengths $\prod_{(i,j)\in R}[h_{ij}]$ (the Schur element of the
block), and the number of hook lengths divisible by $m$ is exactly the $m$-weight $w$ \cite{Mur37,Nak40,Mac95,JK81}; hence $F_R$ has a pole of order $w$,
and $F_{\rm core}$ is regular at $q=\zeta$. In the proof of the Theorem below $e_R$ and $e_{\rm core}$ are replaced by $F_R$ and $F_{\rm core}$.

To summarize what is singular and what is not, here are the orders of the pole at $[m]=0$ of $\tr_qF_R=\tr_qe_R$ (the quantum dimension),
which by Lemma 3 are the orders of the pole of $Q_R$, for small diagrams; they are equal to the number of hook lengths divisible by $m$,
i.e. to $w$. The last column shows that the rank-one idempotents of (\ref{projectors}) may have extra poles:
\be
\begin{array}{c|c|cc|cc|c}
R & {\rm hooks} & w_{m=2} & {\rm core}_2 & w_{m=3} & {\rm core}_3 & {\rm denominator\ of\ } e_R {\rm\ in\ (\ref{projectors})}\\ \hline
{}[2] & 2,1 & 1 & \varnothing & 0 & [2] & [2]\\
{}[3] & 3,2,1 & 1 & [1] & 1 & \varnothing & [2][3]\\
{}[2,1] & 3,1,1 & 0 & [2,1] & 1 & \varnothing & [2][3]\ \ ({\rm extra}\ [2]\ {\rm at}\ m=2)\\
{}[2,2] & 3,2,2,1 & 2 & \varnothing & 1 & [1] & -\\
{}[3,1] & 4,2,1,1 & 2 & \varnothing & 0 & [3,1] & -\\
{}[4] & 4,3,2,1 & 2 & \varnothing & 1 & [1] & [2][3][4]
\end{array}
\ee

\medskip
To prove (\ref{adams}) we need a convenient picture of the neighbourhood of the band: the {\it cut tube}. Fig.\ref{fig:tubeknot} shows
how it arises from the cable, and Fig.\ref{fig:tube} defines the elements we will manipulate -- the open strands $e$, $a$, $a^m$, which
cross the tube winding $0$, $1$, $m$ times, and the insertions $L(Y)$, $R(Y)$ -- and shows what the Adams operation looks like there.

\medskip
\begin{figure}[!t]\centering
\begin{tikzpicture}[scale=.9]
\begin{scope}[scale=.85]
\newcommand{\clos}[3]{(#1,3.3) -- (#1,{3.3+#2}) arc (180:90:.3) -- ({1.5+#2},{3.6+#2}) arc (90:0:.3) -- ({1.8+#2},{3.3+#2})
   -- ({1.8+#2},{-.3-#2}) arc (0:-90:.3) -- ({#1+.3},{-.6-#2}) arc (-90:-180:.3) -- (#1,-.3)}
\foreach \x/\d in {0/.9,.45/.6}{\draw[wire] (\x,-.3)--(\x,3.3); \draw[wire] \clos{\x}{\d}{};}
\draw[wire] (1.2,-.3)--(1.2,.8); \draw[wire] \clos{1.2}{.3}{};
\draw[white,line width=7pt] (1.2,.8)--(-.5,1.2); \draw[wire] (1.2,.8)--(-.5,1.2); \draw[wire] (-.5,1.2)--(-.5,1.9);
\draw[wire] (-.5,1.9)--(1.2,2.7); \foreach \x in {0,.45}{\draw[white,line width=7pt] (\x,1.95)--(\x,2.7); \draw[wire] (\x,1.9)--(\x,2.75);}
\draw[wire] (1.2,2.7)--(1.2,3.3);
\draw[fill=white,draw=strand] (-.2,2.85) rectangle (.65,3.2); \node[font=\scriptsize] at (.22,3.02) {$X_2$};
\draw[accent,densely dashed,rounded corners=2pt] (-.7,.55) rectangle (1.0,1.45);
\node[font=\scriptsize,accent] at (-.95,.4) {(b)};
\node[font=\scriptsize] at (1.7,1.0) {$e$}; \node[font=\scriptsize] at (-.25,-.6) {band};
\node[lab,align=center] at (.8,-2.4) {(a) the cable: band $X_2$\\ and a strand $e$ linked with it};
\end{scope}
\node at (3.6,1.3) {$\longrightarrow$};
\begin{scope}[shift={(6.0,-.6)}]
\fill[strand!8] (-.8,0) -- (-.8,3.6) arc (180:360:.8 and .28) -- (.8,0) arc (0:-180:.8 and .28);
\draw[quiet] (-.8,0)--(-.8,3.6); \draw[quiet] (.8,0)--(.8,3.6);
\draw[quiet] (0,3.6) ellipse (.8 and .28);
\draw[quiet] (-.8,0) arc (180:360:.8 and .28); \draw[quiet,densely dashed] (-.8,0) arc (180:0:.8 and .28);
\foreach \x in {-.18,.18}{\draw[wire] (\x,-.26)--(\x,3.86);}
\node[font=\scriptsize,quiet] at (0,4.15) {band, along the knot};
\draw[wire] (-1.9,2.05)--(-.6,1.8);
\draw[white,line width=7pt] (-.6,1.8)--(.6,1.55); \draw[wire] (-.6,1.8)--(.6,1.55);
\draw[wire,densely dotted,arr] (.6,1.55)--(1.9,1.3);
\node[font=\scriptsize] at (-2.1,2.3) {$e$};
\node[lab,align=center] at (.3,-1.5) {(b) a piece of the tube around the band:\\ $e$ enters it, crosses the band and leaves};
\end{scope}
\node at (9.3,1.3) {$\longrightarrow$};
\begin{scope}[shift={(10.4,.3)}]
\draw[boundary] (0,0) rectangle (3,2.4);
\draw[quiet,line width=.6pt] (1.4,2.4)--(1.6,2.4); \draw[quiet,line width=.6pt] (1.4,0)--(1.6,0);
\draw[quiet,line width=.6pt] (1.4,2.33)--(1.6,2.33); \draw[quiet,line width=.6pt] (1.4,.07)--(1.6,.07);
\foreach \x in {1.35,1.65}{\draw[wire] (\x,0)--(\x,2.4);}
\draw[fill=white,draw=strand] (1.1,1.7) rectangle (1.9,2.1); \node[font=\scriptsize] at (1.5,1.9) {$X_2$};
\draw[white,line width=6pt] (0,1.0)--(3,1.0); \draw[wire,arr] (0,1.0)--(3,1.0);
\node[lab,align=center] at (1.5,-.7) {(c) the tube seen from above and cut open:\\ $L(X_2)$};
\node[font=\scriptsize,quiet] at (1.5,2.65) {top $=$ bottom $=$ the rest of the knot};
\end{scope}
\end{tikzpicture}
\caption{\footnotesize How the tube pictures arise. (a) The cable with the band $X_2$ and a strand $e$ crossing it (step (B)); the dashed
box is enlarged in (b). (b) A piece of the tubular neighbourhood of the band -- a solid torus along the knot: the band runs along its axis,
$e$ enters through the wall, crosses the band and leaves. (c) The same piece seen from above and cut open (Fig.\ref{fig:tube}): the band is
vertical, $e$ horizontal, oriented from the inner to the outer boundary; top and bottom edges are glued through the rest of the knot.}
\label{fig:tubeknot}
\end{figure}
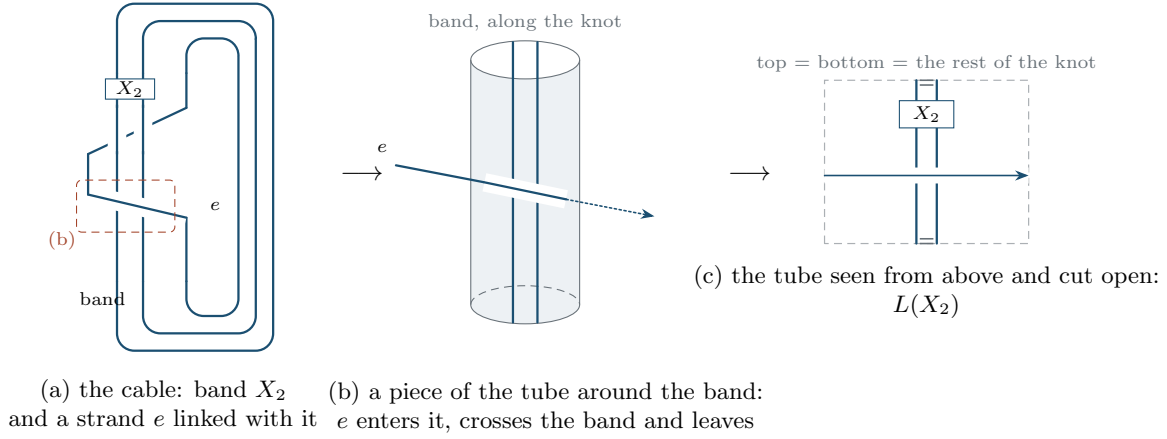
\begin{figure}[!t]\centering
\begin{tikzpicture}[scale=.8]
\newcommand{\tube}{\draw[boundary] (0,0) rectangle (3,2.4);
  \draw[quiet,line width=.6pt] (1.4,2.4)--(1.6,2.4); \draw[quiet,line width=.6pt] (1.4,0)--(1.6,0);
  \draw[quiet,line width=.6pt] (1.4,2.33)--(1.6,2.33); \draw[quiet,line width=.6pt] (1.4,.07)--(1.6,.07);}
\begin{scope}
\tube \draw[wire,arr] (0,1.2)--(3,1.2); \node[lab] at (1.5,-.45) {$e$};
\node[font=\scriptsize,quiet,rotate=90] at (-.25,1.2) {inner edge};
\node[font=\scriptsize,quiet,rotate=90] at (3.25,1.2) {outer edge};
\node[font=\scriptsize,quiet] at (1.5,2.65) {top $=$ bottom};
\end{scope}
\begin{scope}[shift={(4.2,0)}]
\tube \draw[wire] (0,.6) .. controls (.8,.8) and (1.3,1.8) .. (1.5,2.4); \draw[wire,arr] (1.5,0) .. controls (1.7,.6) and (2.3,1.4) .. (3,1.6);
\node[lab] at (1.5,-.45) {$a$};
\end{scope}
\begin{scope}[shift={(8.4,0)}]
\tube \draw[wire] (0,.5) .. controls (.5,.7) and (.8,1.8) .. (1,2.4); \draw[wire] (1,0) .. controls (1.2,.7) and (1.8,1.8) .. (2,2.4);
\draw[wire,arr] (2,0) .. controls (2.2,.6) and (2.6,1.4) .. (3,1.7);
\node[lab] at (1.5,-.45) {$a^2=\psi^2(a)$};
\end{scope}
\begin{scope}[shift={(12.6,0)}]
\tube \foreach \x in {1.2,1.5,1.8}{\draw[wire] (\x,0)--(\x,2.4);}
\draw[fill=white,draw=strand] (.95,1.7) rectangle (2.05,2.1); \node[font=\scriptsize] at (1.5,1.9) {$Y$};
\draw[white,line width=6pt] (0,1.0)--(3,1.0); \draw[wire,arr] (0,1.0)--(3,1.0);
\node[lab] at (1.5,-.45) {$L(Y)$: $e$ over $Y$};
\end{scope}
\begin{scope}[shift={(16.8,0)}]
\tube \draw[wire,arr] (0,1.0)--(3,1.0);
\foreach \x in {1.2,1.5,1.8}{\draw[white,line width=3.6pt] (\x,.6)--(\x,1.4); \draw[wire] (\x,0)--(\x,2.4);}
\draw[fill=white,draw=strand] (.95,1.7) rectangle (2.05,2.1); \node[font=\scriptsize] at (1.5,1.9) {$Y$};
\node[lab] at (1.5,-.45) {$R(Y)$: $e$ under $Y$};
\end{scope}
\begin{scope}[shift={(1.2,-4.3)}]
\tube \draw[wire] (1.5,0)--(1.5,2.4); \node[lab] at (1.5,-.45) {$P_1=h_1$};
\node at (3.4,1.2) {$\stackrel{\psi^2}{\longrightarrow}$};
\end{scope}
\begin{scope}[shift={(8.2,-4.3)}]
\tube \foreach \x in {1.2,1.8}{\draw[wire] (\x,0)--(\x,.8); \draw[wire] (\x,1.6)--(\x,2.4);}
\draw[wire] (1.8,.8)--(1.2,1.6); \draw[white,line width=3pt] (1.2,.8)--(1.8,1.6); \draw[wire] (1.2,.8)--(1.8,1.6);
\node[lab] at (1.5,-.45) {$\hat\CR_1$};
\node[anchor=east] at (-.1,1.2) {$P_2\ =\ \frac{1}{[2]}\Big($}; \node at (3.6,1.2) {$+$};
\end{scope}
\begin{scope}[shift={(12.4,-4.3)}]
\tube \foreach \x in {1.2,1.8}{\draw[wire] (\x,0)--(\x,.8); \draw[wire] (\x,1.6)--(\x,2.4);}
\draw[wire] (1.2,.8)--(1.8,1.6); \draw[white,line width=3pt] (1.8,.8)--(1.2,1.6); \draw[wire] (1.8,.8)--(1.2,1.6);
\node[lab] at (1.5,-.45) {$\hat\CR_1^{-1}$};
\node at (3.4,1.2) {$\Big)$};
\node[font=\small] at (-.6,-1.1) {each term is one strand winding twice around the tube};
\end{scope}
\end{tikzpicture}
\caption{\footnotesize The cut tube: the unknotted tube around a cabled strand, seen from above, is an annulus; it is cut along a radius
into a rectangle whose top and bottom edges are glued back (double ticks). Vertical is along the braid (bottom to top; the gluing is the
closure), horizontal is across the tube. Closed insertions, like the band $P_m$, run vertically. An open strand goes from the inner to
the outer edge: $e$ directly, $a$ with one winding (it crosses the glued edge once), $a^2=\psi^2(a)$ with two. $L(Y)$, $R(Y)$: the closed
insertion $Y$ below or above $e$; for $Y=P_m$ their difference is (\ref{adamsP}). Bottom row: $\psi^2(P_1)=P_2=\cl(X_2)/[2]$, the strand
winding once replaced by the strand winding twice, with both resolutions of the self-crossing.}
\label{fig:tube}
\end{figure}
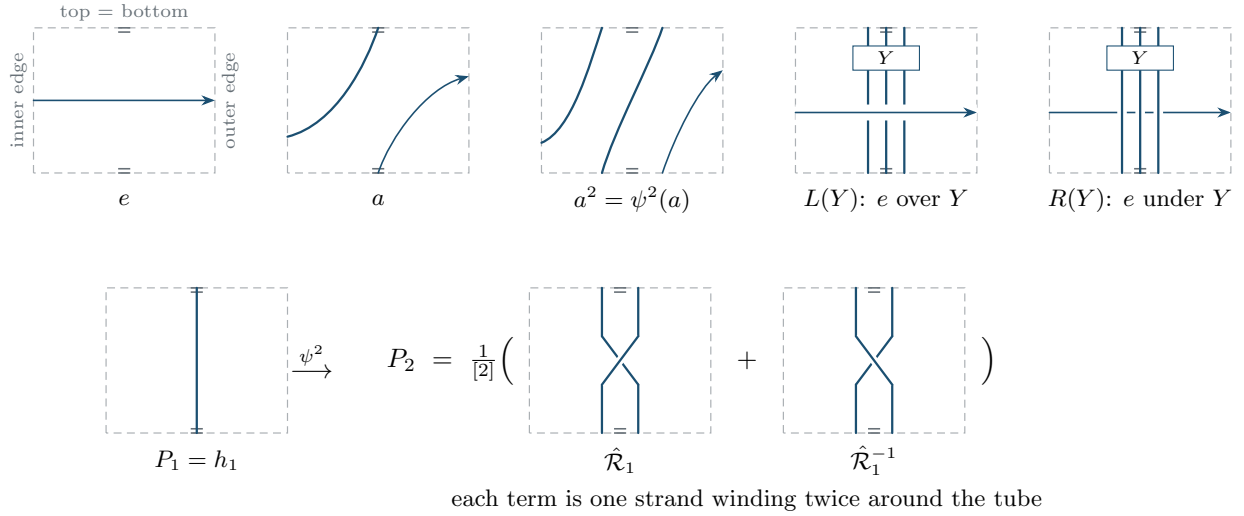
\noindent{\it The shifted wiring.} Diagrams in the cut tube with the same two endpoints are multiplied by joining them radially ($e$ is the unit);
$L$ and $R$ are multiplicative, and all these diagrams commute with each other. Indeed, $a^n$ is a spiral, which has no self-crossings,
so that $a^ia^j=a^{i+j}$; a closed insertion of type $L$ can be pushed into the lowest layer (and of type $R$ into the highest one),
where it slides radially below (above) everything else; and $R(Y)$ is expressed through $L$'s and $a$'s by the skein relation. The main tool is the following auxiliary construction.
For a box $B$ on $n+1$ strands let $\mathcal W_n(B)$ be its closure with the shifted
returns: upper endpoint $j$ is joined to lower endpoint $j+1$ around the tube ($1\le j\le n$), lower endpoint $1$ is the
inner boundary and upper endpoint $n+1$ the outer one (Fig.\ref{fig:wire}). With $B=1$ the strands form one open arc
winding $n$ times, $\mathcal W_n(1)=a^n$; a box on the first $n$ strands only adds one winding, $\mathcal W_n(B\otimes1_1)=a\,\mathcal W_{n-1}(B)$.
\begin{figure}[!t]\centering
\begin{tikzpicture}[scale=.8]
\begin{scope}
\draw[boundary] (-.5,-.4) rectangle (1.5,2.6);
\draw[wire] (0,-.4)--(0,2.6); \draw[wire] (1,-.4)--(1,2.6);
\node[font=\scriptsize,below] at (0,-.45) {in}; \node[font=\scriptsize,below] at (1,-.45) {$2$};
\node[font=\scriptsize,above] at (0,2.65) {$1$}; \node[font=\scriptsize,above] at (1,2.65) {out};
\node[lab] at (.5,-1.2) {$\mathcal W_1(1)=a$};
\end{scope}
\begin{scope}[shift={(4,0)}]
\draw[boundary] (-.5,-.4) rectangle (1.5,2.6);
\draw[wire] (0,-.4)--(0,.7); \draw[wire] (1,-.4)--(1,.7);
\draw[wire] (1,.7)--(0,1.5); \draw[white,line width=3pt] (0,.7)--(1,1.5); \draw[wire] (0,.7)--(1,1.5);
\draw[wire] (0,1.5)--(0,2.6); \draw[wire] (1,1.5)--(1,2.6);
\node[font=\scriptsize,below] at (0,-.45) {in}; \node[font=\scriptsize,below] at (1,-.45) {$2$};
\node[font=\scriptsize,above] at (0,2.65) {$1$}; \node[font=\scriptsize,above] at (1,2.65) {out};
\node[lab] at (.5,-1.2) {$\mathcal W_1(\CR_1)=L(h_1)$};
\end{scope}
\begin{scope}[shift={(8,0)}]
\draw[boundary] (-.5,-.4) rectangle (2.5,2.6);
\draw[wire] (0,-.4)--(0,2.6); \draw[wire] (1,-.4)--(1,2.6); \draw[wire] (2,-.4)--(2,2.6);
\draw[fill=strand!9,draw=strand] (-.3,.9) rectangle (2.3,1.7); \node at (1,1.3) {$B$};
\node[font=\scriptsize,below] at (0,-.45) {in}; \node[font=\scriptsize,below] at (1,-.45) {$2$}; \node[font=\scriptsize,below] at (2,-.45) {$3$};
\node[font=\scriptsize,above] at (0,2.65) {$1$}; \node[font=\scriptsize,above] at (1,2.65) {$2$}; \node[font=\scriptsize,above] at (2,2.65) {out};
\node[lab] at (1,-1.2) {$\mathcal W_2(B)$};
\end{scope}
\node[font=\footnotesize,align=center] at (5,-2.3) {upper $j$ is identified with lower $j+1$ (the return goes once around the tube);\\
in $\mathcal W_1(\CR_1)$ the return from upper $1$ to lower $2$ closes a loop which winds once and passes {\it below} the open arc};
\begin{scope}[shift={(13.5,1.1)}]
\fill[strand!8] (0,0) circle (1.45); \fill[white] (0,0) circle (.45);
\draw[quiet,line width=.4pt] (0,0) circle (1.45); \draw[quiet,line width=.4pt] (0,0) circle (.45);
\draw[wire] (0,0) circle (.95);
\draw[white,line width=5pt] (-1.45,0)--(-.45,0); \draw[wire,arr] (-1.45,0)--(-.45,0);
\node[lab,align=center] at (0,-2.0) {$\mathcal W_1(\CR_1)$ glued:\\ a loop below the arc $=L(h_1)$};
\end{scope}
\begin{scope}[shift={(17.6,1.1)}]
\fill[strand!8] (0,0) circle (1.45); \fill[white] (0,0) circle (.45);
\draw[quiet,line width=.4pt] (0,0) circle (1.45); \draw[quiet,line width=.4pt] (0,0) circle (.45);
\draw[wire] (0,0) circle (.8); \draw[wire] (0,0) circle (1.1);
\draw[fill=white,draw=strand] (-.3,.65) rectangle (.3,1.25); \node[font=\scriptsize] at (0,.95) {$e_{[2]}$};
\draw[white,line width=5pt] (-1.45,0)--(-.45,0); \draw[wire,arr] (-1.45,0)--(-.45,0);
\node[lab,align=center] at (0,-2.0) {$\mathcal W_2(\CR_1\CR_2e_{[2]})$ glued:\\ $=L(h_2)$};
\end{scope}
\end{tikzpicture}
\caption{\footnotesize The shifted wiring, and the result of the gluing for two terms: the returns close into loops around the tube,
which pass below the open arc.}
\label{fig:wire}
\end{figure}

\noindent{\it Example: $m=2$, i.e. (\ref{adams}) for $X_2$.} Before the general statement, let us see how (\ref{adams}) comes about in the
simplest case, where everything can be followed by hand.

\noindent{\it One step: $e_{[2]}$.} From $q[2]e_{[2]}=1+q\CR_1$ and Fig.\ref{fig:wire}, $\mathcal W_1(1)=a$ and
$\mathcal W_1(\CR_1)=L(h_1)$, where $h_1$ is a single closed strand winding once:
\be
[2]\,\mathcal W_1(e_{[2]})=q^{-1}a+L(h_1).
\label{w1}
\ee
The mirror gives $[2]\mathcal W_1(e_{[2]})=q\,a+R(h_1)$, and subtracting, $L(h_1)-R(h_1)=z\,a$, which is just the skein relation for one strand crossing
the open arc: a single strand is {\it not} transparent.

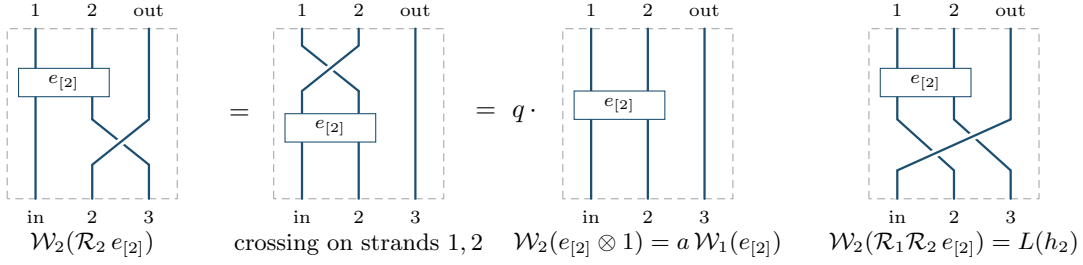
\begin{figure}[!t]\centering
\begin{tikzpicture}[scale=.75]
\newcommand{\frameW}{\draw[boundary] (-.5,-.4) rectangle (2.5,2.6);
 \node[font=\scriptsize,below] at (0,-.45) {in}; \node[font=\scriptsize,below] at (1,-.45) {$2$}; \node[font=\scriptsize,below] at (2,-.45) {$3$};
 \node[font=\scriptsize,above] at (0,2.65) {$1$}; \node[font=\scriptsize,above] at (1,2.65) {$2$}; \node[font=\scriptsize,above] at (2,2.65) {out};}
\newcommand{\ebox}[1]{\draw[fill=white,draw=strand] (-.3,#1) rectangle (1.3,#1+.5); \node[font=\scriptsize] at (.5,#1+.25) {$e_{[2]}$};}
\begin{scope}
\frameW
\draw[wire] (0,-.4)--(0,2.6); \draw[wire] (1,1.0)--(1,2.6); \draw[wire] (2,1.0)--(2,2.6);
\draw[wire] (2,-.4)--(2,.2)--(1,1.0); \draw[white,line width=3pt] (1,.2)--(2,1.0); \draw[wire] (1,-.4)--(1,.2)--(2,1.0);
\ebox{1.4}
\node[lab] at (1,-1.2) {$\mathcal W_2(\CR_2\,e_{[2]})$};
\end{scope}
\node at (3.6,1.1) {$=$};
\begin{scope}[shift={(4.7,0)}]
\frameW
\draw[wire] (0,-.4)--(0,1.5); \draw[wire] (1,-.4)--(1,1.5); \draw[wire] (2,-.4)--(2,2.6);
\ebox{.6}
\draw[wire] (1,1.5)--(0,2.3)--(0,2.6); \draw[white,line width=3pt] (0,1.5)--(1,2.3); \draw[wire] (0,1.5)--(1,2.3)--(1,2.6);
\node[lab] at (1,-1.2) {crossing on strands $1,2$};
\end{scope}
\node at (8.3,1.1) {$=\ q\,\cdot$};
\begin{scope}[shift={(9.8,0)}]
\frameW
\draw[wire] (0,-.4)--(0,2.6); \draw[wire] (1,-.4)--(1,2.6); \draw[wire] (2,-.4)--(2,2.6);
\ebox{1.0}
\node[lab] at (1,-1.2) {$\mathcal W_2(e_{[2]}\otimes1)=a\,\mathcal W_1(e_{[2]})$};
\end{scope}
\begin{scope}[shift={(15.2,0)}]
\frameW
\draw[wire] (1,-.4)--(1,.1)--(0,1.0)--(0,2.6); \draw[wire] (2,-.4)--(2,.1)--(1,1.0)--(1,2.6);
\draw[white,line width=3pt] (0,-.4)--(0,.1)--(2,1.0); \draw[wire] (0,-.4)--(0,.1)--(2,1.0)--(2,2.6);
\ebox{1.4}
\node[lab] at (1,-1.2) {$\mathcal W_2(\CR_1\CR_2\,e_{[2]})=L(h_2)$};
\end{scope}
\end{tikzpicture}
\caption{\footnotesize The step $n=2$ of the recursion (\ref{Wrec}). Left: the term $\CR_2e_{[2]}$ (crossing below the box); the crossing of strands $2,3$ is
pushed down through the returns (lower $2\equiv$ upper $1$, lower $3\equiv$ upper $2$) and reappears between strands $1,2$, next to $e_{[2]}$,
which absorbs it with eigenvalue $q$ (\ref{Pprop}); what remains is $\mathcal W_2(e_{[2]}\otimes1)=a\,\mathcal W_1(e_{[2]})$. Right: in the term
$\CR_1\CR_2e_{[2]}$ the strand from the inner boundary passes above the other two and exits, while they close, with $e_{[2]}$, into a
loop below it: $L(h_2)$.}
\label{fig:wstep}
\end{figure}

\noindent{\it Two steps: $e_{[3]}$.} From $q^2[3]e_{[3]}=(1+q\CR_2+q^2\CR_1\CR_2)\,e_{[2]}$ (the last strand sorted out of the
symmetrizer, see (\ref{projectors}) and (\ref{Pprop}); the crossings sit below the box). Under $\mathcal W_2$ (Fig.\ref{fig:wstep}): the term $1$ gives $a\,\mathcal W_1(e_{[2]})$; in the term $\CR_2$ the crossing between
strands $2,3$ can be pushed down through the returns (lower $2\equiv$ upper $1$, lower $3\equiv$ upper $2$) onto
strands $1,2$, where it meets $e_{[2]}$ and is absorbed with eigenvalue $q$, so this term gives $q\,a\,\mathcal W_1(e_{[2]})$;
in the term $\CR_1\CR_2$ the strand from the inner boundary passes across both others and exits, while strands $1,2$ with
$e_{[2]}$ close onto themselves beside the open arc: $L(h_2)$. Hence $q^2[3]\mathcal W_2(e_{[3]})=(1+q^2)a\mathcal W_1(e_{[2]})+q^2L(h_2)$, i.e.
\be
[3]\,\mathcal W_2(e_{[3]})=q^{-1}[2]\,a\,\mathcal W_1(e_{[2]})+L(h_2),\ \ \ \ \ \ {\rm and\ by\ the\ mirror}\ \ \ \ \
[3]\,\mathcal W_2(e_{[3]})=q\,[2]\,a\,\mathcal W_1(e_{[2]})+R(h_2).
\ee
Subtracting and using (\ref{w1}):
\be
L(h_2)-R(h_2)=z[2]\,a\,\mathcal W_1(e_{[2]})=z\,a\big(q^{-1}a+L(h_1)\big).
\label{dh2}
\ee
\noindent{\it Assembling $X_2$.} As insertions, $X_2=[2]P_2$ with $P_2=2h_2-h_1^2$ (this is $X_2=\CR_1+\CR_1^{-1}=2\CR_1-z$ and
$[2]e_{[2]}=q^{-1}+\CR_1$, $h_1^2=1_2$: indeed $[2](2e_{[2]}-1)=2\CR_1+2q^{-1}-[2]=2\CR_1-z$). Therefore, by multiplicativity,
\be
L(P_2)-R(P_2)&=&2\big(L(h_2)-R(h_2)\big)-\big(L(h_1)-R(h_1)\big)\big(L(h_1)+R(h_1)\big)
\nn\\
&=&2z\,a\big(q^{-1}a+L(h_1)\big)-z\,a\big(2L(h_1)-z\,a\big)=z\,(2q^{-1}+z)\,a^2=z[2]\,a^2,
\ee
\be
\boxed{\ L(X_2)-R(X_2)=z[2]^2\,a^2\ \stackrel{q=i}{=}\ 0\ }
\label{LR2}
\ee
The two terms $L(h_1)$, which are not small at $q=i$, cancel between $2(L(h_2)-R(h_2))$ and $(L(h_1)-R(h_1))(L(h_1)+R(h_1))$;
what remains is proportional to $[2]$ because $P_2$ contains $h_2=e_{[2]}$ with its denominator $[2]$, and to a second $[2]$
from $X_2=[2]P_2$. In the language of (\ref{hecke2}): the trace of $\frac z2X^{(45)}X^{(34)}$ along the closed band is
$z[2]^2a^2-\frac{z[2]^2}{2}\,\widehat{1}$, proportional to $[2]^2$, although the element itself is not.

For $m=3$ one more step gives in the same way $L(h_3)-R(h_3)=z\,a\big(q^{-2}a^2+q^{-1}aL(h_1)+L(h_2)\big)$, and with
$P_3=3h_3-3h_2h_1+h_1^3$ the same algebra yields $L(X_3)-R(X_3)=z[3]^2a^3$, which vanishes at $q=e^{i\pi/3}$ -- the step (b) of
sec.\ref{examples} for $[4]$ at this $q$; for $[4]$ at $q=i$ the layer is $X_2\otimes X_2$ and (\ref{LR2}) is applied twice, one band at a time.

\medskip

\noindent{\bf Lemma 4 (the passage relation).} {\it In the cut tube of Fig.\ref{fig:tube}, with one open arc from the inner to the outer
boundary and closed insertions placed below ($L$) or above ($R$) the arc,}
\be
L(P_m)-R(P_m)=\{q^m\}\,a^m\ \ \ \ \ {\rm for\ all}\ m,\ \ \ \ \ \ \ \ \ {\rm hence}\ \ \ \ \ L(X_m)-R(X_m)=\{q\}[m]^2\,a^m\ \stackrel{q^{2m}=1}{=}\ 0,
\label{lem4}
\ee
{\it where $a^m$ is the open arc winding $m$ times around the tube.}

\noindent{\it Proof.} Put $W_n:=[n]\,\mathcal W_{n-1}(e_{[n]})$,
$W_1=1$ (the radial arc). The row projector satisfies $q^n[n+1]e_{[n+1]}=\big(1+q\CR_n+q^2\CR_{n-1}\CR_n+\cdots+q^n\CR_1\cdots\CR_n\big)\,e_{[n]}$
(the general form of the decomposition used in the example for $n=2$). Let us apply $\mathcal W_n$ to each term. The term $1$ gives
$a\,\mathcal W_{n-1}(e_{[n]})$: the last strand adds one winding. In the term $q^k\CR_{n-k+1}\cdots\CR_n$ with $1\le k\le n-1$, the
$k$ crossings of the last strand with the block are pushed down through the returns onto the block itself, where each is absorbed
by $e_{[n]}$ with eigenvalue $q$ (\ref{Pprop}), giving $q^{2k}a\,\mathcal W_{n-1}(e_{[n]})$; and the term $q^n\CR_1\cdots\CR_n$ is the one in
which the strand from the inner boundary crosses the whole block and exits, while the block with $e_{[n]}$ closes onto itself
below the arc: $q^nL(h_n)$. (Fig.\ref{fig:wstep} shows the case $n=2$, see the example above.) Thus $q^n[n+1]\mathcal W_n(e_{[n+1]})=(1+q^2+\cdots+q^{2(n-1)})\,a\,\mathcal W_{n-1}(e_{[n]})+q^nL(h_n)$, i.e.
\be
W_{n+1}=q^{-1}\,a\,W_n+L(h_n)\ \ \ \Longrightarrow\ \ \ W_{n+1}=\sum_{j=0}^{n}q^{-j}a^{j}L(h_{n-j}),\ \ \ \ h_0=1 .
\label{Wrec}
\ee

Now let us change all crossings and replace $q\to q^{-1}$: $e_{[n]}$ is unchanged (it is characterized by $\CR_ie_{[n]}=qe_{[n]}$ and by
normalization; under the mirror this condition becomes $\CR_i^{-1}e_{[n]}=q^{-1}e_{[n]}$, which is the same condition), the wiring has no crossings, and ``below'' becomes ``above'': $W_{n+1}=q\,a\,W_n+R(h_n)$.
The derivation of (\ref{Wrec}) uses only (\ref{skein}), (\ref{Pprop}) and isotopies in the tube, no curls are removed, so $A$ does not enter
and need not be changed under the mirror.
Subtracting the two forms,
\be
L(h_n)-R(h_n)=\{q\}\,a\,W_n=\{q\}\,a\sum_{j=0}^{n-1}q^{-j}a^jL(h_{n-1-j}),
\label{dhn}
\ee
(for $n=1,2$ this is (\ref{w1}), (\ref{dh2}) above). To go to power sums, it is convenient to use generating functions. $L$ and $R$ are
multiplicative and commute with sums, so with $\HH(u)=\sum_nh_nu^n$ eq.(\ref{dhn}) reads
\be
L(\HH(u))-R(\HH(u))=\{q\}\,a\,u\sum_{j\ge0}(q^{-1}au)^j\cdot L(\HH(u))=\frac{\{q\}\,au}{1-q^{-1}au}\,L(\HH(u)),
\nn\\
{\rm i.e.}\ \ \ \ \
\frac{R(\HH(u))}{L(\HH(u))}=1-\frac{\{q\}au}{1-q^{-1}au}=\frac{1-q\,au}{1-q^{-1}au} .
\ee
Taking the logarithm and using (\ref{hP}), $\log\HH(u)=\sum_m P_mu^m/m$,
\be
\sum_{m\ge1}\frac{u^m}{m}\big(R(P_m)-L(P_m)\big)=\log(1-qau)-\log(1-q^{-1}au)=-\sum_{m\ge1}\frac{u^m}{m}\,(q^m-q^{-m})\,a^m,
\ee
which is the first equality of (\ref{lem4}); the second follows from Lemma 2, $\{q^m\}[m]=\{q\}[m]^2$. $\Box$

\noindent This is what we meant by the Adams operation in (\ref{adamsP}): the skein relation (\ref{skein1}) goes
to itself with $q\to q^m$ and $a\to a^m$, and (\ref{Wrec}) explains why the terms $L(h_j)$, which are not small at the root of unity, cancel for all $m$.

\medskip
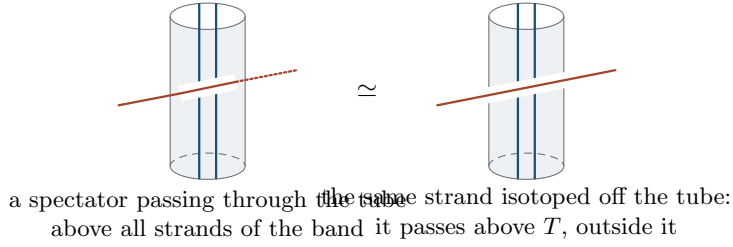
\begin{figure}[!t]\centering
\begin{tikzpicture}[scale=.62]
\newcommand{\cyl}{\fill[strand!8] (-.8,0) -- (-.8,3.2) arc (180:360:.8 and .28) -- (.8,0) arc (0:-180:.8 and .28);
  \draw[quiet] (-.8,0)--(-.8,3.2); \draw[quiet] (.8,0)--(.8,3.2); \draw[quiet] (0,3.2) ellipse (.8 and .28);
  \draw[quiet] (-.8,0) arc (180:360:.8 and .28); \draw[quiet,densely dashed] (-.8,0) arc (180:0:.8 and .28);
  \foreach \x in {-.18,.18}{\draw[wire] (\x,-.26)--(\x,3.46);}}
\begin{scope}
\cyl
\draw[wire,draw=accent] (-1.9,1.3)--(-.6,1.55);
\draw[white,line width=6pt] (-.6,1.55)--(.6,1.8); \draw[wire,draw=accent] (-.6,1.55)--(.6,1.8);
\draw[wire,draw=accent,densely dotted] (.6,1.8)--(1.9,2.05);
\node[lab,align=center] at (0,-1.0) {a spectator passing through the tube\\ above all strands of the band};
\end{scope}
\node at (3.4,1.6) {$\simeq$};
\begin{scope}[shift={(6.8,0)}]
\cyl
\draw[white,line width=6pt] (-1.9,1.3)--(1.9,2.05); \draw[wire,draw=accent] (-1.9,1.3)--(1.9,2.05);
\node[lab,align=center] at (0,-1.0) {the same strand isotoped off the tube:\\ it passes above $T$, outside it};
\end{scope}
\end{tikzpicture}
\caption{\footnotesize Lemma 5(i): a strand which crosses the band entirely above all its strands can be lifted above the tube $T$
around the band (Reidemeister II, III), after which it belongs to the exterior of $T$; similarly for a strand passing below.}
\label{fig:spectator}
\end{figure}
\noindent{\bf Lemma 5 (spectators and separation).} {\it (i) Relation (\ref{lem4}) holds inside the cable of any knot, for any strand
of the cable which crosses the closed band $X_m$ as a whole, with the rest of the cable as a fixed exterior. (ii) After every strand
not belonging to a given band has been passed below it, the cable is a conjugate, by (\ref{YB}) alone, of the product of the cable of
the band and the cable of the rest.}

\noindent{\it Proof.} (i) Let $\mathbb A$ be the annulus (the tube of Fig.\ref{fig:tube} before cutting), with one marked point on each
of the two boundary circles of $\mathbb A$ -- the inner and the outer one -- and let $T=\mathbb A\times I$ be its thickening, a solid
torus around the whole closed band. Relation (\ref{lem4}) is an identity in the skein module of $T$ relative to the two marked points,
i.e. between linear combinations of tangles in $T$ with these two endpoints; it is proved entirely inside $T$ (proof of Lemma 4). Skein
relations are local, so an identity in the skein of $T$ remains true after any embedding of $T$ into $S^3$, with any link diagram outside
$T$ attached to the two endpoints. We now explain how to choose $T$ so thin that it contains nothing but the band and the one segment of
$e$ at the chosen crossing, which enters through the inner and leaves through the outer boundary circle; all other segments of $e$,
including its other crossings with the band, will then belong to the exterior. At the place where $e$ crosses the band, the
other strands of the cable either run parallel to the band outside its tube, or cross the band as a whole, entirely above or entirely
below all its strands (this is the structure of a cable block, and it is preserved by the preceding steps (A) and (B): (A) only inserts
boxes on the band, (B) only reverses crossings of a whole strand with a whole band). A strand passing above all strands of the band
is isotopic, by Reidemeister II and III moves which do not touch $e$, to a strand lying above the tube $T$, i.e. outside it
(Fig.\ref{fig:spectator}); the same below. After this, a thin enough tube around the band contains only the band and the chosen segment of $e$, and (\ref{lem4}) applies. The isotopies used are identities
in the skein, so the trace is unchanged.

(ii) In a block $B_r$ all strands of the first cabled strand pass over all strands of the second. If in every block the strands of
a band pass over all other strands wherever the two meet, the band and the rest form two cables which lie one above the other.
Let $g$ be the braid which moves the strands of the band to the first positions and the other strands to the last ones, keeping the
order inside each group, with no other crossings than the necessary ones -- every crossing of $g$ is a crossing of a band strand with a
strand of the rest, the band strand passing over (the standard regrouping braid).
Then the whole cabled braid $S$ satisfies $S=g\,(S_{\rm band}\otimes S_{\rm rest})\,g^{-1}$, by the braid relations only -- this is the
argument of sec.\ref{bor}, where the two cables, one lying above the other, are pulled apart; the bottom of $g$ is the original order of
strands, its top is the grouped order. The insertion $Y=X_m^{\otimes w}\otimes e_{\rm core}$ sits at the bottom, on the first cabled strand;
it is not required to commute with $g$. Instead, $g$ can be written as a sequence of elementary moves, in each of which a whole group of
strands carrying one box (a band with its $X_m$, or the core strands with $e_{\rm core}$) passes over or under a whole group carrying
another box, i.e. as a product of cable crossings between the groups; the sliding rule (\ref{slide}) applies to every such crossing and
moves each box, as a whole, along with its group. Therefore $g^{-1}\,Y\,g=Y_{\rm band}\otimes Y_{\rm rest}$ -- the same boxes on the same
strands, now grouped. Hence, by cyclicity (\ref{xyclic}) and multiplicativity (\ref{multr}) of the trace,
$\tr_q\big(Y\,S\big)=\tr_q\big(g^{-1}Yg\,(S_{\rm band}\otimes S_{\rm rest})\big)=\tr_q\big(Y_{\rm band}S_{\rm band}\big)\,\tr_q\big(Y_{\rm rest}S_{\rm rest}\big)$. $\Box$

\medskip
\noindent{\bf Theorem.} {\it For every knot ${\cal K}$, every $R$ and every $q$ such that $q^2$ is a primitive $m$-th root of unity, (\ref{RMfact}) holds.}

\noindent{\it Proof.} Write $H_R$ as in (\ref{Hdef}), with the insertion $F_R\simeq e_R$ of sec.\ref{not} on the first cabled strand
(this does not change $H_R$), and multiply numerator and denominator by $[m]^w$. (A): by Lemma 3, and by (\ref{cyc}),
$[m]^wF_R\simeq\gamma_R\,X_m^{\otimes w}\otimes F_{\rm core}$; both sides are regular at $q=\zeta$, and the equality is an equality of
closures, i.e. of traces in the cable of any knot, at $q=\zeta$. So $[m]^wF_R$ may be replaced in both numerator and denominator by
the right hand side. The denominator does not vanish: with $C={\rm core}_m(R)$ and $D_C=\tr_qF_C$, the regular quantum dimension of the core,
multiplicativity (\ref{multr}) and $\tr_qX_m=\tr_q\cl(X_m)=[m]\{A^m\}/\{q^m\}=\{A^m\}/\{q\}$ give
$[m]^w\tr_qF_R\big|_{q=\zeta}=\gamma_R\big(\{A^m\}/\{\zeta\}\big)^wD_C(A,\zeta)\neq0$ for generic $A$ -- which also confirms that the pole
of $\tr_qF_R$ is of order exactly $w$. After the replacement the cable carries $w$ bands of $m$ strands, each with $X_m$, and a core of $|{\rm core}|$ strands
with $F_{\rm core}$;
$\gamma_R$ cancels. (B): take the first band. Every other strand of the cable -- of the core or of another band -- crosses it as a whole
in every block, above it or below it; by Lemma 5(i) and (\ref{lem4}) each of these crossings may be reversed at no cost, one strand at a
time, until the first band lies above everything else. (C): by Lemma 5(ii) the trace factorizes, (\ref{multr}), into the trace of the
$m$-strand cable with $X_m$ -- which is, up to the normalization fixed by the unknot and the framing factor, $H_{[m]}$ -- and the trace
of the rest. Repeat for the remaining $w-1$ bands. What remains of the rest is the cable with $F_{\rm core}\simeq e_{\rm core}$, i.e. $H_{\rm core}$. The
unknot normalizations match because the same steps apply to the unknot, and the framing factors match by (\ref{Fprop}):
$\theta_R=\theta_{[m]}^w\theta_{\rm core}$ at $q^{2m}=1$. Note that besides $[m]=0$ and $\{q^m\}=0$ we used only that $[j]\ne0$ for $j$ not divisible by $m$
(in Lemmas 2 and 3), i.e. that $q^2$ is a primitive root. $\Box$

\noindent{\it Remark.} If $q^{2m}=1$ but $q^2$ is not primitive, Lemma 3 fails (e.g. at $q=i$ the element $[4]e_{[4]}$ is still singular),
but (\ref{Rpm}) and (\ref{RMfact}) with this $m$ remain true: they follow from the Theorem for $m'={\rm ord}(q^2)$, which divides $m$,
because removing an $m$-ribbon does not change the $m'$-core, and $H_{[m]}=H_{[m']}^{m/m'}$ at such $q$.
Relation (\ref{sigmam}), on the contrary, requires $q^2$ to be primitive: at $q=1$ one has $H_{[m]}=\sigma^m\neq\sigma(A^m)$.

\medskip
\noindent We did {\it not} prove (\ref{sigmam}), i.e. that $H_{[m]}$ at $q=\zeta$ is the special polynomial at
$A^m$. Lemma 4 is about a strand crossing the band, while for (\ref{sigmam}) one needs the band crossing itself.
For the time being (\ref{sigmam}) remains a conjecture, supported by all the examples which we computed.

\section{Conclusion}

To summarize, we suggested the following explanation of factorization properties
for HOMFLY polynomials at the roots of unity:
At $q=1$ singular is the cable itself, $\sim z^{-r}$, and only the identity term of the projector survives. Mixed crossings can be changed,
since their smoothings lower the pole, and after that the copies are separated by ``blue over red'' and Yang-Baxter,
so that $H_R=H_{[1]}^{|R|}=\sigma(A)^{|R|}$. The reason is just one singularity, $D\sim z^{-1}$.

At $q=\zeta$, when $q^2$ is a primitive $m$-th root of unity, singular is the projector, because of $[m]=0$, and what survives of it is
$[m]^we_R\to X_m^w\otimes e_{\rm core}$ -- this follows from the Murnaghan-Nakayama rule. Other strands can be moved through the bands,
because $L(X_m)-R(X_m)\propto[m]^2=0$ -- this is the Adams operation. Then the copies are separated in the same way as at $q=1$,
and $H_R=H_{[m]}^{w}H_{\rm core}$. In all examples which we checked $H_{[m]}=\sigma(A^m)$.

Eq.(\ref{RMfact}) is true for reduced polynomials at generic $A$, where $m$ is the order of $q^2$.
For links with differently colored components our argument can not be applied directly, and this remains to be understood.
As to (\ref{sigmam}), it is checked in many examples, but still not proved.

\section*{Acknowledgements}

The AI models Claude Opus 5.5 and Fable 5.1 were used to write computer programs
for the numerical verification of the statements in this paper,
and to generate the figures.
All results were checked by the authors, who take full responsibility for the content.


{\small
\begin{thebibliography}{DMMSS13}

\bibitem[Ais97]{Ais97}
A.K.~Aiston,
{\it Adams operators and knot decorations},
preprint (1997), arXiv:q-alg/9711015.

\bibitem[AM98]{AM98}
A.K.~Aiston and H.R.~Morton,
{\it Idempotents of Hecke algebras of type A},
J.~Knot Theory Ramifications {\bf 7} (1998) 463--487, arXiv:q-alg/9702017.

\bibitem[AM14]{AM14}
A.~Anokhina and An.~Morozov,
{\it Cabling procedure for the colored HOMFLY polynomials},
Theor.~Math.~Phys. {\bf 178} (2014) 1--58, arXiv:1307.2216.

\bibitem[CS74]{CS74}
S.-S.~Chern and J.~Simons,
{\it Characteristic forms and geometric invariants},
Ann.~of Math. {\bf 99} (1974) 48--69.

\bibitem[DMMSS13]{DMMSS13}
P.~Dunin-Barkowski, A.~Mironov, A.~Morozov, A.~Sleptsov and A.~Smirnov,
{\it Superpolynomials for toric knots from evolution induced by cut-and-join operators},
JHEP {\bf 03} (2013) 021, arXiv:1106.4305.

\bibitem[JK81]{JK81}
G.~James and A.~Kerber,
{\it The representation theory of the symmetric group},
Addison-Wesley (1981), Sec.~2.7.

\bibitem[KMS25]{KMS25}
D.~Khudoteplov, A.~Morozov and A.~Sleptsov,
{\it Can Yang-Baxter imply Lie algebra?},
Phys.~Lett.~B {\bf 867} (2025) 139586, arXiv:2503.13437.

\bibitem[KM15]{KM15}
Ya.~Kononov and A.~Morozov,
{\it Factorization of colored knot polynomials at roots of unity},
Phys.~Lett.~B {\bf 747} (2015) 500--510, arXiv:1505.06170.

\bibitem[Luk05]{Luk05}
S.G.~Lukac,
{\it Idempotents of the Hecke algebra become Schur functions in the skein of the annulus},
Math.~Proc.~Cambridge Philos.~Soc. {\bf 138} (2005) 79--96, arXiv:math/0110119.

\bibitem[Mac95]{Mac95}
I.G.~Macdonald,
{\it Symmetric functions and Hall polynomials},
2nd ed., Oxford University Press (1995), Ch.~I.7, Ex.~11.

\bibitem[MMM16]{MMM16}
A.~Mironov, R.~Mkrtchyan and A.~Morozov,
{\it On universal knot polynomials},
JHEP {\bf 02} (2016) 078, arXiv:1510.05884.

\bibitem[MS10]{MS10}
A.~Morozov and A.~Smirnov,
{\it Chern-Simons theory in the temporal gauge and knot invariants through the universal quantum R-matrix},
Nucl.~Phys.~B {\bf 835} (2010) 284--313, arXiv:1001.2003.

\bibitem[Mor02a]{Mor02a}
H.R.~Morton,
{\it Skein theory and the Murphy operator},
J.~Knot Theory Ramifications {\bf 11} (2002) 475--492, arXiv:math/0102098.

\bibitem[Mor02b]{Mor02b}
H.R.~Morton,
{\it Power sums and Homfly skein theory},
Geom.~Topol.~Monogr. {\bf 4} (2002) 235--244, arXiv:math/0111101.

\bibitem[Mur37]{Mur37}
F.D.~Murnaghan,
{\it The characters of the symmetric group},
Amer.~J.~Math. {\bf 59} (1937) 739--753.

\bibitem[Nak40]{Nak40}
T.~Nakayama,
{\it On some modular properties of irreducible representations of a symmetric group, I, II},
Japan.~J.~Math. {\bf 17} (1940) 165--184, 411--423.

\bibitem[Vog99]{Vog99}
P.~Vogel,
{\it The universal Lie algebra},
preprint (1999), \url{https://webusers.imj-prg.fr/~pierre.vogel/grenoble-99b.pdf}.

\bibitem[Vog11]{Vog11}
P.~Vogel,
{\it Algebraic structures on modules of diagrams},
J.~Pure Appl.~Algebra {\bf 215} (2011) 1292--1339.

\bibitem[Wit89]{Wit89}
E.~Witten,
{\it Quantum field theory and the Jones polynomial},
Commun.~Math.~Phys. {\bf 121} (1989) 351--399.

\bibitem[Zhu13]{Zhu13}
S.~Zhu,
{\it Colored HOMFLY polynomials via skein theory},
JHEP {\bf 10} (2013) 229, arXiv:1206.5886.

\bibitem[Zhu23]{Zhu23}
S.~Zhu,
{\it New structures for colored HOMFLY-PT invariants},
Sci.~China Math. {\bf 66} (2023) 341--392, arXiv:2105.02037.

\end{thebibliography}
}
\end{document}